\documentclass[%
10pt,
 twocolumn,
nofootinbib,
 amsmath,amssymb,
 aps,
prb,
longbibliography
]{revtex4-2}
\usepackage[T1,T5]{fontenc}
\usepackage[utf8]{inputenc}
\usepackage[vietnamese,english]{babel}
\usepackage{natbib}
\usepackage{hyperref}
\hypersetup{
	colorlinks,%
	citecolor=blue,%
	filecolor=blue,%
	linkcolor=blue,%
	urlcolor=blue
}
\usepackage{graphicx}
\usepackage{svg}
\usepackage{bm}
\usepackage{mathtools}
\usepackage{amsthm}
\usepackage{listings}
\usepackage{epsfig}
\usepackage{amsfonts}
\usepackage{bbold}
\usepackage{url}
\usepackage{booktabs}
\usepackage{multirow}
\usepackage{footnote}
\usepackage[]{appendix}
\usepackage[makeroom]{cancel}
\makeatletter
\def\cantox@vector#1#2#3#4#5#6#7#8{%
  \dimen@.5\p@
  \setbox\z@\vbox{\boxmaxdepth.5\p@
   \hbox{\kern-1.2\p@\kern#1\dimen@$#7{#8}\m@th$}}%
  \ifx\canto@fil\hidewidth  \wd\z@\z@ \else \kern-#6\unitlength \fi
  \ooalign{%
    \canto@fil$\m@th \CancelColor
    \vcenter{\hbox{\dimen@#6\unitlength \kern\dimen@
      \multiply\dimen@#4\divide\dimen@#3 \vrule\@depth\dimen@\@width\z@
      \vector(#3_-#4){#5}%
    }}_{\raise-#2\dimen@\copy\z@\kern-\scriptspace}$%
    \canto@fil \cr
    \hfil \box\@tempboxa \kern\wd\z@ \hfil \cr}}
\def\bcancelto#1#2{\let\canto@vector\cantox@vector\cancelto{#1}{#2}}
\makeatother
\usepackage{soul} 

\begin{document}

\title{Heavy-Hole--Light-Hole Mixing and Spin--Photon Coupling in Germanium Flopping-Mode Qubits}

\author{Jose Reina-G{\'{a}}lvez}
\email{jose.reina-galvez@uni-konstanz.de}
\affiliation{Department of physics, University of Konstanz, D-78457, Konstanz, Germany}

\author{Guido Burkard}
\email{guido.burkard@uni-konstanz.de}
\affiliation{Department of physics, University of Konstanz, D-78457, Konstanz, Germany}

\begin{abstract}

Intrinsic spin--orbit interaction in germanium provides electrically active flopping-mode qubits without requiring magnetic-field gradients to generate spin--charge hybridization. We study double quantum dots (DQDs) with a multiband Luttinger--Kohn framework that retains heavy-hole (HH) and light-hole (LH) states on equal footing. Modeling vertical confinement with a finite confinement potential along the growth direction brings selected subbands into a regime of appreciable HH--LH mixing. This admixture activates electric-dipole spin-coupling through LH--LH and LH--HH transitions, complementing the conventional HH--HH pseudospin-flip resonance. For the parameter sets considered, these additional channels can produce larger spin--photon figures of merit than the HH--HH transition. HH--LH mixing thus supplies a tunable ingredient for qubit--cavity coupling in germanium hole-spin qubits and extends the operating regimes available to circuit-QED architectures.
\end{abstract}

\date{\today}

\maketitle

\section{Introduction}

Semiconductor spin qubits combine electrically defined nanostructures with spin states that can maintain coherence over long timescales, making them a well-established platform for quantum information processing~\cite{Loss_DiVincenzo_1998,Hanson_RMP_2007,Zwanenburg_RMP_2013,Burkard_2023}. Fast electrical manipulation and long-range connectivity, however, require coupling the spin degree of freedom to electric fields. Since a bare spin couples only weakly to such fields, electrical driving and coupling to microwave photons generally rely on spin--charge hybridization. Magnetic-field gradients, spin--orbit interaction, exchange coupling, and electric-dipole spin resonance provide different routes to generate this hybridization~\cite{Tokura_PRL_2006,Golovach_PRB_2006,PioroLadriere_NatPhys_2008,Medford_NatNano_2013,Reed_PRL_2016,Medford_prl_2013,Croot_PRR_2020_FloppingEDSR,Hajati_PRR_2025_Crosstalk}. The resulting charge admixture strengthens the electric response but also increases sensitivity to charge noise, making the balance between dipole strength and decoherence a central design principle~\cite{Benito_PRB_2019_FloppingNoise}.

Microwave cavities provide a means to extend qubit interactions over larger distances and to perform dispersive readout within circuit quantum electrodynamics~\cite{Childress_pra_2004,Wallraff_Nature_2004,Blais_RMP_2021}. Strong coupling between quantum dots and cavity electric fields was first established by exploiting charge degrees of freedom~\cite{Frey_PRL_2012,Petersson_Nature_2012,Viennot_Science_2015,Mi_Science_2017,Stockklauser_PRX_2017}. For spin qubits, an analogous interface requires the spin excitation to borrow sufficient charge character to interact with the cavity, which has motivated extensive theoretical and experimental work on spin--photon coupling~\cite{input_output_benito_prb,Samkharadze_Science_2018,Mi_Nature_2018_SpinPhoton,Landig_Nature_2018,Borjans_Nature_2020,HarveyCollard_PRX_2022}. A large electric dipole matrix element alone is therefore not sufficient to characterize the quality of a spin--photon transition; the effective linewidth generated by the same hybridization must also be taken into account~\cite{DAnjou_Burkard_PRB_2019_OptimalReadout,input_output_benito_prb}.

Hole spins in germanium are particularly suitable for exploring this balance because their valence-band spin--orbit interaction is intrinsic and strong. In addition, group-IV materials can exhibit reduced contact hyperfine coupling relative to III--V systems~\cite{Kloeffel_ARCMP_2013,Scappucci_NatRevMater_2021}. Strained Ge/SiGe heterostructures support high-mobility two-dimensional hole gases and high-quality gate-defined quantum dots~\cite{Sammak_AdvFunctMater_2019,Lodari_PRB_2019}, and have enabled rapid progress in single- and two-qubit operation~\cite{Hendrickx_Nature_2020,Hendrickx_Nature_2021,Wang2024}. Germanium double quantum dots are especially appealing for cavity coupling in the flopping-mode regime. In this configuration, interdot motion enhances the electric dipole moment, while spin--orbit interaction converts part of that orbital motion into spin--charge hybridization~\cite{Mutter_PRR_2021_Natural_HHs,input_output_benito_prb,Hajati_Burkard_PRB_2024_DynamicSweetSpot}.

Most theoretical descriptions of germanium hole flopping-mode qubits reduce the valence-band structure to an effective heavy-hole (HH) subspace. This approximation is controlled when light-hole (LH) states remain well separated from the active HH manifold. Confinement, strain, or device design can instead reduce the HH--LH separation, making explicit multiband effects relevant to the low-energy dynamics. Within the Luttinger--Kohn Hamiltonian, angular-momentum components with HH and LH character are directly coupled, while the standard Rashba interaction supplies additional electrically active spin--orbit channels~\cite{Winkler_2003_SpinOrbit,Bulaev_Loss_PRB_2007,Kloeffel_PRB_2013_CircuitQED,Loss_direct_rashba_2018,Adelsberger_Benito_PRB_2022}. HH--LH mixing can therefore alter the available dipole-active states and open coupling mechanisms that are absent from an HH-only treatment. The central question is whether LH admixture merely produces quantitative corrections or instead provides useful channels for coherent cavity coupling.

We study this multiband regime in a germanium double quantum dot using a multiband Luttinger--Kohn model that retains HH and LH states explicitly. A finite-well model describes the confinement along the growth direction, with bound states labeled $n=1,2,\ldots$, and we select the HH state with $n=2$ and the LH state with $n=1$ as the active pair. This choice produces sizable HH--LH mixing without requiring excited in-plane orbitals in the truncated basis. Projection onto the double-dot states yields an effective tunneling Hamiltonian in localized and bonding--antibonding representations. We couple the system to a microwave cavity through the length-gauge dipole operator and calculate its response using input--output theory with a dressed decoherence model. HH--HH, LH--LH, and LH--HH spin-flip transitions can then be assessed within a common framework through their spin--photon coupling and linewidth.

The paper is organized as follows. In Sec.~\ref{sec:theory}, we develop the multiband Luttinger--Kohn description, the finite-well treatment of vertical confinement, and the effective double-dot Hamiltonian, before introducing the cavity dipole interaction and the input--output formalism. Section~\ref{sec:Results} examines the spectrum and eigenstate composition and relates them to the cavity response of representative HH--HH, LH--LH, and LH--HH spin-flip transitions. We conclude by discussing how HH--LH mixing modifies the accessible regimes of spin--photon coupling in germanium.

\section{Theory}\label{sec:theory}

\subsection{Luttinger--Kohn Hamiltonian under a finite potential well}

\begin{figure}
    \centering
    \includegraphics[width=1.\linewidth]{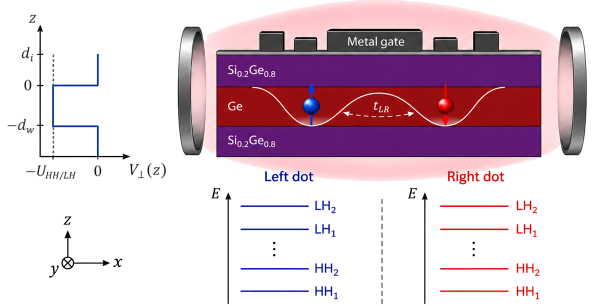}
    \caption{Schematic of the Ge double quantum dot considered in this work. Vertical confinement along the growth direction $z$ is described by the finite potential $V_\perp(z)$, with well depth $U_{HH/LH}$ and widths $d_w$ and $d_i$. The left and right localized hole states are indicated in blue and red, respectively, and are connected by the interdot tunneling processes collectively denoted by $t_{LR}$, including HH--HH, LH--LH, LH--HH, and spin--orbit-assisted channels. The energy ordering of the states is schematic and can vary in the actual simulations depending on the model parameters.}
    \label{fig:scheme_model}
\end{figure}

In this work, we describe a hole in a double quantum dot (DQD) vertically confined in a strained $\mathrm{Si_{0.2}Ge_{0.8}/Ge/Si_{0.2}Ge_{0.8}}$ planar heterostructure and laterally defined by the electrostatic potential of metallic gates. In the absence of a magnetic field, the full Hamiltonian is
\begin{equation}
H = H_{\mathrm{kin}} + V_{\parallel}(x,y) + V_{\perp}(z) + H_{\mathrm{SOI}},
\label{eq:Hamiltonian_T}
\end{equation}
where $H_{\mathrm{kin}}$ is the kinetic energy operator, $V_{\perp}(z)$ describes the vertical confinement, $V_{\parallel}(x,y)$ describes the in-plane confinement, and $H_{\mathrm{SOI}}$ is the standard Rashba spin--orbit interaction (SOI) for holes. 
A schematic overview of the reduced model is shown in Fig.~\ref{fig:scheme_model}: within the Ge layer, $V_{\parallel}(x,y)$ produces the lateral double-well profile defining the left and right dots, with minima centred at $x=\pm a$, while $V_{\perp}(z)$ describes the vertical confinement giving rise to the splitting between the HH and LH subbands. The left and right localized dot states are connected by interdot tunneling.

Since the quantum-dot dimensions considered here are much larger than the lattice spacing and the carrier density is low, the effective-mass approximation is applicable~\cite{Winkler_2003_SpinOrbit,Wang2024}. The valence-band kinetic energy is therefore described by the Luttinger--Kohn (LK) Hamiltonian, $H_{\mathrm{kin}}\equiv H_{\mathrm{LK}}$. In germanium, the split-off band is separated from the valence band by $\Delta_{\mathrm{SO}}=0.29\,\mathrm{eV}$ and does not contribute to the low-energy dynamics. We therefore restrict the description to the $4\times4$ subspace with total angular momentum $J=3/2$. In the basis
\[
|J,m_j\rangle\hspace{-0.035cm}=\hspace{-0.035cm}\{|3/2,3/2\rangle,|3/2,-3/2\rangle,|3/2,1/2\rangle,|3/2,-1/2\rangle\}\hspace{-0.02cm},
\]
the LK Hamiltonian reads
\begin{equation}
H_{\mathrm{LK}}=
\begin{pmatrix}
P+Q & 0 & S & R\\
0 & P+Q & R^\dagger & -S^\dagger\\
S^\dagger & R & P-Q & 0\\
R^\dagger & -S & 0 & P-Q
\end{pmatrix},
\label{eq:LK_H}
\end{equation}
with
\begin{align*}
P &= \frac{\hbar^2\gamma_1}{2m_0}(k_x^2+k_y^2+k_z^2),\quad
Q = \frac{\hbar^2\gamma_2}{2m_0}(k_x^2+k_y^2-2k_z^2),\\
R &= \sqrt{3}\frac{\hbar^2}{2m_0}\Big[-\gamma_2(k_x^2-k_y^2)+i\gamma_3(k_xk_y+k_yk_x)\Big],\\
S &= -\sqrt{3}\frac{\hbar^2}{2m_0}\gamma_3\Big[(k_x-i k_y)k_z+k_z(k_x-i k_y)\Big].
\end{align*}
Here $k_\epsilon=-i\partial_\epsilon$ ($\epsilon=x,y,z$), $m_0$ is the bare electron mass, and $\gamma_{1,2,3}$ are the standard Ge Luttinger parameters describing the bulk valence-band structure: $\gamma_1=13.38$, $\gamma_2=4.24$, and $\gamma_3=5.69$~\cite{Luttinger_Kohn_1955,Winkler_2003_SpinOrbit, Vurgaftman_JAP_2001,Wang2024}. The diagonal HH and LH sectors of the LK Hamiltonian define the in-plane and out-of-plane effective masses,
\[
m_{\parallel}^{HH/LH}=\frac{m_0}{\gamma_1\pm\gamma_2},\qquad m_{\perp}^{HH/LH}=\frac{m_0}{\gamma_1\mp 2\gamma_2},
\]
where the upper sign corresponds to heavy holes (HHs) and the lower sign to light holes (LHs).

Both the lateral and vertical confinement potentials $V_{\parallel}$ and $V_{\perp}$ have 4x4 diagonal structure with two pairs of identical diagonal elements.
Assuming lateral confinement in the form of a quartic double-well potential, we write
\begin{equation}
V_{\parallel}(x,y)=\frac{1}{2}m_{\parallel}^{HH/LH}\omega_{0,HH/LH}^2\left(\frac{(x^2-a^2)^2}{4a^2}+y^2 \right),
\label{eq:Vparallel}
\end{equation}
where $2a$ is the interdot separation and the strength of the harmonic potential is
\[
m_{\parallel}^{HH/LH}\omega_{0,HH/LH}^2=\frac{\hbar^2(\gamma_1+\gamma_2)}{m_0 a_0^4},
\]
with $a_0=50\,\mathrm{nm}$ taken as the characteristic lateral confinement length, which sets the size scale of the HH in-plane ground-state orbital. Due to the different effective mass of the LH in-plane orbital, its dot size is $42\,\mathrm{nm}$ instead. To construct a simple orbital basis, we first consider the diagonal part of Eq.~\eqref{eq:Hamiltonian_T}, for which the HH and LH sectors are decoupled. Near each minimum of the double-well potential, the in-plane confinement is approximated by a harmonic potential. This leads to Fock--Darwin-type states localized in each dot, following the flopping-mode treatment of Ref.~\cite{Mutter_PRR_2021_Natural_HHs},
\begin{equation}
\psi_{jl}(\rho,\varphi)=\sqrt{\frac{j!}{\pi(j+|l|)!}}\left(\frac{\rho}{b}\right)^{|l|}\frac{e^{il\varphi-\rho^2/2b^2}}{b}L_j^{|l|}\left(\frac{\rho^2}{b^2}\right),
\label{eq:FD_wave}
\end{equation}
with the length scale
\[
b=\sqrt{\frac{\hbar}{m_{\parallel}^{HH/LH}\omega_{0,HH/LH}}}.
\]
These states solve the lateral part of the diagonal Hamiltonian for the corresponding HH/LH effective masses. They are then used as basis functions onto which the full multiband Hamiltonian, including the off-diagonal LK and Rashba terms, is projected. In the Results section and in the following, we restrict the in-plane orbital sector to the ground state ($j=l=0$) in order to simplify the numerical calculations and isolate the main physical effects captured by the model.

The vertical confinement and strain are modeled by a finite potential well along the growth direction $z$. This accounts for the band offsets experienced by HH and LH states in a Ge quantum well embedded in SiGe~\cite{Wang2024}. Assuming uniaxial strain, we write
\begin{equation}
V_\perp(z)=
\begin{cases}
0, & 0<z<d_i,\\
-U_{HH/LH}, & -d_w<z<0,\\
0, & z<-d_w.
\end{cases}
\label{eq:perp_potential}
\end{equation}
With this convention, the Ge well region lies lower in energy by $U_{HH/LH}$ than the surrounding barriers. The total wavefunction is then written as the product of the in-plane envelope in Eq.~\eqref{eq:FD_wave} and the corresponding finite-well eigenfunction $\phi_n(z)$ along $z$, obtained from the effective-mass finite-well problem summarized in Appendix~\ref{app:potential_well}, namely
\begin{equation}
\Psi_n^{L/R}(x,y,z)=\psi_{00}^{HH/LH}(x\pm a,y)\phi_n^{HH/LH}(z),
\label{eq:total_wavefunction}
\end{equation}
for each HH and LH state, since both the lateral and vertical envelopes depend on the corresponding effective masses. We orthogonalize Eq.~\eqref{eq:total_wavefunction} to obtain the Wannier states
\begin{eqnarray}
&&|L/R,HH/LH\rangle =\phi_n^{HH/LH}(z)\sqrt{N_{HH/LH}}\times \\ &&
\left(\psi_{00}^{HH/LH}(x\pm a,y)-\gamma_{HH/LH}\psi_{00}^{HH/LH}(x\mp a,y)\right),\nonumber
\label{eq:wavefunction_wannier}
\end{eqnarray}
where the normalization factor depends on the HH and LH sector and is given by
\[
N_{HH/LH}=\left(1+\gamma^2_{HH/LH}-2\gamma_{HH/LH}S_{HH/LH}\right)^{-1},
\]
with
\[
\gamma_{HH/LH}=\left.\left(1-\sqrt{1-S_{HH/LH}^2}\right)\right/S_{HH/LH},
\]
and with the overlap
\begin{eqnarray}
S_{HH/LH}\hspace{-.13cm}&=&\hspace{-.18cm}\int\hspace{-.13cm} dx dy \hspace{-.1cm}\left[\psi_{00}^{HH/LH}(x+ a,y)\right]^*\hspace{-.1cm} \psi_{00}^{HH/LH}(x- a,y)   \nonumber \\ 
&=&\exp\!\left[-\frac{m_0a^2\omega_{0,HH/LH}}{\hbar (\gamma_1 \pm \gamma_2)}\right].
\end{eqnarray}

Returning to Eq.~\eqref{eq:perp_potential}, throughout this work we take $U_{HH}=100~\mathrm{meV}$, $U_{LH}=217~\mathrm{meV}$ or $232~\mathrm{meV}$, and $d_w=d_i/3=18~\mathrm{nm}$. In contrast to the Luttinger parameters $\gamma_{1,2,3}$, the quantities $U_{HH}$ and $U_{LH}$ are not treated as independent microscopic band offsets. Instead, they are effective confinement depths within the reduced finite-well model, introduced to parametrize the net vertical subband alignment arising from band offset, strain, and electrostatics~\cite{Wang2024}. The two values of $U_{LH}$ are used to modify the relative HH--LH subband alignment and thereby tune the character of the low-energy eigenstates. The purpose of this choice is not to identify a unique device realization of these confinement depths, but rather to isolate the physical regime in which LH admixture becomes appreciable and can be tested as a resource for enhancing the spin--photon figure of merit.

Finally, the standard Rashba SOI term is \cite{Loss_direct_rashba_2018}
\[
H_{\mathrm{SOI}}=\alpha_R \langle E_z\rangle (k_x J_y-k_y J_x),
\]
where $J_{x,y}$ are the $x$ and $y$ components of the spin-$3/2$ operator in the LK basis. Its matrix representation is
\begin{eqnarray*}
H_{\mathrm{SOI}} &=& i\frac{\alpha_{R} \langle E_z\rangle}{2}
\begin{pmatrix}
0 & 0 & -\sqrt{3}k_- & 0 \\
0 & 0 & 0 & \sqrt{3}k_+ \\
\sqrt{3}k_+ & 0 & 0 & -2k_- \\
0 & -\sqrt{3} k_- & 2k_+ & 0
\end{pmatrix},
\end{eqnarray*}
where $k_{\pm}=k_x\pm i k_y$, $\alpha_R$ is the Rashba coefficient, and $\langle E_z\rangle$ is the average electric field induced by the structural inversion asymmetry associated with the out-of-plane confinement. The prefactor of the Rashba SOI term is fixed at $\alpha_R\langle E_z\rangle=12~\mathrm{meV\,nm}$, which we use as a representative upper-bound estimate for the standard Rashba-type linear-in-momentum coupling in Ge hole systems. This operator acts on the same pairs of states as the $S$ term in Eq.~\eqref{eq:LK_H}, but without involving $k_z$, and it also directly couples LH states with opposite pseudospin. Consequently, transitions between the $\pm 3/2$ HH states arise only indirectly, either through strong HH--LH mixing or, when the LH states are energetically well separated from the HH manifold, through higher-order spin--orbit terms generated by downfolding the LH sector onto the HH subspace~\cite{Bulaev_Loss_PRB_2007,Winkler_2003_SpinOrbit,Fabrizio_SOI_GaAs_HH_prl_2014}. This reflects the fact that $H_{\mathrm{SOI}}$ does not directly couple the $\pm 3/2$ sector at lowest order.

We do not include atomistic interface-induced HH--LH transition-dipole terms, which lie beyond the present smooth finite-well envelope-function model. Such terms can arise from inversion-symmetry breaking at sharp heterostructure interfaces and, after projection onto a HH-like subspace, may appear as effective linear-in-momentum spin--orbit interactions with Dresselhaus-like symmetry~\cite{Philippopoulos_PRB_2020,Luo_PRL_2010,RodriguezMena_PRB_2023}. They may therefore quantitatively modify the coupling strengths in real Ge/GeSi devices, but are not expected to change the main physical message of the present reduced model: isolating the role of explicit HH--LH admixture in the spin--photon figure of merit.

Because we are interested in a DQD, the total Hamiltonian is written in terms of left ($L$) and right ($R$) localized contributions together with the interdot coupling block,
\begin{equation}
\tilde{H}=\begin{pmatrix}
{H}_{L} & {H}_{\rm LR} \\
{H}_{\rm LR}^{\dagger} & {H}_{R}
\end{pmatrix}.
\label{eq:H_total}
\end{equation}
Each block in Eq.~\eqref{eq:H_total} contains both diagonal and off-diagonal interdot matrix elements, and its dimension depends on the number of HH and LH states retained in the truncated Hilbert space, as determined by the bound states of the finite well along $z$. For instance, the minimal basis that includes HH--LH coupling and both pseudospin projections for each state leads to $8\times8$ matrices $H_\xi$ with $\xi=L,R,LR$.

The HH--LH pairs are ordered according to the finite-well quantum number $n$ in each sector, as illustrated schematically in Fig.~\ref{fig:scheme_model}. We find that the coupling depends strongly on the parity of the selected HH and LH states: pairs with opposite parity exhibit qualitatively different behavior from those with the same parity. Same-parity pairs can also be brought close in energy in principle, but in the parameter regimes explored here, their spin-flip matrix elements are typically weaker, because the dominant $S$-type HH--LH tunneling channels are suppressed and the coupling relies mainly on smaller $R$-type matrix elements. We therefore focus on an opposite-parity pair with sizable HH--LH mixing, namely the HH state with $n=2$ and the LH state with $n=1$, while the HH $n=1$ state remains lower in energy. The lower HH $n=1$ level is assumed to be filled and inert, analogous to the standard shell-filling picture in multihole quantum dots, where lower-lying orbitals form filled shells and the active spin degree of freedom is carried by the outermost hole~\cite{Lawrie_NanoLett_2020,Liles_NatCommun_2018, John_NatCommun_2025_10SpinGe}. 


To suppress the involvement of excited in-plane orbitals, the characteristic energy scales within the retained HH--LH manifold should remain smaller than the corresponding in-plane orbital excitation energies. These scales include the HH--LH level separation, the tunneling and hybridization matrix elements, and the Zeeman splitting. In the low-magnetic-field regime, the Zeeman contribution is small, while the relevant tunneling and HH--LH hybridization amplitudes remain below approximately $40~\mu\mathrm{eV}$. The dominant energy scale controlling the separation of the retained HH and LH sectors is therefore their effective gap. Owing to the reparametrization of the two sectors in terms of a common detuning parameter, this separation is not given by the bare vertical splitting $\Delta_{HL}$ alone, but by $\Delta_{HL}^{\mathrm{eff}}=\Delta_{HL}+\delta\epsilon_{HL}$, where $\delta\epsilon_{HL}$ denotes the in-plane energy difference introduced by the reparametrization; see Appendix~\ref{app:LK_elements}. We therefore require $|\Delta_{HL}^{\mathrm{eff}}|<\min\!\left(\hbar\omega_{0,\mathrm{LH}},\hbar\omega_{0,\mathrm{HH}}\right)$, which is satisfied for the parameter sets considered here; see Appendix~\ref{app:potential_well}.

\subsection{Introducing magnetic field}

Equation~\eqref{eq:Hamiltonian_T} describes the system in the absence of a magnetic field. In the presence of a magnetic field, the DQD Hamiltonian acquires a Zeeman term that describes the coupling between the hole spin and the external field,
\begin{eqnarray}
H_{Z}&=&2\mu_B \bar{\kappa} \boldsymbol{J}\cdot \mathbf{B}+2\mu_B q \left(J_x^3 B_x+J_y^3 B_y+J_z^3 B_z \right)\nonumber \\
&=&2\mu_B \bar{\kappa} J_z B_z+2\mu_B q J_z^3 B_z,
\label{eq:Zee}
\end{eqnarray}
where $\bar{\kappa}=3.41$, $q=0.067$, $\mu_B=e\hbar/2m_0$ is the Bohr magneton, $\mathbf{B}$ is the magnetic field, and $\boldsymbol{J}$ is the spin-$3/2$ vector operator. In this work, we retain only the $z$ component of the Zeeman interaction, motivated by the strong out-of-plane anisotropy in Ge, although in-plane fields can also be useful for identifying sweet spots, i.e., operating points that are less sensitive to charge noise~\cite{Wang2024}. Therefore, Eq.~\eqref{eq:Zee} acts diagonally in the $L/R$ basis.

The magnetic field also modifies the LK operators in Eq.~\eqref{eq:LK_H} through the generalized momentum substitution $\hbar \mathbf{k}\rightarrow \hbar \mathbf{k}+e\mathbf{A}$, which leads to magnetic-field-dependent matrix elements in the localized basis, as shown in Appendix~\ref{app:LK_elements}. Following Ref.~\cite{Mutter_PRR_2021_Natural_HHs}, this leads to left- and right-localized in-plane wavefunctions that acquire a magnetic phase,
\begin{equation}
\Psi_n^{L/R}(x,y,z)=e^{i\,y\frac{(1-2c_0)x\pm a}{2l_B^2}}
e^{i\frac{c_xx+c_yy}{l_B}}
\psi_{00}(x\pm a,y)\phi_n(z),
\label{eq:wavefunction_pm_phase}
\end{equation}
for each HH and LH state. Equation~\eqref{eq:wavefunction_pm_phase} must therefore be used instead of Eq.~\eqref{eq:total_wavefunction}. The corresponding Wannier states are then constructed as in Eq.~\eqref{eq:wavefunction_wannier}, while Eq.~\eqref{eq:FD_wave} is updated through the magnetic-field-dependent Fock--Darwin length,
\[
b^2=\frac{\hbar}{m_{\parallel}^{HH/LH}\sqrt{\omega_{0,HH/LH}^2+\omega_{L,HH/LH}^2}},
\]
where $\omega_{L,HH/LH}=eB/2m_{\parallel}^{HH/LH}$ denotes the orbital Larmor frequency and $l_B=\sqrt{\hbar/eB}$ is the magnetic length. Since observable quantities are gauge invariant, we set $c_x=0$, $c_y=0$, and $c_0=1/2$ throughout the remainder of this work, following Ref.~\cite{Mutter_PRR_2021_Natural_HHs}. It is important to note that the Wannier states also include total angular-momentum character,
\[
|L/R,HH/LH,J,m_j\rangle=|L/R,HH/LH\rangle |J,m_j\rangle,
\]
which is already incorporated in the structure of the LK Hamiltonian, the Rashba SOI term, and the Zeeman interaction.

\subsection{Hamiltonian in the orbital ground-state basis}

Restricting the description to the \(8\times 8\) subspace spanned by the left (\(L\)) and right (\(R\)) orbital ground states for a selected HH--LH pair, as obtained from Eq.~\eqref{eq:wavefunction_wannier} including the magnetic phase of Eq.~\eqref{eq:wavefunction_pm_phase} and the Zeeman contribution, the DQD Hamiltonian can be written as
\begin{equation}
H_T=H_{0}+\mathcal{V}+\mathcal{V}^\dagger ,
\label{eq:HT}
\end{equation}
where the diagonal part reads
\begin{eqnarray}
H_{0}
&=&\tau_0\otimes\Big[H_Z+(\delta\epsilon_{HL}+\epsilon^z_{HH})P_{HH}
+\epsilon^z_{LH} P_{LH}\Big]\nonumber\\
&&+\frac{\epsilon}{2}\,\tau_z\otimes \mathcal{I}_4
= \tau_0\otimes H_{\rm loc} +\frac{\epsilon}{2}\,\tau_z\otimes \mathcal{I}_4.
\label{eq:diag_LR_basis}
\end{eqnarray}
Here, \(\tau_{i=x,y,z}\) are Pauli matrices acting in the \((L,R)\) dot subspace, \(\epsilon\) is the detuning parameter, and \(\delta\epsilon_{HL}\) denotes the additional HH--LH energy offset introduced by the reparameterization discussed above, acting only in the HH sector. The quantities \(\epsilon^z_{HH}\) and \(\epsilon^z_{LH}\) represent the vertical confinement energies of the selected HH and LH states, respectively, as determined by the finite-well problem along the growth direction. The matrices \(\mathcal{I}_4\) and \(\tau_0\) denote the identity operators in the \(4\times4\) LK basis and the \(2\times2\) dot subspace, respectively. We neglect off-diagonal terms in \(H_0\), since their contribution to the dynamics is negligible compared with the interdot tunneling terms.

The term which collects the interdot tunnelings is written as
\begin{align}
\mathcal{V}={}&
\tau_+\otimes
\Big[
t_{HH}P_{HH}+t_{LH}P_{LH}
+(t_{S_+}+t_{SOI_+})S_+
\notag\\
&+(t_{S_-}-t_{SOI_-})S_-
+t_{SOI,LH}D +t_{R_+}R_+\notag\\
&+t_{R_-}R_-\Big]
=\tau_+\otimes \mathcal{T}=\tau_+\otimes (\mathcal{T}_H+\mathcal{T}_A),
\label{eq:off_LR_basis}
\end{align}
with \(\tau_\pm=(\tau_x\pm i\tau_y)/2\), where \(\mathcal{T}_H\) is the Hermitian part of the operator appearing on the right-hand side of the tensor product,
$$
\mathcal{T}_H=t_{HH}P_{HH}+t_{LH}P_{LH}
+t_{R_+}R_+ +t_{R_-}R_-,
$$
while \(\mathcal{T}_A\) is the anti-Hermitian part,
$$
\mathcal{T}_A=(t_{S_+}+t_{SOI_+})S_+ +(t_{S_-}-t_{SOI_-})S_- +t_{SOI,LH}D.
$$
We also define the projectors onto the heavy-hole and light-hole sectors in the LK subspace as
\begin{equation*}
P_{HH}=\frac{1}{2}\left(J_z^2-\frac{1}{4}\mathcal{I}_4\right),
\qquad
P_{LH}=\mathcal{I}_4-P_{HH}.
\end{equation*}
In addition, we introduce the non-Hermitian transition operators
\begin{align*}
S_\pm &=
\frac{1}{\sqrt{3}}
\left(
P_{HH}J_\pm P_{LH}-P_{LH}J_\mp P_{HH}
\right), \\
R_\pm &=
\frac{1}{2\sqrt{3}}
\left(
P_{HH}J_\pm^2P_{LH}+P_{LH}J_\mp^2 P_{HH}
\right), \\
D &=
\frac{1}{2}
\left(
P_{LH}J_+P_{LH}-P_{LH}J_-P_{LH}
\right),
\end{align*}
with \(J_\pm=J_x\pm iJ_y\). The local term \(H_0\) contains the Zeeman splitting, the vertical confinement energies, the additional HH--LH offset \(\delta\epsilon_{HL}\), and the detuning \(\epsilon\). By contrast, \(\mathcal V\) collects the interdot hybridization processes, including the spin-conserving tunneling amplitudes \(t_{HH}\) and \(t_{LH}\), the HH--LH mixing channels \(t_{S_\pm}\) and \(t_{R_\pm}\), and the associated standard Rashba-assisted terms. In particular, the SOI contributes both to the same matrix elements as the \(S_\pm\) operators and to the LH spin-flip channel described by \(t_{SOI,LH}\). The explicit expressions for the tunneling matrix elements and $\delta\epsilon_{HL}$ are given in Appendix~\ref{app:LK_elements}. In the \((L,R)\) basis, the full Hamiltonian may equivalently be written as
\begin{equation}
H_T
=\tau_0\otimes H_{\mathrm{loc}}
+\frac{\epsilon}{2}\tau_z\otimes\mathcal{I}_4
+\tau_x\otimes \mathcal T_H
+i\tau_y\otimes \mathcal T_A.
\label{eq:HT_LR_compact}
\end{equation}

To express Eq.~\eqref{eq:HT_LR_compact} in the bonding--antibonding basis, we introduce the unitary transformation acting on the orbital \((L,R)\) subspace,
\begin{equation}
U_\tau=\frac{1}{\sqrt{2}}
\begin{pmatrix}
1 & 1\\
1 & -1
\end{pmatrix},
\end{equation}
which defines the bonding and antibonding states as
\begin{equation}
|+\rangle=\frac{|L\rangle+|R\rangle}{\sqrt{2}},
\qquad
|-\rangle=\frac{|L\rangle-|R\rangle}{\sqrt{2}}.
\end{equation}

In the truncated \(8\times 8\) space, the corresponding transformation is \(U=U_\tau\otimes \mathcal{I}_4\). The total Hamiltonian then becomes:
\begin{equation}
H_T^{\rm orb}
=\tau_0\otimes H_{\mathrm{loc}}
+\frac{\epsilon}{2}\tau_x\otimes\mathcal{I}_4
+\tau_z\otimes \mathcal T_H
-i\tau_y\otimes \mathcal T_A,
\label{eq:HT_BA_real}
\end{equation}
showing that the Hermitian interdot hybridization processes become diagonal in the orbital pseudospin through \(\tau_z\), while the detuning term \(\epsilon\) mixes the bonding and antibonding sectors through \(\tau_x\). At zero detuning, \(\epsilon=0\), the bonding--antibonding splitting is therefore controlled directly by the hybridization processes encoded in \(\mathcal{T}_H\) and \(\mathcal{T}_A\), with the latter connecting bonding and antibonding states belonging to different bands.

\subsection{Dipole operator}

A classical electric drive or a cavity electric field can be coupled to the DQD either through minimal coupling in the generalized momentum (velocity gauge) or through the dipole operator (length gauge). In the present work, we adopt the \emph{length gauge}. This choice is particularly convenient for the truncated HH--LH DQD Hilbert space considered here, because gauge invariance is exact only in a complete basis, while in a finite basis the velocity gauge can generate spurious or strongly basis-dependent direct couplings through the explicit time dependence of the LK and Rashba operators. By contrast, in the length gauge the static LK Hamiltonian remains unchanged and the electric field enters only through the dipole operator, which makes the physical origin of the driven transitions more transparent and avoids artificially enhancing momentum-induced matrix elements within the reduced model. 

In the present case, the field must be polarized along the DQD axis (the \(x\) direction), since this is the relevant direction for coupling bonding and antibonding states of both HH and LH character at zero detuning, \(\epsilon=0\). Therefore, we write the coupling of the DQD to the electric field of a microwave cavity as
\begin{equation}
H_{\rm I}=-e\mathcal{E}x \cos(\omega_R t).
\label{eq:HI}
\end{equation}
Here, \(\mathcal{E}\) is the field amplitude and \(f_R=\omega_R/2\pi\) is the driving frequency. In the Wannier basis \(|L/R,HH/LH\rangle|m_j\rangle\), and for a single cavity photon, the drive term for electric-dipole spin resonance (EDSR) becomes
\begin{equation}
H_{\rm I}=\hbar g_c (b+b^\dagger)\tau_z\otimes\mathcal{I}_4,
\label{eq:HI_cavity}
\end{equation}
and the total Hamiltonian acquires the photon term \(\hbar\omega_c b^\dagger b\), where \(b\) and \(b^\dagger\) are the cavity photon annihilation and creation operators. In the bonding/antibonding basis we  have instead $H_{\rm I}^{\rm orb}=\hbar g_c (b+b^\dagger)\tau_x\otimes\mathcal{I}_4.$

Strictly speaking, in Eq.~\eqref{eq:HI_cavity}, one should distinguish the charge couplings \(g_c^{HH}\) and \(g_c^{LH}\) in the HH and LH sectors, since
\begin{equation*}
\hbar g_c^{HH/LH}=e\mathcal{E}_{\rm zpf}\, a\, N_{HH/LH}(1-\gamma_{HH/LH}^2),
\end{equation*}
with \(\mathcal{E}_{\rm zpf}\) the single-photon electric field. However, given the dot sizes in the HH and LH sectors, \(50\) nm and \(42\) nm, respectively, and for the interdot separation \(a=100\) nm considered here (see Results in Sec.~\ref{sec:Results}), we have \(N_{HH/LH}(1-\gamma_{HH/LH}^2)\to 1\), so that the two charge couplings become approximately equal. We therefore fix \(g_c/2\pi=50\) MHz, consistent with values used in theoretical descriptions of cavity-coupled double quantum dots and hole flopping-mode qubits~\cite{Guido_prb_2016_valley, input_output_benito_prb,Mutter_PRR_2021_Natural_HHs}.

To describe the dissipative dynamics of the DQD--cavity system, including its steady state, it is convenient to work in the eigenbasis of Eq.~\eqref{eq:HT_LR_compact}, which implies transforming the interaction Hamiltonian to
\begin{equation}
\tilde{H}_{\rm I} = \hbar g_c (b + b^\dagger)\sum_{m,m'=0}^{7} d_{mm'}\, |m\rangle\langle m'|,
\end{equation}
where \(|m\rangle\) and \(|m'\rangle\) are the eigenstates of the \(8\times8\) truncated Hilbert space, including HH--LH mixing and the left/right character, ordered by increasing energy.


\subsection{Input--output theory and decoherence model}

To describe the DQD--cavity device as an open quantum system, we adopt the Heisenberg picture and formulate the dynamics in terms of quantum Langevin equations (QLEs), following standard input--output theory~\cite{Gardiner_Collett_1985,Gardiner_Zoller_2004,Blais_RMP_2021}. The relevant system operators are the cavity photon operators \(b,b^\dagger\) and the transition operators \(\sigma_{mm'} = |m\rangle\langle m'|\). Within this framework, the outgoing fields at the two cavity ports, \(b_{\mathrm{out},1}\) and \(b_{\mathrm{out},2}\), are related to the incoming weak probe fields \(b_{\mathrm{in},1}\) and \(b_{\mathrm{in},2}\) through the standard input--output relations.

Assuming that only the ground state of the truncated Hilbert space is populated, \(p_0=\langle \sigma_{00}\rangle\approx1\), the linear response to a weak probe is governed by the coherences \(\sigma_{0m}\) with \(m>0\). For a cavity driven by a near-resonant microwave field of frequency \(\omega_R\), the QLEs in the rotating frame read
\begin{eqnarray} 
\dot{b}&=& -i\Delta_0 b - \frac{\kappa}{2}b + \sqrt{\kappa_1}\, b_{\mathrm{in},1} + \sqrt{\kappa_2}\, b_{\mathrm{in},2}\nonumber\\ && - i g_c \sum_{m=1}^{7} d_{0m}\sigma_{0m},\label{eq:adot_eq} \\ 
\dot{\sigma}_{0m}&=& -i(\omega_{m0}-\omega_R)\sigma_{0m} - \sum_{m'} \gamma_{0m,0 m'} \sigma_{0 m'}\nonumber\\ &&+\sqrt{2\gamma}\,\mathcal{F} - i g_c d_{m0}\Delta P_{0m} b,
\label{eq:sigmadot_eq} 
\end{eqnarray}
where \(\Delta_0=\omega_c-\omega_R\) is the detuning between the cavity and the probe, \(\kappa\) is the total cavity linewidth, and \(\hbar\omega_{m0}=\Delta_{m0}=E_m-E_0\) is the transition frequency between the ground state and the excited state \(|m\rangle\) with $E_m$ the corresponding eigenvalue. Moreover, \(\Delta P_{0m}=p_0-p_m\approx 1\) within the ground-state approximation. The rates \(\kappa_{1,2}\) describe the coupling to the two cavity ports; throughout this work we assume symmetric coupling and neglect internal cavity losses, such that \(\kappa_1=\kappa_2=\kappa/2\). The operator $\mathcal{F}$ represents the quantum noise associated with the DQD, while $b_{\mathrm{in},i}$ are the incoming fields at the two ports. The outgoing fields are then obtained from the standard input--output relation $b_{\mathrm{out},i}=\sqrt{\kappa_i}\,b-b_{\mathrm{in},i}$.

The decoherence kernel $\gamma$ with matrix elements \(\gamma_{0m,0m'}\) describes the dissipative dynamics within the coherence subspace and, in general, can couple different transition operators \(\sigma_{0m}\). In the present work we retain the dominant decoherence channels associated with charge relaxation and charge-noise-induced dephasing, following the input--output treatment of spin--photon coupling in double quantum dots~\cite{input_output_benito_prb}. These processes define the characteristic charge-coherence decay rate $\gamma_c=\gamma_{\rm rel}/2+\gamma_\phi,$ where \(\gamma_{\rm rel}\) denotes the relaxation rate induced by the environment and \(\gamma_\phi\) is the pure-dephasing rate. In analogy with Ref.~\cite{input_output_benito_prb}, we further decompose the latter as $\gamma_\phi=\gamma_\phi^{(0)}+\gamma_\phi^{(1)},$ where \(\gamma_\phi^{(0)}\) is a residual constant dephasing floor and $\gamma_\phi^{(1)}$ describes the additional dephasing associated with the sensitivity of the transition energy to detuning fluctuations. The latter is proportional to $\partial(E_m-E_0)/\partial\epsilon$, which is zero at the sweet spot $\epsilon=0$. In the calculations below we take \(\gamma_{\rm rel}/2\pi=100~\mathrm{MHz}\), \(\gamma_\phi^{(0)}/2\pi=50~\mathrm{MHz}\), and assign a negligible residual decay rate \(\gamma_s=0~\mathrm{MHz}\) to predominantly spin-like coherences.

To make contact with the physical origin of the linewidths, it is useful to introduce a bare coherence basis separating charge-like and spin-like sectors. In analogy with Ref.~\cite{input_output_benito_prb}, we denote by \(\sigma_\tau\) a coherence connecting opposite bonding/antibonding sectors within a fixed band and pseudo-spin channel, e.g.,\footnote{Here and in the following, we distinguish HH and LH states by their angular-momentum components, $m_j=\pm 3/2$ and $m_j=\pm 1/2$, rather than by an explicit HH/LH label.}
\begin{equation*}
|-,\pm 3/2\rangle\langle +,\pm 3/2|,
\qquad
|-,\pm 1/2\rangle\langle +,\pm 1/2|.
\end{equation*}
and by \(\sigma_s\) a coherence differing primarily in pseudo-spin while remaining within the same charge sector, e.g.,
\begin{equation*}
|-,\pm 3/2\rangle\langle -,\mp 3/2|,
\qquad
|-,\pm 1/2\rangle\langle -,\mp 1/2|,
\end{equation*}
but also mixed heavy-hole--light-hole channels in the same charge sector. Their dissipative dynamics is modeled as
\begin{equation}
\dot{\sigma}_\tau = -\gamma_c\,\sigma_\tau,
\qquad
\dot{\sigma}_s = -\gamma_s\,\sigma_s=0.
\label{eq:tau_sigma_decay}
\end{equation}

In the multiband problem, not all coherences are purely charge-like or purely spin-like. Mixed heavy-hole--light-hole channels, such as $|-,-3/2\rangle\langle +,-1/2|$, simultaneously connect opposite bonding--antibonding sectors and different angular-momentum projections. Therefore, they cannot be regarded, in principle, as purely spin coherences. In the present model, however, the charge-relaxation channels are determined by the dipole operator, which connects bonding and antibonding states while preserving both the band and pseudospin indices, see Eq.~\eqref{eq:HI_cavity}. Consequently, mixed HH--LH coherences such as $|-,-3/2\rangle\langle +,-1/2|$ have no direct matrix element of the charge-relaxation operator and therefore acquire no direct bare charge-relaxation contribution. Their corresponding bare decay rate is thus limited to the residual spin-decoherence rate $\gamma_s$. However, a finite linewidth emerges in the dressed-state basis because the corresponding physical coherences contain admixtures of charge-active intraband channels.

The physical eigenstate coherences \(\sigma_{0m}\) are linear combinations of these bare coherence channels,
\begin{equation}
\sigma_{0m}=\sum_{\mu} T_{\mu m}\,\sigma_\mu,
\label{eq:T_rotation}
\end{equation}
where \(\mu\) runs over the chosen bare coherence basis and \(T_{\mu m}\) is the corresponding rotation matrix determined by the eigenstate composition. In this representation, the dressed decoherence matrix entering Eq.~\eqref{eq:sigmadot_eq} has components
\begin{equation}
\gamma_{0m,0m'} =
\sum_{\mu,\nu}
T_{\mu m}^{*}\,
\gamma^{\mathrm{bare}}_{\mu\nu}\,
T_{\nu m'} .
\label{eq:gamma_dressed}
\end{equation}
Here, \(\gamma^{\mathrm{bare}}_{\mu\nu}\) is diagonal in the bare coherence basis dictated by \(\sigma_\tau\) and \(\sigma_s\), and contains the relaxation and dephasing rates associated with each bare coherence channel.

Because the dressed decoherence matrix $\gamma_{0m,0m'}$ generally couples different coherences, the linewidth associated with a given transition cannot, in general, be described by an independent scalar decay rate. Instead, the response of a selected coherence must be obtained from the full coherence-space linear-response matrix, as in standard input--output treatments of cavity-coupled open quantum systems~\cite{Gardiner_Collett_1985,Gardiner_Zoller_2004, Blais_RMP_2021,input_output_benito_prb}. We therefore define, for each transition, an effective complex detuning by projecting the full response matrix onto the selected channel, or equivalently by downfolding the remaining coherences through a Schur-complement reduction in the spirit of projection methods~\cite{Feshbach_1958}. The corresponding effective linewidth is obtained from the imaginary part of this reduced complex detuning. This procedure yields a transition-dependent and frequency-dependent quantity $\gamma_{0m}^{\mathrm{eff}}(\omega_R)$, which is used in the susceptibility entering the cavity response. For a predominantly spin-like transition, we identify the effective spin-decoherence rate on resonance as $\gamma_{s,0m}=\gamma_{0m}^{\mathrm{eff}}(\omega_R=\omega_{m0})$. When this quantity is plotted as a function of magnetic field, the probe and cavity frequencies are adjusted to follow the selected transition.

Equations~\eqref{eq:adot_eq} and \eqref{eq:sigmadot_eq} yield the expectation values $(\bar{b},\bar{\sigma}_{mm'})$ in the stationary regime. This defines the susceptibilities through $\bar\sigma_{0m}
=\chi_{0m}(\omega_R)\,\bar b,$ with 
\begin{equation}
\chi_{0m}(\omega_R)
=-\frac{g_c\,d_{m0}}{\omega_{m0}-\omega_R-i\gamma^{\mathrm{eff}}_{0m}(\omega_R)},
\label{eq:susceptibility_eff}
\end{equation}
where we set \(\Delta P_{0m}=1\). Substituting this expression into the stationary version of the cavity equation, Eq.~\eqref{eq:adot_eq}, and using the input--output relation $A=\bar{b}_{\mathrm{out},2}/\bar{b}_{\mathrm{in},1}$ we find the transmission amplitude,
\begin{equation}
A=\frac{-i\kappa/2}
{\Delta_0-i\kappa/2
-g_c^2\displaystyle\sum_{m=1}^{7}
\frac{|d_{0m}|^2}
{\omega_{m0}-\omega_R-i\gamma^{\mathrm{eff}}_{0m}(\omega_R)} }.
\label{eq:cavity_A}
\end{equation}
which is, in general, complex. Equation~\eqref{eq:cavity_A} includes the reduced contribution of all transitions in the truncated $8\times8$ manifold. We have assumed $\langle b_{\mathrm{in},2}\rangle=0$ and $\langle\mathcal F\rangle=0$. Taking a loaded cavity quality factor $Q=\omega_c/\kappa=2500$ and cavity frequencies in the range $\omega_c/2\pi=5$--$20~\mathrm{GHz}$ gives $\kappa/2\pi=2$--$8~\mathrm{MHz}$. For simplicity, throughout this work, we fix $\kappa/2\pi=2~\mathrm{MHz}$.

\begin{figure*}[t]
    \centering
    \includegraphics[width=0.745\textwidth]{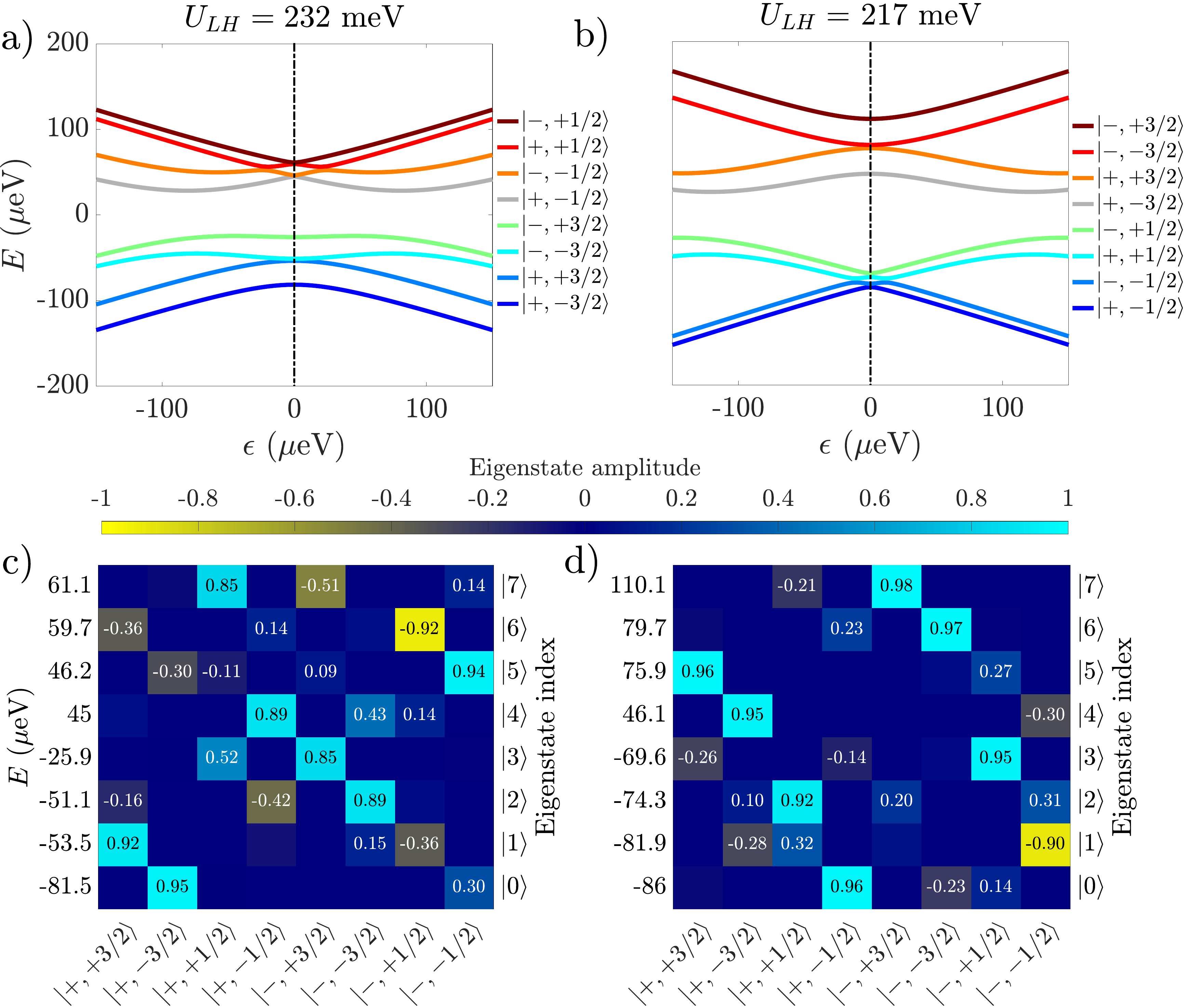}
    \caption{(a,b) Energy spectrum as a function of detuning \(\epsilon\) for two representative values of \(U_{LH}\), \(U_{LH}=232~\mathrm{meV}\) and \(U_{LH}=217~\mathrm{meV}\), respectively. The corresponding eigenstate amplitudes at \(\epsilon=0\), indicated by the vertical black dash-dotted lines in the top panels, are shown in panels (c,d) in the bonding--antibonding HH--LH basis for a low magnetic field \(B=25~\mathrm{mT}\). At low magnetic field, panels (c,d) exhibit appreciable HH--LH mixing, and the states contain substantial admixtures of HH- and LH-like components. Increasing the magnetic field leads to nearly pure bonding and antibonding eigenstates, with the HH--LH admixture significantly reduced. The legends in the top panels correspond to the dominant eigenstate character at \(\epsilon=0~\mu\mathrm{eV}\).}
    \label{fig:energy_and_eigenvector_main_2}
\end{figure*}

\section{Results} 
\label{sec:Results}

\subsection{General considerations}

In the following, we focus on the opposite-parity HH--LH pair illustrated schematically in Fig.~\ref{fig:scheme_model}, namely the HH state with $n=2$ and the LH state with $n=1$. For this choice, the double quantum dot contains the corresponding HH and LH states on both the left and right sites, leading to an eight-dimensional truncated Hilbert space. Because the two states have opposite parity along $z$, matrix elements that do not explicitly connect them through $k_z$ vanish by orthogonality. As a result, the off-diagonal tunneling terms satisfy $t_{SOI_\pm}=t_{R_\pm}=0$ in Eqs.~\eqref{eq:off_LR_basis} and \eqref{eq:H_total}, which yields
\begin{eqnarray}
H_{\rm LR}=
\begin{pmatrix}
t_{HH} & 0 & t_{S_+} & 0\\
0 & t_{HH} & 0 & t_{S_{-}}\\
-t_{S_{+}} & 0 & t_{LH} & t_{SOI,LH}\\
0 & -t_{S_{-}} & -t_{SOI,LH} & t_{LH}
\end{pmatrix}.
\label{eq:matrix_off_pair12}
\end{eqnarray}

\begin{figure}
    \centering
    \includegraphics[width=0.494\linewidth]{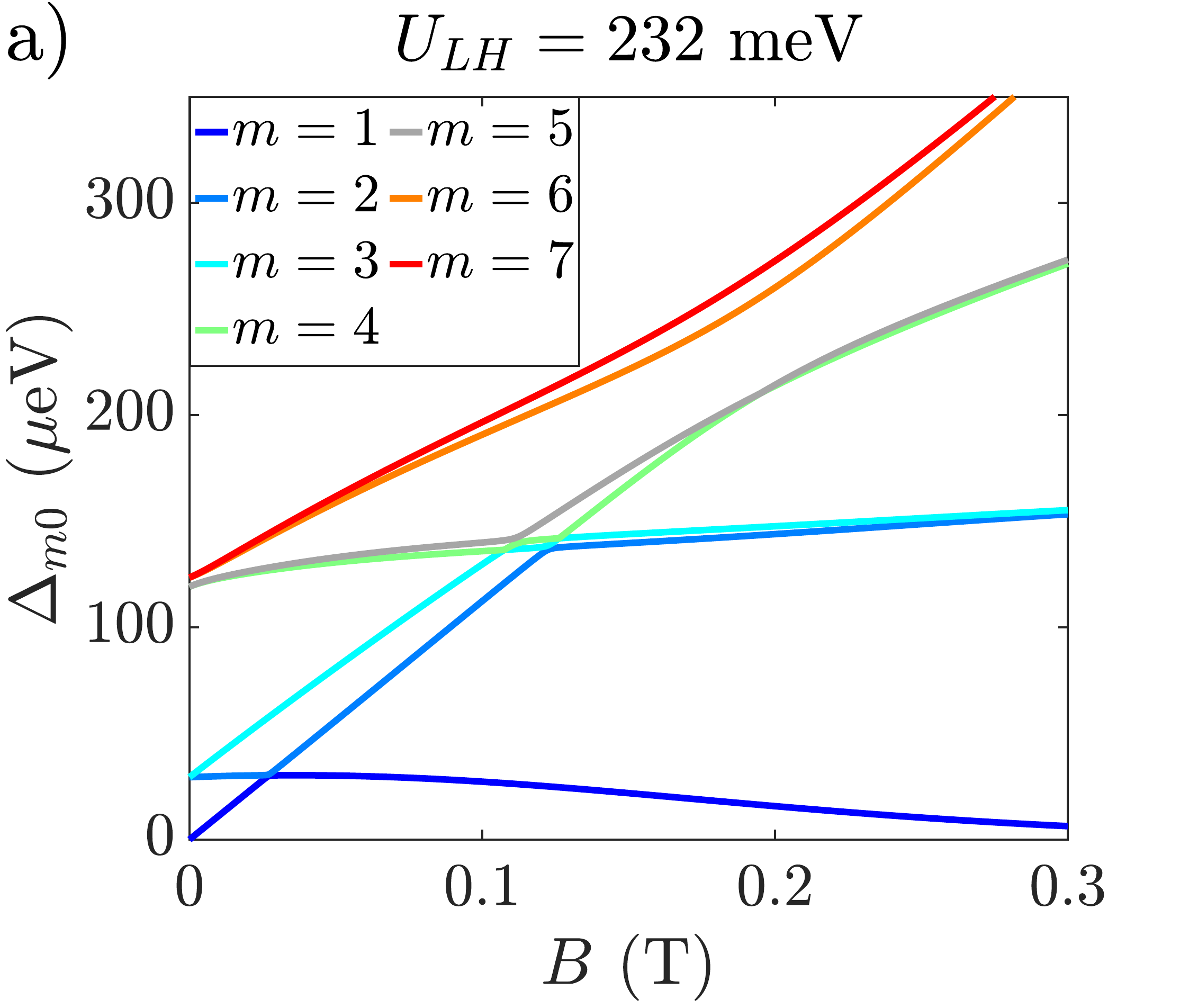}
    \includegraphics[width=0.494\linewidth]{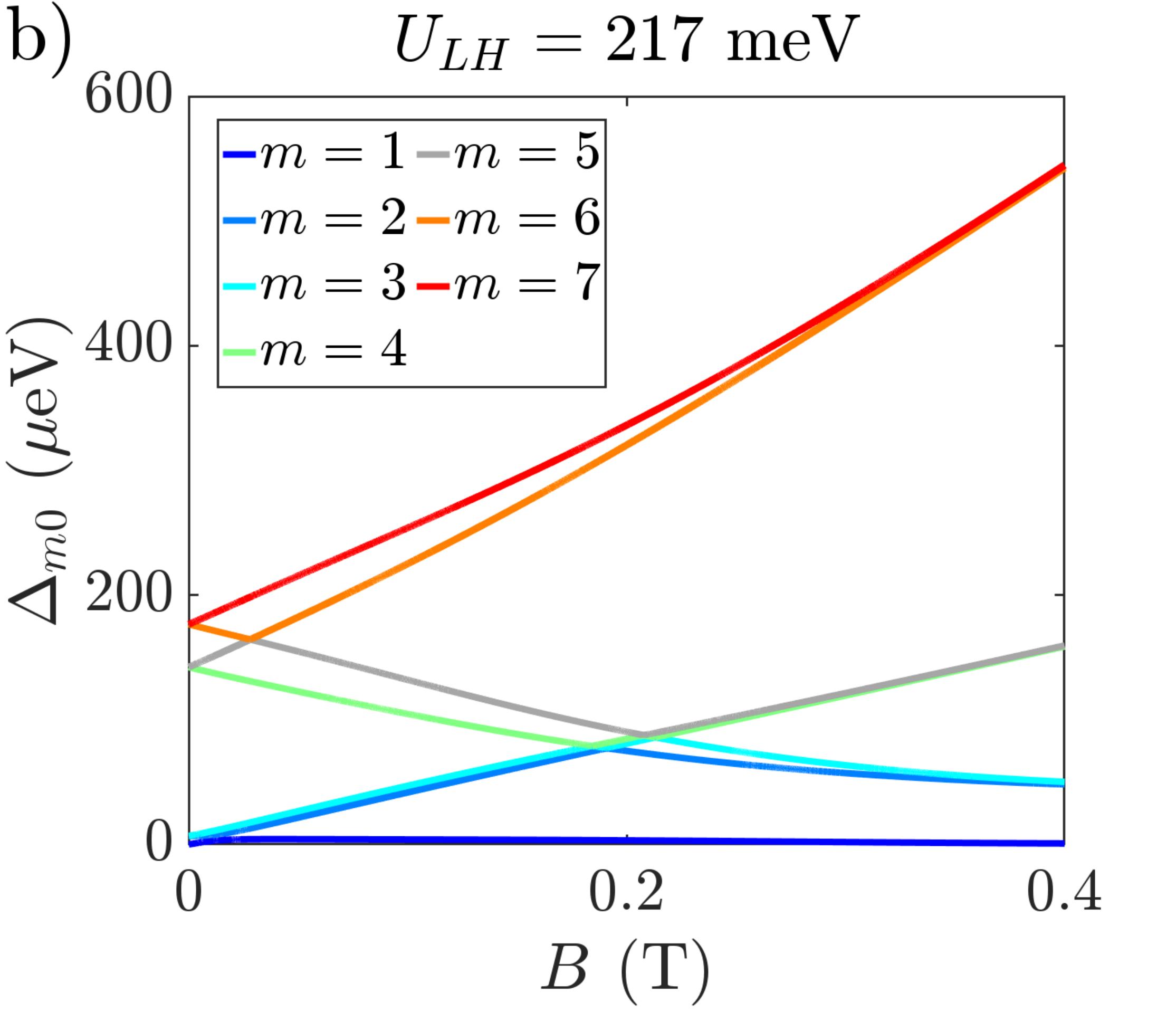}
    \includegraphics[width=\linewidth]{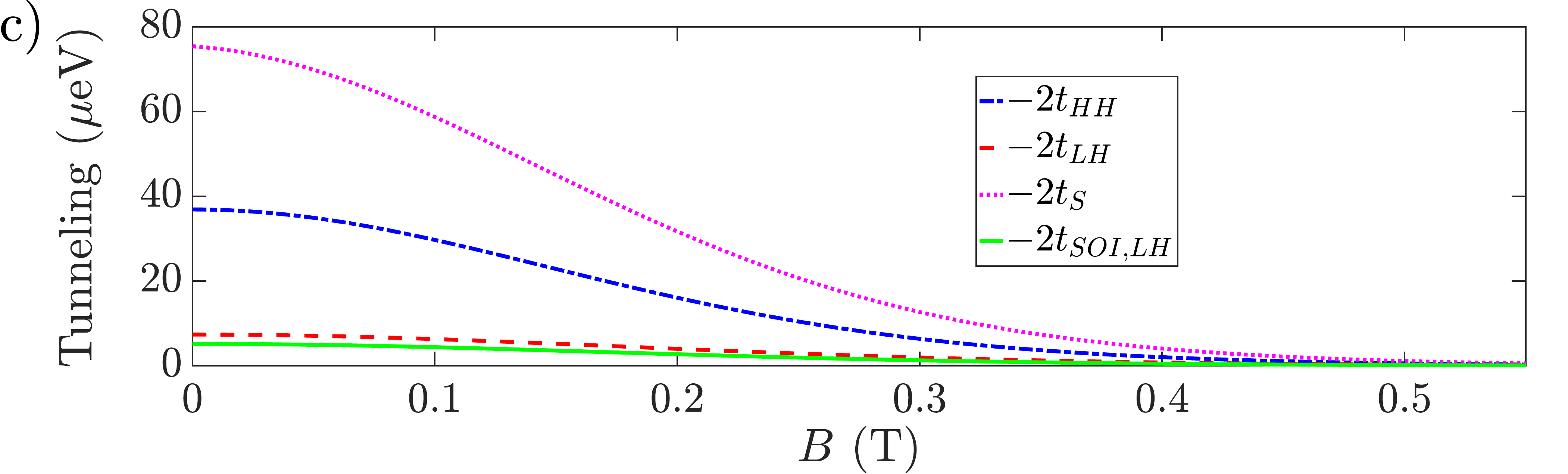}
    \caption{(a,b) Excitation energies \(\Delta_{m0}=E_m-E_0\) at \(\epsilon=0\) as a function of magnetic field for the eight-state truncated model, using \(U_{LH}=232~\mathrm{meV}\) and \(U_{LH}=217~\mathrm{meV}\), respectively. (c) Relevant tunneling matrix elements as a function of magnetic field. Their decrease with increasing magnetic field reflects the reduction of the effective dot size.}
    \label{fig:diffE_vs_B}
\end{figure}

The resulting energy spectra for two representative values of $U_{LH}$ are shown in Fig.~\ref{fig:energy_and_eigenvector_main_2}(a,b) as a function of detuning $\epsilon$. For $U_{LH}=232~\mathrm{meV}$, the ground state is predominantly associated with the HH $n=2$ sector, whereas $U_{LH}=217~\mathrm{meV}$ shifts the ground state toward stronger LH character. For $|\epsilon|\gg |t_i|$, with $i=HH,LH,S_\pm, SOI$, the eigenstates in both cases are predominantly localized on the left or right dot. At $\epsilon=0$, however, the eigenstates are not simple bonding and antibonding states, as shown in Fig.~\ref{fig:energy_and_eigenvector_main_2}(c,d), because of the pronounced HH--LH hybridization at low magnetic field. This mixing is progressively reduced as the magnetic field increases, and the eigenstate amplitudes become close to purely bonding or antibonding character. Physically, this follows from the fact that the Zeeman splitting eventually exceeds the HH--LH tunneling amplitudes $t_{S_\pm}$, since the splitting increases while the tunneling amplitudes decrease, as shown in Fig.~\ref{fig:diffE_vs_B}(c). The latter decrease is a consequence of the reduction of the effective dot size by the magnetic field~\cite{Mutter_PRR_2021_Natural_HHs}; see also Appendix~\ref{app:LK_elements}. Once the magnetic field has reduced the tunneling amplitudes to nearly zero, the eigenstates evolve into states that are effectively localized on the left and right dots, as expected.

Despite the multiband character of the model, some general trends can still be identified. For the parameter set considered here and \(U_{LH}=232~\mathrm{meV}\), the ground state is predominantly a bonding HH state,
\[
|0\rangle \simeq \beta_0^+|+,{-3/2}\rangle+\upsilon_0^-|-,{-1/2}\rangle,
\]
with \(|\beta_0^+| > |\upsilon_0^-| \), consistent with the negative HH tunneling amplitude \(t_{HH}\). In the same regime, the first and second excited states may be approximated as
\begin{eqnarray*}
|1\rangle &\simeq& \alpha_1^+|+,{3/2}\rangle + \beta_1^-|-,{-3/2}\rangle \\ &&-\eta_1^+|+,{-1/2}\rangle + \nu_1^-|-,{1/2}\rangle,\\[5pt]
|2\rangle &\simeq& -\alpha_2^+|+,{3/2}\rangle + \beta_2^-|-,{-3/2}\rangle \\ &&- \eta_2^+|+,{-1/2}\rangle + \nu_2^-|-,{1/2}\rangle,
\end{eqnarray*}
with all coefficients taken real and positive. Before the first avoided crossing visible in Fig.~\ref{fig:diffE_vs_B}(a), the dominant weights satisfy 
$\alpha_1^+>\beta_1^-,\eta_1^+,\nu_1^-$ and 
$\beta_2^->\alpha_2^+,\eta_2^+,\nu_2^-.$

Beyond this crossing, the ground state remains essentially unchanged, while the first excited state evolves toward a predominantly $|-,{-3/2}\rangle$ character. Therefore, the resonance between the ground state and the first excited state becomes predominantly charge-like, with almost no spin character and hence no visible vacuum Rabi splitting. In addition, in the large-field limit, $\Delta_{10}\approx -2t_{HH}$ at zero detuning, as expected. A similar analysis can be performed for $U_{LH}=217~\mathrm{meV}$, although in this case a crossing occurs at lower magnetic field and additional crossings appear around $B=0.2$~T but at lower energies than in the case where the HH state forms the ground state; see Fig.~\ref{fig:diffE_vs_B}(a,b). Because the effective offset $\delta\epsilon_{HL}$ also increases with magnetic field, it compensates the larger Zeeman slope of the HH branch at very large fields. As a result, the HH state does not become the ground state, even though the tunneling amplitudes are strongly suppressed and the eigenstates become effectively localized on the left and right dots.

A useful feature of the bonding--antibonding representation is that mixing between $|+,\pm 3/2\rangle$ and $|+,\pm 1/2\rangle$ would require an imaginary part of the corresponding $S$-type tunneling amplitude. Here, $t_{S_\pm}$ are always real, so this same-orbital-sector mixing is absent. Instead, $\mathrm{Re}(t_{S_\pm})$ couples $|+,\pm 3/2\rangle$ to $|-,\pm 1/2\rangle$ through the $-i\tau_y\mathcal{T}_A$ term in Eq.~\eqref{eq:HT_BA_real}. This structure explains, for instance, why \(|1\rangle\) acquires a \(|-,{1/2}\rangle\) component rather than a \(|+,{1/2}\rangle\) one, together with a weaker \(|+,{-1/2}\rangle\) admixture arising from the tunneling term \(t_{SOI,LH}\), which connects \(|-,{1/2}\rangle\) and \(|+,{-1/2}\rangle\). In turn, this induces a \(|-,{-3/2}\rangle\) component through the coupling \(t_{S_-}\). The same mechanism explains why \(|0\rangle\) contains a \(|-,{-1/2}\rangle\) admixture at low magnetic field, as shown in Fig.~\ref{fig:energy_and_eigenvector_main_2}(c), together with extremely small \(|+,{1/2}\rangle\) and \(|-,{-3/2}\rangle\) components.

\begin{table}[t]
\centering
\begin{tabular}{ccc}
\toprule
$U_{LH}$ & Transition & Character/message \\
\midrule
$232~\mathrm{meV}$ & HH--HH & Spin-like at low $B$; HH flopping-mode. \\
$232~\mathrm{meV}$ & HH--HH & Charge-like at larger $B$; broad $\gamma_s$. \\
$217~\mathrm{meV}$ & LH--LH & Spin-like; larger dipole than HH--HH. \\
$217~\mathrm{meV}$ & LH--HH & Mixed spin-like; largest $r$. \\
\bottomrule
\end{tabular}
\caption{Representative transition regimes.}
\label{tab:transition_regimes}
\end{table}

The dipole matrix elements, however, follow a simpler structure: only bonding-to-antibonding transitions with the same band character are directly coupled. For example, the strength of the dipole transition between \(|0\rangle\) and \(|1\rangle\) is essentially controlled by the product \(\beta_0^+\beta_1^-\). Maximizing this product enhances the charge character of the transition and therefore also its susceptibility to charge decoherence, making it more charge-like than spin-like, as occurs at large magnetic field. What we seek is not simply to maximize the dipole matrix element, but rather to maximize the ratio between the dipole coupling and the effective linewidth of the two dressed-state resonances, or more specifically of the upper and lower polariton branches. We define this ratio as~\cite{input_output_benito_prb} 
\begin{equation}
r=\frac{g_c d}{\sqrt{(\gamma_s^2+(\kappa/2)^2)/2}}=\frac{g_s}{\Gamma},
\label{eq:Gamma_strong_coupling}
\end{equation}
where $d$ and $\gamma_s$ are evaluated for a selected transition on resonance, i.e., with $\omega_c=\omega_{m0}(B)$ at each magnetic field. Thus, in plots of $r$ versus $B$, the cavity frequency is adjusted to follow the chosen transition. This should be distinguished from the cavity-transmission maps, where $\omega_c$ is fixed and resonance occurs only at the magnetic fields satisfying $\omega_{m0}(B)=\omega_c$. Values of $r$ larger than one indicate the strong-coupling regime. This highlights the central trade-off in the problem: one must borrow enough charge character to make the transition dipole-allowed and sufficiently strong, while preserving enough spin character to limit the associated decoherence and ultimately access the strong-coupling regime.

We now turn to the representative transition regimes summarized in Table~\ref{tab:transition_regimes}. For \(U_{LH}=232~\mathrm{meV}\), the low-energy spectrum allows us to recover a HH--HH pseudospin-flip response in line with previous two-level-like flopping-mode descriptions~\cite{input_output_benito_prb,Mutter_PRR_2021_Natural_HHs}. At larger magnetic fields, however, the same HH--HH transition becomes increasingly charge-like and therefore suffers from a broader linewidth. We then set \(U_{LH}=217~\mathrm{meV}\) to investigate LH--LH and LH--HH transitions, which can provide larger dipole matrix elements while retaining a more favorable spin-like character. The remaining parameters are kept as introduced in the previous sections and are summarized in Table~\ref{tab:simulation_parameters} of Appendix~\ref{app:parameters}. It is important to note that an HH--LH transition can also be achieved when the HH state is the ground state; however, in that case the cavity frequency generally has to be increased to about \(30~\mathrm{GHz}\) or more, which is less favorable for current experimental setups.

\subsection{HH--HH spin-flip}

\begin{figure}
\centering
\includegraphics[width=1.\linewidth]{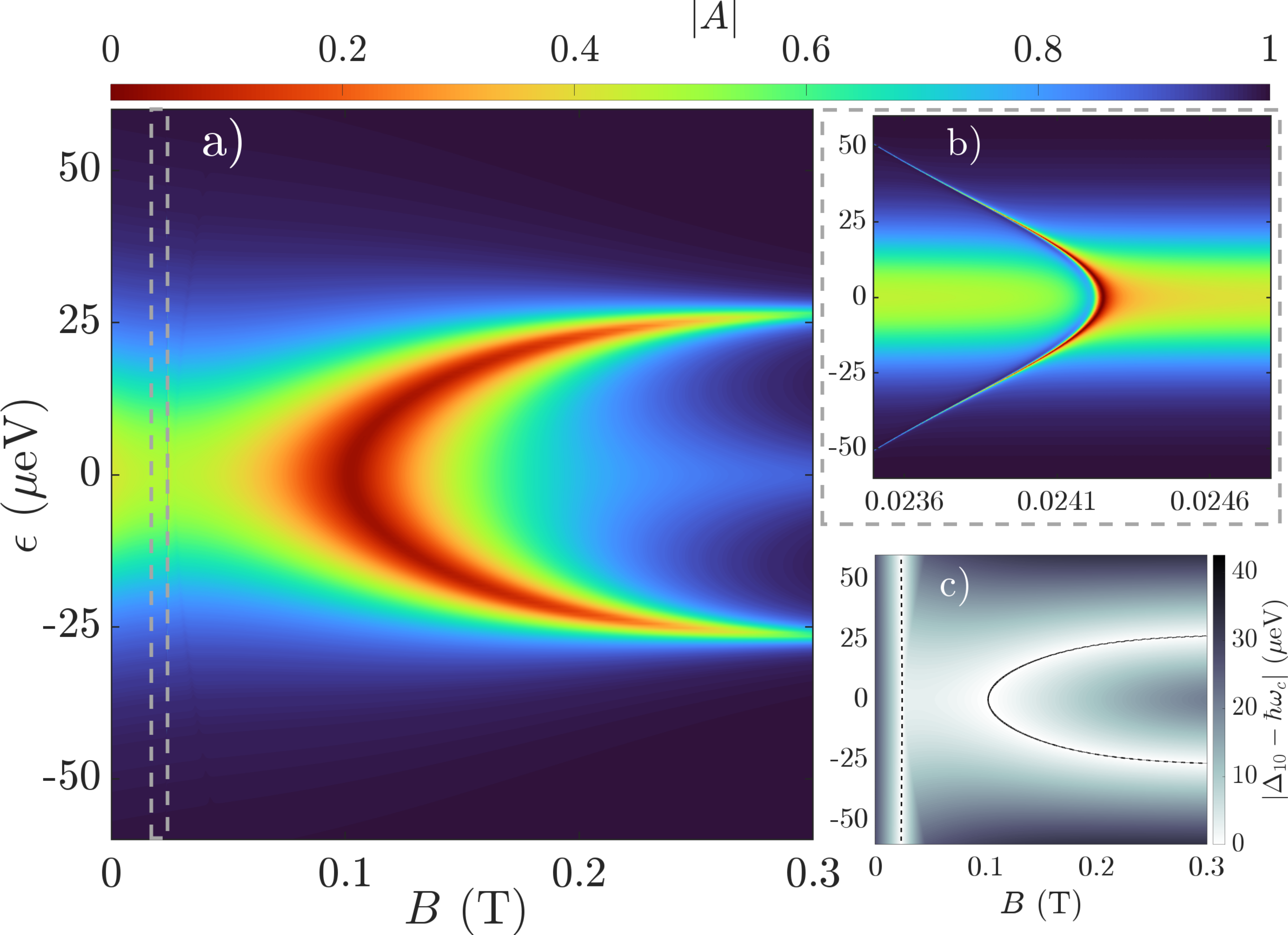}
\caption{(a) Cavity transmission amplitude $|A|$ as a function of magnetic field $B$ and detuning $\epsilon$ for the HH--HH pseudospin-flip regime, $U_{LH}=232~$meV, for $\Delta_0=0$. (b) Zoom of the low-field resonance marked by the gray rectangle in panel (a). (c) Resonance condition for the relevant transition energies relative to the cavity frequency. The black dashed lines indicate the boundaries of a $1\%$ tolerance window around the resonance condition. The C-shaped resonance has a predominantly charge-like character, and therefore no vacuum Rabi splitting is observed. We chose $\omega_c/2\pi=6.54~\mathrm{GHz}$.}
\label{fig:HH-HH_spin_flip}
\end{figure}

In this regime, the relevant low-energy dynamics are captured by the first three eigenstates, so that the dominant excitation energies are $\Delta_{10}$ and $\Delta_{20}$, as shown in Fig.~\ref{fig:diffE_vs_B}(a). We do not consider the third excited state, since it is not dipole-coupled to the ground state for the HH--LH pair considered here. Indeed, from Fig.~\ref{fig:energy_and_eigenvector_main_2}(c),
$$
|3\rangle \simeq \beta_3|-,{3/2}\rangle + \eta_3|+,{1/2}\rangle,
$$
so that no matrix element connects $|+,{-3/2}\rangle$ to the dominant components of $|3\rangle$, and therefore $d_{03}=0$.

\begin{figure}
\centering
\includegraphics[width=1.\linewidth]{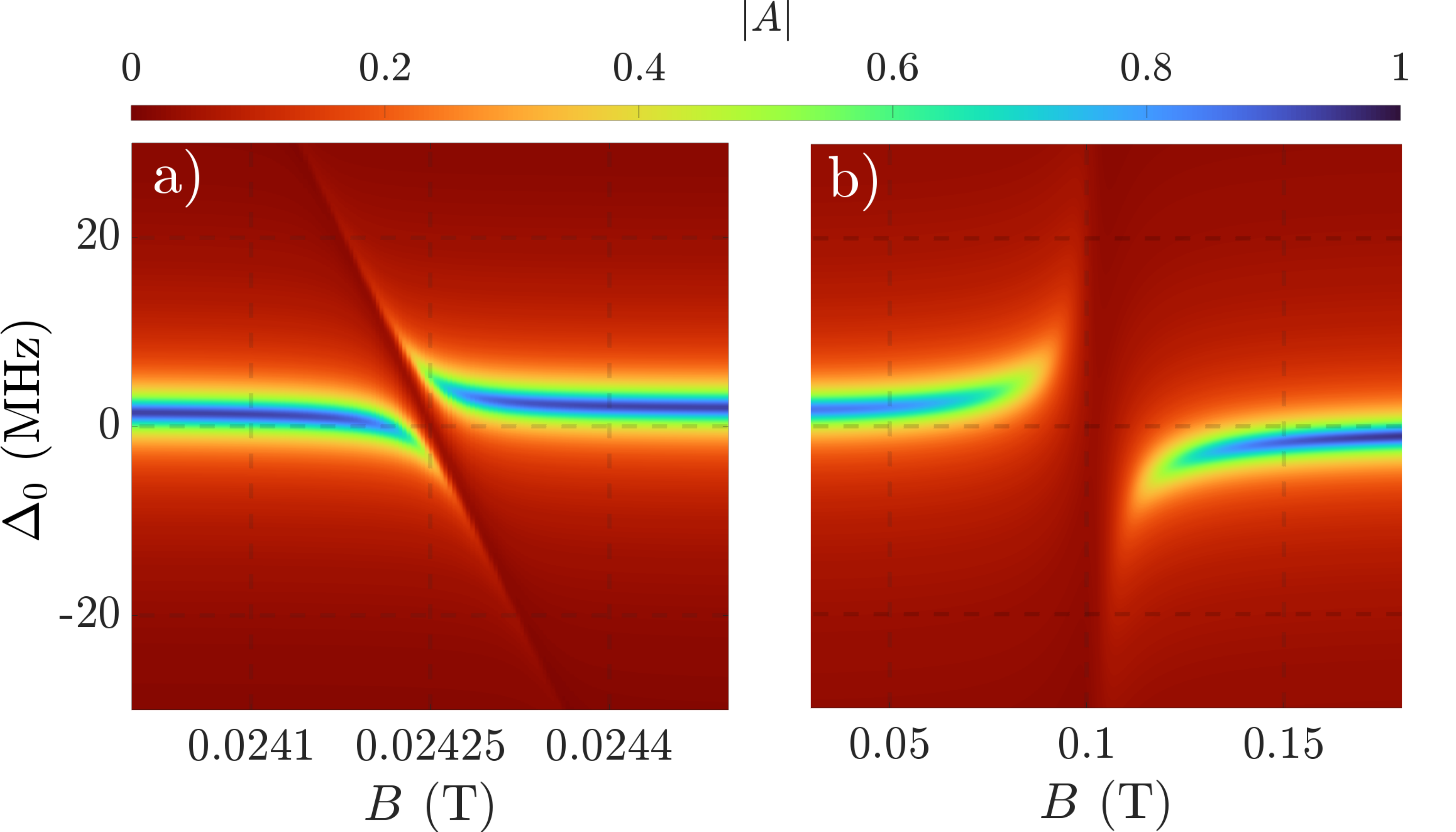}
\caption{Cavity response amplitude $|A|$ as a function of probe--cavity detuning $\Delta_0=\omega_c-\omega_R$ and magnetic field $B$ for the HH pseudospin-flip transition with $U_{LH}=232~\mathrm{meV}$ and $\epsilon=0$. The left panel shows a narrow magnetic-field window around the low-field resonance, where the cavity response exhibits an avoided crossing associated with the hybridization between the cavity mode and the spin-like transition. The right panel shows the response at the intermediate-field resonance, where the charge character dominates and the vacuum Rabi splitting is no longer visible because of strong charge decoherence.}
\label{fig:HH-HH_spin_flip_2}
\end{figure}

\begin{figure}
\centering
\includegraphics[width=0.8\linewidth]{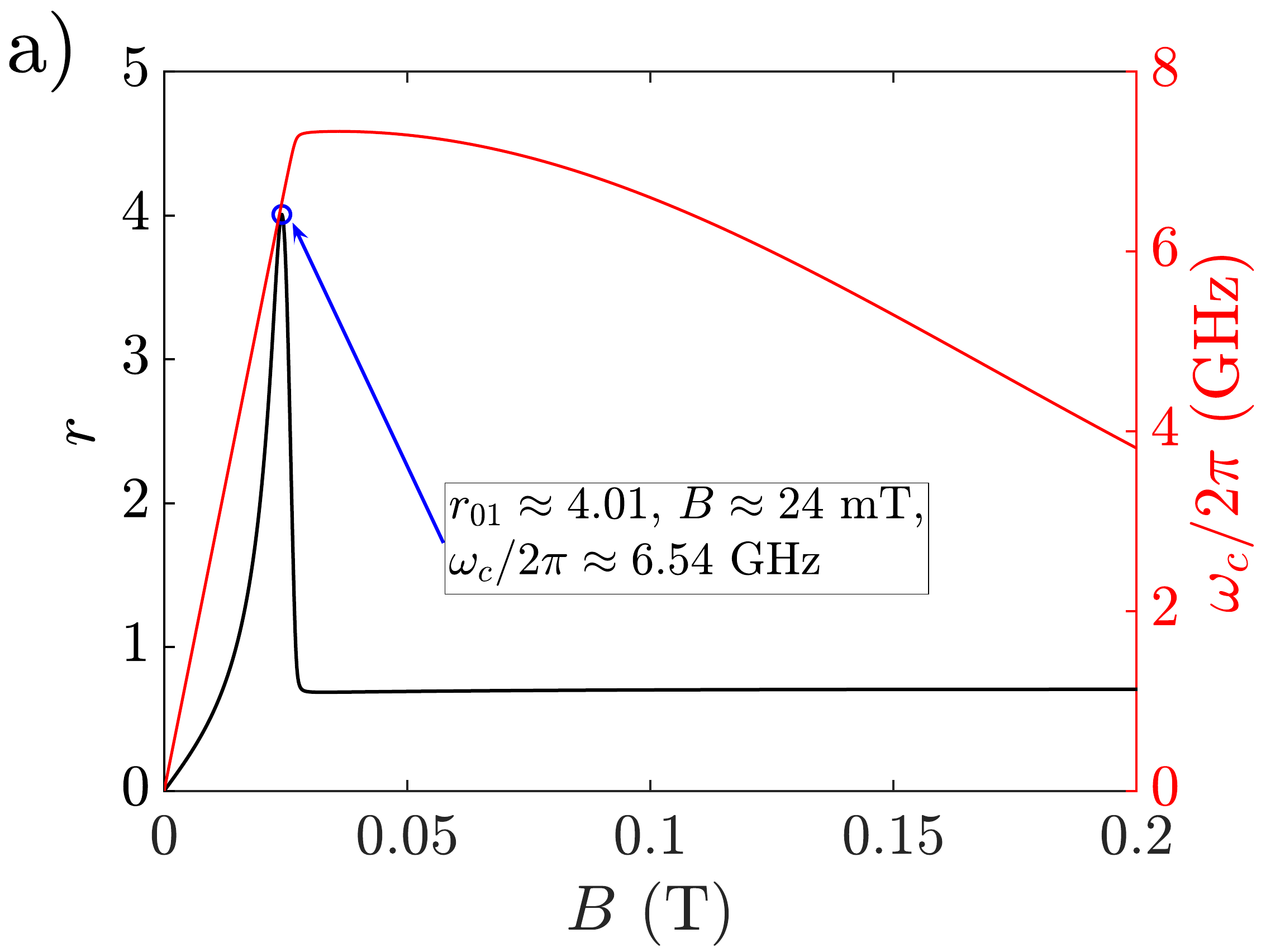}
\hspace*{-1.1cm}
\includegraphics[width=0.8\linewidth]{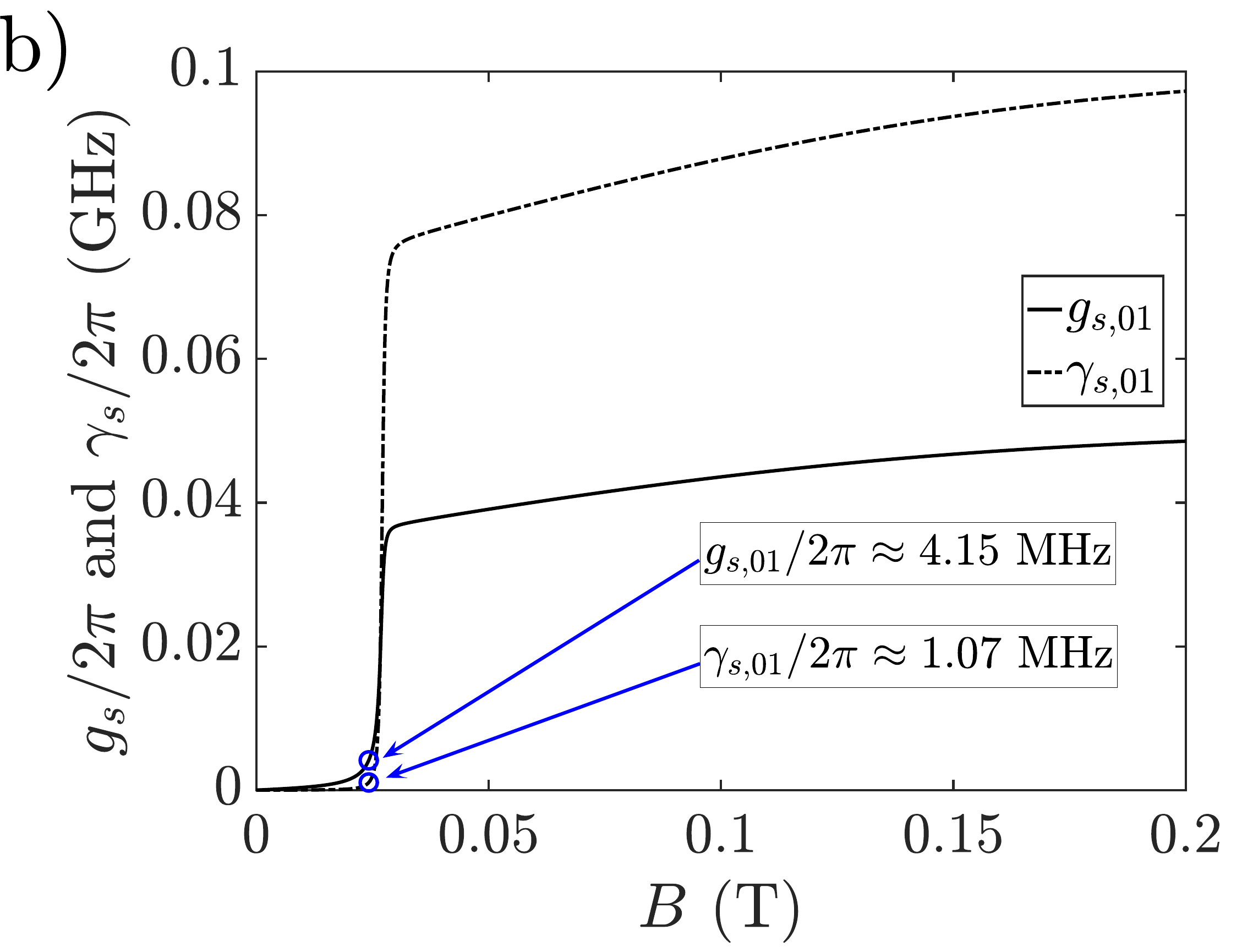}
\caption{(a) Ratio $r$ for $\epsilon=0$ and $\Delta_0=0$, defined in Eq.~\eqref{eq:Gamma_strong_coupling}, as a function of magnetic field for the $|0\rangle\rightarrow |1\rangle$ transition. Values $r>1$ indicate the strong-coupling regime. The blue circle marks the maximum value of $r$, while the right axis shows the cavity frequency required to satisfy the resonance condition at each magnetic field. (b) Spin-photon coupling $g_{s,01}=g_c d_{01}$ and effective on-resonance spin-decoherence rate $\gamma_{s,01}$ as functions of magnetic field. The blue circles indicate the values of $g_{s,01}$ and $\gamma_{s,01}$ at the magnetic field where $r$ is maximized.}
\label{fig:d_gamma_Gamma_vs_B_01}
\end{figure}

For magnetic fields slightly below the first avoided crossing in Fig.~\ref{fig:diffE_vs_B}(a), where $\Delta_{10}\approx 30~\mu\mathrm{eV}$, the dipole matrix element satisfies $d_{01}\neq 0$ and the transition approaches resonance with the cavity. We choose a cavity frequency $\omega_c/2\pi\simeq 6.54~\mathrm{GHz}$, since this maximizes $r$ for the transition between states $|0\rangle$ and $|1\rangle$ at low magnetic field. The corresponding transmission map is shown in Fig.~\ref{fig:HH-HH_spin_flip}(a), with a zoom of the low-field resonance around $B=22~\mathrm{mT}$ shown in Fig.~\ref{fig:HH-HH_spin_flip}(b), corresponding to the gray rectangle in Fig.~\ref{fig:HH-HH_spin_flip}(a). This low-field resonance lies in the strong-coupling regime and exhibits a clear vacuum Rabi splitting, as shown in Fig.~\ref{fig:HH-HH_spin_flip_2}(a). Quantitatively, Fig.~\ref{fig:d_gamma_Gamma_vs_B_01}(b) gives a spin-photon coupling $g_{s,01}=g_c d_{01}\approx 4.21~\mathrm{MHz}$ at the optimal point, corresponding to about $10\%$ of the bare charge coupling $g_c$. This value is consistent with previous estimates for electrically driven HH pseudospin-flip in flopping-mode hole qubits and related input--output descriptions~\cite{Mutter_PRR_2021_Natural_HHs,input_output_benito_prb}.

In addition to this main spin-like resonance, Fig.~\ref{fig:HH-HH_spin_flip}(a) shows a broad, distorted C-shaped resonance at intermediate magnetic fields. This feature is absent in a purely HH description, where the LH states are down-folded into the HH $n=1 $ ground-state manifold. In the present model, it arises from explicit HH--LH mixing. However, although this additional resonance is a genuine multiband feature, it is predominantly charge-like. The dipole matrix element is close to unity, but the charge decoherence is also strongly enhanced, with $\gamma_{01}^{\rm eff}\sim\gamma_c=100~\mathrm{MHz}$, leading to $r<1$; see Fig.~\ref{fig:d_gamma_Gamma_vs_B_01}. The same conclusion is visible in Fig.~\ref{fig:HH-HH_spin_flip_2}(b), where the linewidths of the two dressed-state resonances exceed their separation. As a result, the two branches fade in the color map as the resonance magnetic field is approached, indicating substantial decoherence, in agreement with the interpretation of Ref.~\cite{input_output_benito_prb}. Therefore, for this parameter set, the strong-coupling regime is reached only at low magnetic field.

The intermediate-field resonance can be traced back to the lowering of the second excited state after the first avoided crossing in Fig.~\ref{fig:diffE_vs_B}(a). This is confirmed in Fig.~\ref{fig:HH-HH_spin_flip}(c), where we display the resonance condition relative to the cavity frequency. These additional structures show that explicit HH--LH mixing opens regimes that are absent in simplified HH-only models. They also suggest that further tuning of the HH--LH composition could make such C-shaped resonances less charge-like and therefore more favorable for strong coupling.

\subsection{LH--LH and LH--HH spin-flips}

We now turn to spin-flip transitions in the configuration with $U_{LH}=217~\mathrm{meV}$, for which the ground state has stronger LH character. This choice allows us to investigate transitions involving LH-like ground-state components and either LH-like or HH-like excited-state components. In particular, we focus on the transition between the ground state and the second excited state, since the transition from the ground state to the first excited state is predominantly charge-like and is therefore not the optimal candidate for coherent spin-photon coupling.

\begin{figure}
\centering
\includegraphics[width=1\linewidth]{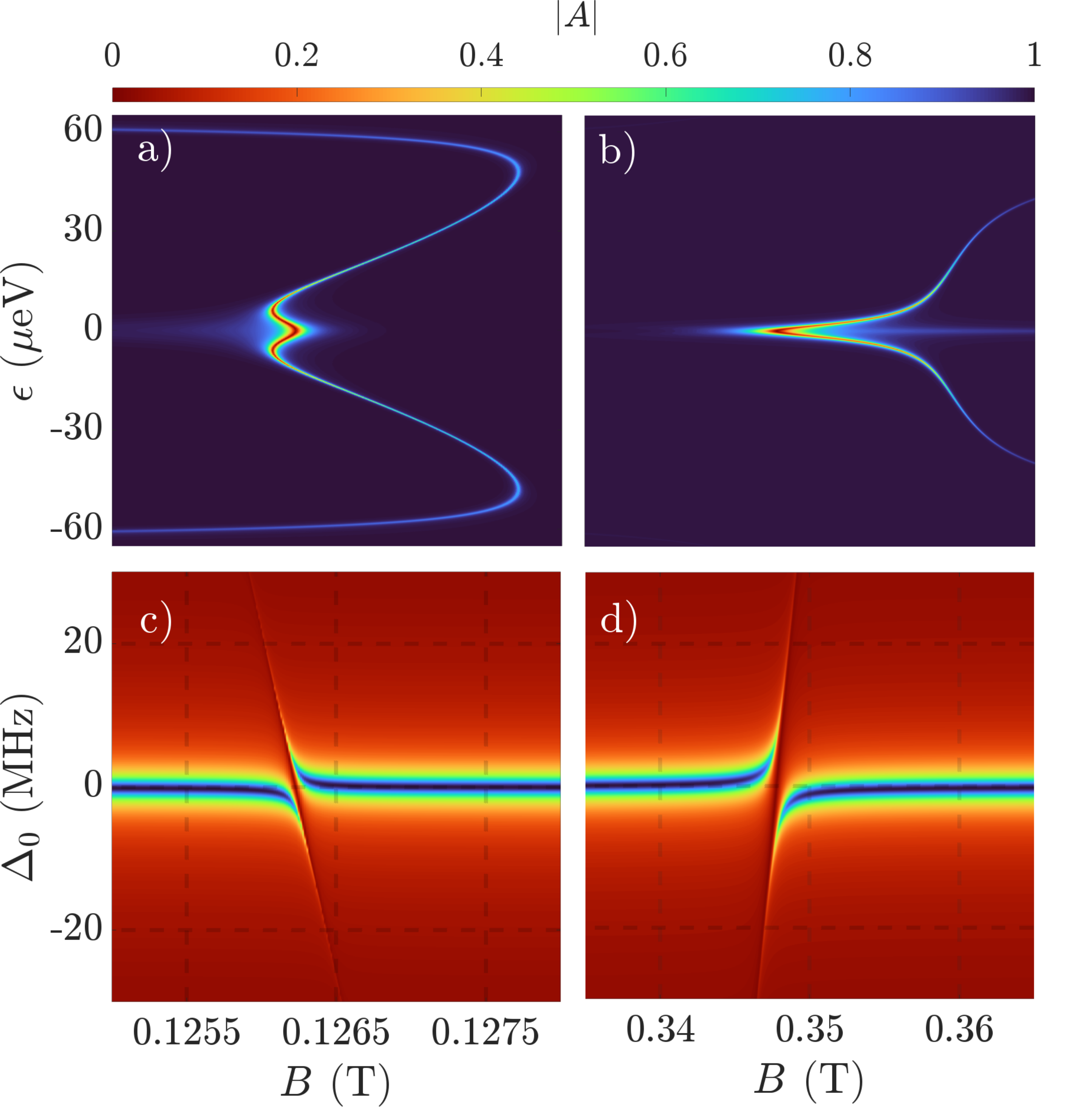}
\caption{Cavity transmission amplitude $|A|$ for the $U_{LH}=217~\mathrm{meV}$ configuration, where the ground state has stronger LH character. (a,b) Transmission as a function of magnetic field $B$ and detuning $\epsilon$ for a fixed cavity frequency $\omega_c/2\pi=12.58~\mathrm{GHz}$ and $\Delta_0=0$. Panel (a) shows the lower-field resonance associated mainly with an LH--LH spin-flip transition, while panel (b) highlights the higher-field resonance associated with an LH--HH spin-flip transition. (c,d) Corresponding cavity response as a function of probe--cavity detuning $\Delta_0=\omega_c-\omega_R$ and magnetic field $B$ around the two selected resonances at $\epsilon=0$. In both cases, the dressed-state resonances remain well resolved, showing that the corresponding transitions can reach the strong-coupling regime.}
\label{fig:LH-LH_spin_flip}
\end{figure}

The cavity transmission for a fixed cavity frequency $\omega_c/2\pi=12.58~\mathrm{GHz}$ is shown in Fig.~\ref{fig:LH-LH_spin_flip}(a,b) near the relevant resonance conditions. Both of them are a product of transition of the type $|0\rangle\rightarrow |2\rangle$. This interpretation was checked by plotting the resonance condition for $\Delta_{20}$, in analogy with Fig.~\ref{fig:HH-HH_spin_flip}(c). Figure~\ref{fig:LH-LH_spin_flip}(a) shows a resonance associated mainly with an LH--LH spin-flip transition. Additional weak resonance features appear at finite detuning but they are strongly suppressed due to $\gamma_\phi^{(1)}$. 

\begin{figure}
\centering
\includegraphics[width=0.8\linewidth]{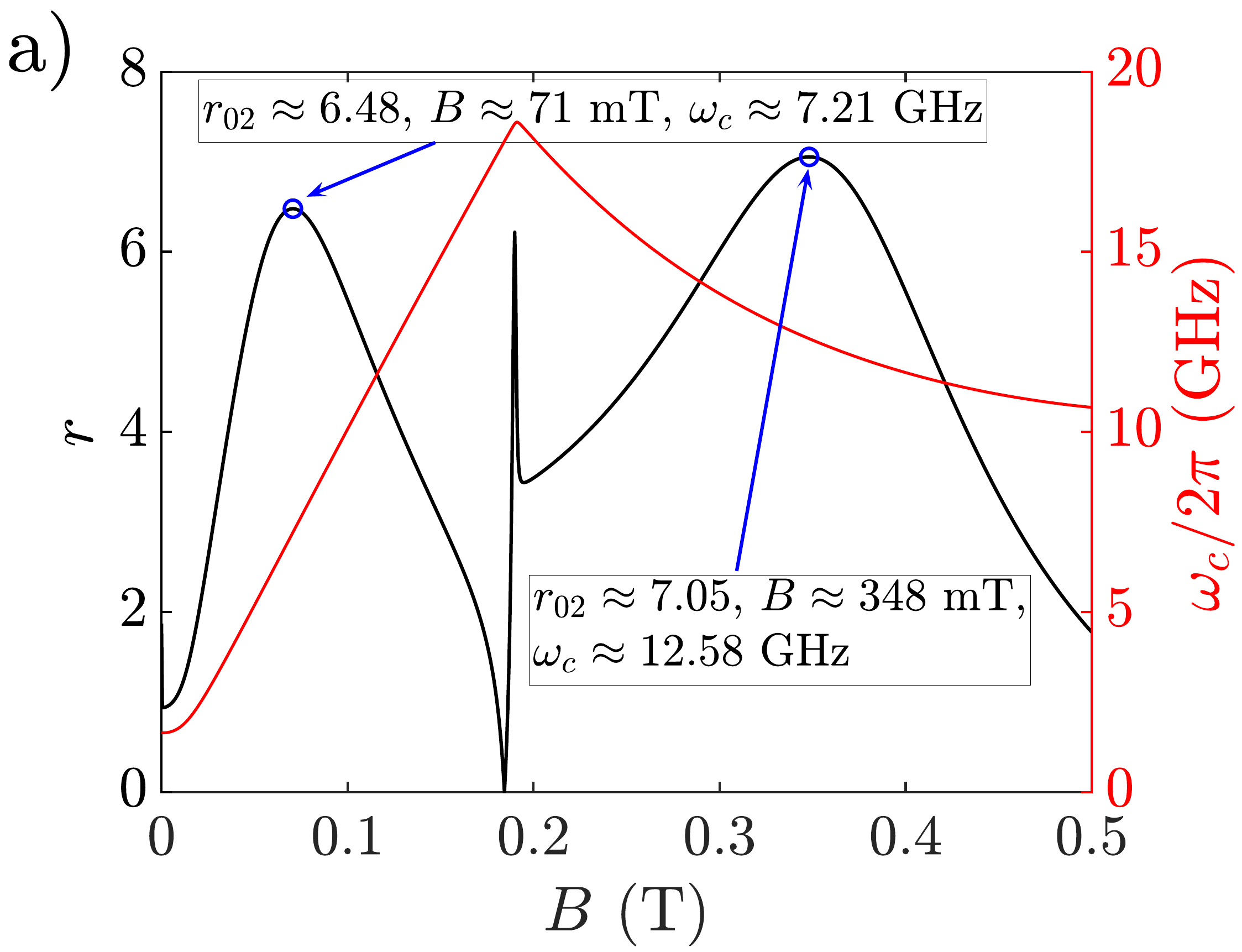}
\hspace*{-1.2cm}
\includegraphics[width=0.8\linewidth]{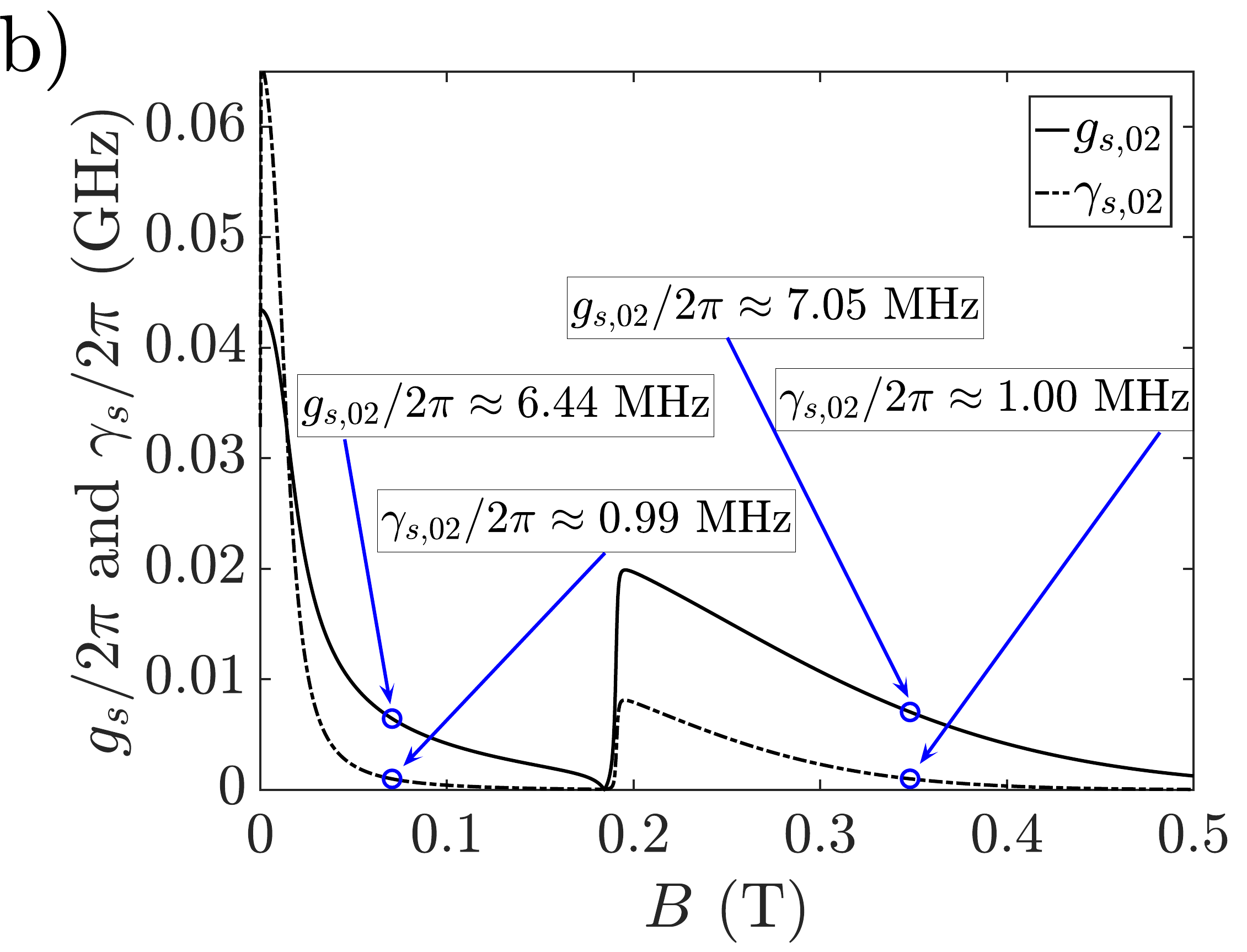}
\caption{(a) Ratio $r$ at $\epsilon=0$  and $\Delta_0=0$, defined in Eq.~\eqref{eq:Gamma_strong_coupling}, as a function of magnetic field for the $|0\rangle\rightarrow |2\rangle$ transition in the $U_{LH}=217~\mathrm{meV}$ configuration. Values $r>1$ indicate the strong-coupling regime. The curve shows two maxima: the first corresponds mainly to an LH--LH spin-flip transition, while the second corresponds mainly to an LH--HH spin-flip transition. The blue circle marks the maximum value of $r$, and the right axis shows the cavity frequency required to satisfy the resonance condition at each magnetic field. 
For the cavity frequency chosen in Fig.~\ref{fig:LH-LH_spin_flip}, the LH--HH resonance is maximized, blue circle at $B=348~$mT, while the LH-LH transition at that frequency gives $r_{02}\approx 4.14$, $g_{s,02}/2\pi\approx 3~\mathrm{MHz}$ and $\gamma_{s,02}/2\pi\approx 0.22~\mathrm{MHz}$ at $B=126~$mT. (b) Spin-photon coupling $g_{s,02}=g_c d_{02}$ and effective on-resonance spin-decoherence rate $\gamma_{s,02}$ as functions of magnetic field. The blue circles indicate the values of $g_{s,02}$ and $\gamma_{s,02}$ at the magnetic fields where $r$ is maximized for both the LH--LH and LH--HH transitions.}
\label{fig:d_gamma_Gamma_vs_B_02}
\end{figure}

The zero-detuning LH--LH resonance can have a much larger dipole matrix element and reach a larger value of $r$ than the HH--HH counterpart discussed above, provided that the cavity frequency is chosen appropriately, as shown in Fig.~\ref{fig:d_gamma_Gamma_vs_B_02}(a). However, for the cavity frequency that maximizes the LH--HH transition, the corresponding LH--LH resonance yields a slightly smaller $g_s$ than the LH--HH resonance shown in Fig.~\ref{fig:d_gamma_Gamma_vs_B_01}(a). This is partially compensated for by its smaller $\gamma_s$, resulting in a comparable value of $r$. This behaviour is visible in Fig.~\ref{fig:LH-LH_spin_flip}(c) and is indicated more explicitly in the caption of Fig.~\ref{fig:d_gamma_Gamma_vs_B_02}(a).

Figure~\ref{fig:LH-LH_spin_flip}(b) shows a second resonance condition, now associated with an LH--HH spin-flip transition. This transition appears because the $|+,{+1/2}\rangle$-like state is pushed upward by the Zeeman energy, allowing the $|+,{-3/2}\rangle$-like state to move down in energy and become the new state $|2\rangle$ as the magnetic field increases. As shown in Fig.~\ref{fig:d_gamma_Gamma_vs_B_02}(a) and in the cavity-response map of Fig.~\ref{fig:LH-LH_spin_flip}(d), this LH--HH transition gives the largest value of $r$ in the parameter space considered here.

At intermediate magnetic fields, Fig.~\ref{fig:d_gamma_Gamma_vs_B_02}(a) also shows an additional sharp maximum with a comparable value of $r$. This feature occurs near a near-degeneracy involving the $|+,{1/2}\rangle$- and $|+,{-3/2}\rangle$-like states. Although the resulting hybridization can produce a strong-coupling value of $r$, the transition is no longer easily classified as a clean $|-1/2\rangle\rightarrow |+1/2\rangle$ LH--LH spin flip or as a $|-1/2\rangle\rightarrow |-3/2\rangle$ LH--HH spin flip. We therefore regard this intermediate-field maximum as a less useful operating point: it confirms that strong coupling can be enhanced near multilevel hybridization, but it does not provide a well-isolated transition with a clear spin character. For this reason, the cleaner LH--LH and LH--HH resonances discussed above are more suitable for identifying practical operating regimes.

Interestingly, the avoided crossing in Fig.~\ref{fig:LH-LH_spin_flip}(d) has the opposite orientation compared with the LH--LH resonance in Fig.~\ref{fig:LH-LH_spin_flip}(c). This apparent inversion does not originate from the sign of the dipole matrix element, since the observable splitting is controlled by $|g_c d_{02}|$. Instead, it follows from the opposite magnetic-field dependence of the corresponding transition energies, as shown in Fig.~\ref{fig:diffE_vs_B}(b), i.e. the two transitions have effective transition $g$ factors with opposite signs. The corresponding spin-photon coupling and effective spin-decoherence rate are also improved compared with the HH--HH case, as shown in Fig.~\ref{fig:d_gamma_Gamma_vs_B_02}(b). These results indicate that using an explicitly mixed LH--HH pair can provide a more favorable balance between dipole strength and decoherence than a purely HH-like pseudospin transition.

\section{Conclusions}

We investigated spin--photon coupling in a germanium double quantum dot operated in the flopping-mode regime using a multiband Luttinger--Kohn framework that retains heavy-hole and light-hole states explicitly. Finite confinement along the vertical direction brings a selected opposite-parity HH--LH pair into a regime where hybridization becomes appreciable. The resulting LH admixture modifies the low-energy eigenstates and introduces electrically active spin-coupling channels that are absent from an HH-only description.

The two effective LH confinement depths considered here lead to qualitatively different regimes. For $U_{LH}=232~\mathrm{meV}$, the ground state remains predominantly HH-like, and the low-field response contains a HH--HH pseudospin-flip resonance consistent with the conventional flopping-mode picture. As the magnetic field increases, this transition acquires stronger charge character. Its electric dipole matrix element increases, but the accompanying linewidth broadens as well, reducing the spin--photon figure of merit. For $U_{LH}=217~\mathrm{meV}$, the low-energy eigenstates contain a larger LH contribution, bringing LH--LH and LH--HH transitions into the relevant spectral range. Some of these channels support larger electric dipoles without becoming predominantly charge-like and therefore achieve a more favorable ratio of coherent spin--photon coupling to decoherence.

This comparison shows that the electric dipole matrix element cannot be used on its own to identify the most useful spin--photon transition. The effective dressed linewidth is equally decisive. HH--LH mixing can improve the balance between coupling and decoherence by opening additional spin-active electric-dipole channels, whereas excessive charge admixture produces broader resonances and weakens coherent cavity coupling. Controlled HH--LH hybridization consequently provides a tunable degree of freedom for optimizing Ge hole-spin qubits in circuit-QED architectures rather than acting only as a perturbative correction to the HH-only model.

More broadly, our results suggest that vertical confinement, strain, electrostatics, and magnetic field offer complementary ways to control this multiband character. A natural extension of the present model would include additional in-plane orbitals, microscopic strain profiles, interface-induced HH--LH dipole terms, and device-specific noise sources. These ingredients would allow a more quantitative assessment of the parameter regimes in which multiband hole physics improves coherent electrical control and long-range spin--photon coupling in germanium quantum-dot devices.

\begin{acknowledgments}
J. R.-G. acknowledges fruitful discussions with H. T. Bui during the preparation of this manuscript. J. R.-G. and G. B. thank W. A. Coish for insightful discussions and helpful comments.
This research is supported by the German Research
Foundation [Deutsche Forschungsgemeinschaft (DFG)] under
Projects No.~450396347 and No.~425217212 - SFB 1432.
\end{acknowledgments}

\appendix

\section{Finite-well envelope functions}
\label{app:potential_well}

The vertical envelope functions used in the main text are obtained from the standard effective-mass treatment of a finite quantum well along the growth direction~\cite{bastard1988wave,davies1998physics}. We solve
\begin{equation}
\left[-\frac{\hbar^2}{2m_{\perp}}\frac{d^2}{dz^2}+V_{\perp}(z)\right]\phi_n(z)=E_n\phi_n(z),
\end{equation}
with

\begin{equation}
V_{\perp}(z)=\begin{cases}0, & 0<z<d_i,\\
-U, & -d_w<z<0,\\
0, & z<-d_w.
\end{cases}
\end{equation}
Here $d_w$ is the well width, $d_i$ is the upper barrier region included in the model, see Fig.~\ref{fig:scheme_model}, and $U>0$ is the confinement depth. For bound states, $-U<E_n<0$, and we define
\begin{equation}
\lambda_n=\frac{\sqrt{-2m_\perp E_n}}{\hbar},\qquad
k_n=\frac{\sqrt{2m_\perp(U+E_n)}}{\hbar}.
\end{equation}
The allowed energies are obtained from the finite-well quantization condition
\begin{equation}
\tan(k_n d_w)=\frac{2\lambda_n k_n}{k_n^2-\lambda_n^2},
\label{eq:finite_well_quantization}
\end{equation}
together with
$k_n^2+\lambda_n^2=2m_\perp U/\hbar^2.$
The matching at the interfaces follows the BenDaniel--Duke boundary conditions~\cite{bendandaniel1966space}. In the present model, these reduce to continuity of $\phi_n$ and $\partial_z\phi_n$ at each interface because the effective mass is taken to be the same across the corresponding barrier and well regions for a given HH or LH sector.

The normalized wavefunction per region is
\begin{equation}
\phi_n(z)=
\begin{cases}
\mathcal{N}_n e^{-\lambda_n z}, & 0<z<d_i\\[8pt]
\mathcal{N}_n\hspace*{-0.1cm}\left[\cos(k_n z)-\dfrac{\lambda_n }{k_n}\sin(k_n z)\right]\hspace*{-0.05cm},& -d_w<z<0\\[8pt]
\mathcal{N}_n\dfrac{\cos(k_n d_w)+\dfrac{\lambda_n }{k_n}\sin(k_n d_w)}
{e^{-\lambda_n (d_w- z)}}, & z<-d_w
\end{cases}
\label{eq:phi_piecewise_appendix}
\end{equation}
this defines the full piecewise finite-well envelope used in the numerical evaluation of the vertical matrix elements entering the tunneling terms in Eq.~\eqref{eq:off_LR_basis}; see Appendix~\ref{app:LK_elements}. The normalization constant $\mathcal{N}_n$ is fixed by normalizing the full finite-well wavefunction over the vertical domain. The HH and LH sectors are described by the same form, but with their corresponding effective masses $m_\perp$ and confinement depths $U$.

For the parameters used in the main text, the resulting HH--LH energy shift at zero magnetic field is $\delta\epsilon_{HL}(B=0)\simeq150~\mu\mathrm{eV}$. For $U_{\mathrm{LH}}=217~\mathrm{meV}$ and $232~\mathrm{meV}$, respectively, the bare vertical separations are $\Delta_{HL}=-0.01~\mathrm{meV}$ and $-0.08~\mathrm{meV}$, giving
$\Delta_{HL}^{\mathrm{eff}}
=\Delta_{HL}+\delta\epsilon_{HL}(B=0)
\simeq
0.14~\mathrm{meV}
\quad\text{and}\quad
0.07~\mathrm{meV}.$
These values remain smaller than the characteristic in-plane orbital excitation energies,
$\min(\hbar\omega_{0,LH},\hbar\omega_{0,HH})
\approx
\min(0.39,0.57)~\mathrm{meV},$
justifying the truncation to the selected HH--LH pair.

\onecolumngrid

\section{Generalized tunnel couplings and detuning under magnetic field}
\label{app:LK_elements}

Here we summarize the tunneling matrix elements and detuning terms used to obtain the results presented in the main text. We first consider the diagonal energy contribution. For the HH and LH sectors, the on-site energies can be written as
\begin{eqnarray}
\epsilon_{HH/LH,nn'}^{L/R}&=&\Bigg[\hbar\omega_{0,HH/LH}\left(-3\omega_{0,HH/LH}\left(\frac{a^2 m_0 }{\hbar(\gamma_{1} \pm \gamma_{2})}+\frac{1}{\omega_{HH/LH}}\right)\frac{S_{HH/LH}^2}{8(1-S_{HH/LH}^2)}\right.\nonumber \\
&&+\left.\frac{3}{32}\frac{\hbar\omega_{0,HH/LH}(\gamma_{1} \pm \gamma_{2})}{a^2 m_0\omega_{HH/LH}^2}+\frac{\omega_{HH/LH}}{\omega_{0,HH/LH}} \right) +\epsilon^z_{HH/LH,nn'}\Bigg]\delta_{n,n'} = \epsilon_{HH/LH}^{L/R}+\epsilon^z_{HH/LH,nn}.
\end{eqnarray}
This contribution is independent of the site, $L$ or $R$, and depends on the quantum number $n$ associated with the vertical confinement energies. In the main text, we simplify the notation by dropping the $n$ label, but here we keep it in order to present the generalized expressions. In addition, we introduce the overlap between the left- and right-localized Fock--Darwin orbitals in a perpendicular magnetic field,
\begin{equation}
S_{HH/LH}=\exp\!\left[-\frac{m_0a^2\big(\omega_{L,HH/LH}^2+\omega_{HH/LH}^2\big)}{\hbar(\gamma_1 \pm \gamma_2)\omega_{HH/LH}}\right].
\end{equation}
We introduce the detuning $\epsilon$ by rewriting the HH and LH on-site energies in terms of a common detuning parameter:
\begin{eqnarray}
\epsilon_{HH,nn}^{L/R}
&=& \epsilon_{LH,nn}^{L/R}+\hbar\omega_{0,HH}\left[\omega_{0,LH}\left(\frac{a^2 m_0 }{\hbar(\gamma_{1} - \gamma_{2})}+\frac{1}{\omega_{LH}}\right)\frac{3S_{LH}^2}{8(1-S_{LH}^2)}-\omega_{0,HH}\left(\frac{a^2 m_0 }{\hbar(\gamma_{1} + \gamma_{2})}+\frac{1}{\omega_{HH}}\right)\frac{3S_{HH}^2}{8(1-S_{HH}^2)}\right. \nonumber \\[6pt]
&&+\left. \frac{3}{32}\frac{\hbar\omega_{0,HH}(\gamma_{1} + \gamma_{2})}{a^2 m_0\omega_{HH}^2}+\frac{\omega_{HH}}{\omega_{0,HH}}-\frac{3}{32}\frac{\hbar\omega_{0,LH}(\gamma_{1} -\gamma_{2})}{a^2 m_0\omega_{LH}^2}-\frac{\omega_{LH}}{\omega_{0,LH}} \right]\nonumber \\[6pt]
&=& \epsilon_{LH}^{L/R}+\epsilon^z_{HH,nn}+\delta\epsilon_{HL}=E_{avg}^{LH}\pm\epsilon/2+\delta\epsilon_{HL}+\epsilon^z_{HH,nn}.
\end{eqnarray}

Therefore, $\epsilon_{LH,nn}^{L/R}=E_{avg}^{LH}\pm\epsilon/2+\epsilon^z_{LH,nn},$ where the $\pm$ sign is chosen for the $L$ and $R$ sites, respectively. The constant offset $E_{avg}^{LH}$ can be dropped, yielding Eq.~\eqref{eq:diag_LR_basis} once the magnetic field is included. The quantity $\delta\epsilon_{HL}$ is therefore not an independent constant once the magnetic-field-dependent Fock--Darwin orbitals are included. Rather, it represents the relative in-plane energy shift between the HH and LH sectors generated by the reparametrization to a common detuning variable. In the main text, $\delta\epsilon_{HL}(B=0)$ is used to characterize the low-field HH--LH alignment, while its magnetic-field dependence is retained in the numerical calculations.

The remaining terms to calculate are the tunneling matrix elements appearing in Eq.~\eqref{eq:off_LR_basis}. We first consider the usual left-to-right tunneling for each HH or LH state, which is diagonal in the $z$ basis:
\begin{eqnarray}
t_{HH/LH,nn'} &=& -\frac{3}{4}N_{HH/LH}\gamma_{HH/LH}\Bigg[\frac{\hbar}{\omega_{HH/LH}}+\frac{a^2 m_0}{\gamma_1 \pm \gamma_2}\Bigg]\omega_{0,HH/LH}^2\delta_{n,n'}.
\end{eqnarray}
Next, we consider the tunneling terms that induce an angular-momentum change from $\pm 3/2$ to $\pm 1/2$ between HH and LH states with different parity:
\begin{equation}
t_{S_\pm,nn'}=\pm\frac{\sqrt{3}\hbar\sqrt{N_{LH}N_{HH}}\langle\phi^{\mathrm{HH}}_n(z)|\partial_z|\phi^{\mathrm{LH}}_{n'}(z)\rangle\gamma_{3}}{(\gamma_{1}-\gamma_{2})\omega_{HH}+(\gamma_{1}+\gamma_{2})\omega_{LH}}S_{HL}\Bigg[2a(\gamma_{HH}\gamma_{LH}-1)\big(\omega_{L,HH}\pm\omega_{HH}\big)\big(\omega_{L,LH}\mp\omega_{LH}\big)\Bigg].
\end{equation}
The vertical matrix element in this expression is computed using the full piecewise finite-well envelope in Eq.~\eqref{eq:phi_piecewise_appendix}. Continuing, the corresponding Rashba term couples HH and LH states with the same parity:
\begin{equation}
t_{SOI_\pm,nn'}=\frac{\sqrt{3}\alpha_{R}\langle E_z\rangle m_0\sqrt{N_{LH}N_{HH}}\langle\phi^{\mathrm{HH}}_n(z)|\phi^{\mathrm{LH}}_{n'}(z)\rangle}{\hbar\left[(\gamma_{1}-\gamma_{2})\omega_{HH}+(\gamma_{1}+\gamma_{2})\omega_{LH}\right]}S_{HL}\Bigg[2a(\gamma_{HH}\gamma_{LH}-1)\big(\omega_{L,HH}\pm\omega_{HH}\big)\big(\omega_{L,LH}\mp\omega_{LH}\big)\Bigg].
\end{equation}
Here we have defined the HH--LH lateral overlap as
\begin{equation}
S_{HL}=\int dxdy\,\left[\psi_{00}^{HH}(x+a,y)\right]^*\psi_{00}^{LH}(x-a,y)=\frac{2\exp\!\left[-\dfrac{2m_0a^{2}\big(\omega_{L,LH}\omega_{L,HH}+\omega_{HH}\omega_{LH}\big)}{\hbar\big((\gamma_1-\gamma_{2})\omega_{HH}+(\gamma_1+\gamma_{2})\omega_{LH}\big)}\right]\sqrt{(\gamma_{1}^2-\gamma_{2}^2)\omega_{HH}\omega_{LH}}}{(\gamma_1-\gamma_{2})\omega_{HH}+(\gamma_1+\gamma_{2})\omega_{LH}}.
\end{equation}
We then consider the tunneling terms that induce an angular-momentum change from $\pm 3/2$ to $\mp 1/2$, again between HH and LH states with the same parity:
\begin{eqnarray}
t_{R_\pm,nn'}&=& \frac{\sqrt{3}\sqrt{N_{LH}N_{HH}}S_{HL}\langle\phi^{\mathrm{HH}}_n(z)|\phi^{\mathrm{LH}}_{n'}(z)\rangle}{\left((\gamma_1-\gamma_{2})\omega_{HH}+(\gamma_1+\gamma_{2})\omega_{LH}\right)^2}\Bigg[(1+\gamma_{HH}\gamma_{LH}) \nonumber\\
&&\times\Bigg(2m_0a^2\Big[\gamma_2(\omega_{L,HH}\omega_{L,LH}-\omega_{HH}\omega_{LH})^2+\gamma_2\frac{\omega_{L,HH}\omega_{L,LH}}{(\gamma_1-\gamma_2)(\gamma_1+\gamma_2)}\left((\gamma_1-\gamma_{2})\omega_{HH}-(\gamma_1+\gamma_{2})\omega_{LH}\right)^2 \nonumber\\
&&\pm 2(\omega_{L,HH}\omega_{L,LH}-\omega_{HH}\omega_{LH})\sqrt{\frac{\omega_{L,HH}\omega_{L,LH}}{(\gamma_1-\gamma_2)(\gamma_1+\gamma_2)}}\gamma_{3}\left((\gamma_1-\gamma_{2})\omega_{HH}-(\gamma_1+\gamma_{2})\omega_{LH}\right)\Big]\Bigg)\Bigg].
\end{eqnarray}
Finally, the Rashba terms acting only on the LH subspace induce spin flips between LH states with the same $z$ quantum number:
\begin{eqnarray}
t_{SOI,LH,nn'}&=&\frac{\alpha_{R}\langle E_z\rangle N_{LH}S_{LH}\Big[a(1-\gamma_{\mathrm{LH}}^{2})m_0\big(\omega_{L,LH}^2-\omega_{\mathrm{LH}}^{2}\big)\Big]}{\hbar(\gamma_{1}-\gamma_{2})\omega_{\mathrm{LH}}}\delta_{nn'}.
\end{eqnarray}
The vertical matrix elements entering the expressions above are evaluated over the full finite-well domain using Eq.~\eqref{eq:phi_piecewise_appendix}. They are given by

\begin{eqnarray}
\langle\phi^{\mathrm{HH}}_n(z)|\partial_z|\phi^{\mathrm{LH}}_{n'}(z)\rangle&=&-\langle\phi^{\mathrm{LH}}_n(z)|\partial_z|\phi^{\mathrm{HH}}_{n'}(z)\rangle \nonumber \\ 
&=& \mathcal{N}_{nn'}\Bigg[\frac{e^{-(\phi_i^{LH}+\phi_i^{HH})}-1}{1+s_{LH}/(rs_{HH})}+\frac{\big(\cos(\varphi_{HH})+s_{HH}\sin(\varphi_{HH})\big)\big(\cos(\varphi_{LH})+s_{LH}\sin(\varphi_{LH})\big)}{1+s_{LH}/(rs_{HH})} \nonumber\\
&&+\frac{r}{r^2-1}\Bigg((r+s_{LH}s_{HH})+\sin(\varphi_{HH})\Big[(s_{LH}-rs_{HH})\cos(\varphi_{LH})-(rs_{LH}s_{HH}+1)\sin(\varphi_{LH})\Big] \nonumber\\
&&+\cos(\varphi_{HH})\Big[(s_{HH}-rs_{LH})\sin(\varphi_{LH})-(r+s_{LH}s_{HH})\cos(\varphi_{LH})\Big]\Bigg)\Bigg],
\label{eq:phiz_partial}
\end{eqnarray}
and
\begin{eqnarray}
\langle\phi^{\mathrm{HH}}_n(z)|\phi^{\mathrm{LH}}_{n'}(z)\rangle&=&\frac{\mathcal{N}_{nn'}}{k_{LH}}\Bigg[\frac{1-e^{-\phi_{i}^{LH}-\phi_{i}^{HH}}}{s_{LH}+rs_{HH}}+\frac{\big(\cos(\varphi_{HH})+s_{HH}\sin(\varphi_{HH})\big)\big(\cos(\varphi_{LH})+s_{LH}\sin(\varphi_{LH})\big)}{s_{LH}+rs_{HH}} \nonumber\\
&&+\frac{1}{r^2-1}\Bigg[(rs_{HH}-s_{LH})+\cos(\varphi_{HH})\Big[(s_{LH}-rs_{HH})\cos(\varphi_{LH})-(rs_{LH}s_{HH}+1)\sin(\varphi_{LH})\Big] \nonumber\\ 
&&+\sin(\varphi_{HH})\Big[(r+s_{LH}s_{HH})\cos(\varphi_{LH})+(rs_{LH}-s_{HH})\sin(\varphi_{LH})\Big]\Bigg]\Bigg].
\label{eq:phiz_braket}
\end{eqnarray}
Here $\mathcal{N}_{nn'}=\mathcal{N}_{n}^{HH}\mathcal{N}_{n'}^{LH}$ is the product of the global normalization constants of the full piecewise finite-well wavefunctions, with
\begin{align}
\mathcal{N}_{n}^{LH} &= \frac{2\sqrt{\lambda_{LH}}}{\sqrt{\left(1-3s_{LH}^2\right)\cos\bigl(2\varphi_{LH}\bigr)-2e^{-2\phi_i^{LH}}+\left(3+2\phi_w^{LH}\right)\left(s_{LH}^2+1\right)-s_{LH}\left(s_{LH}^2-3\right)\sin\bigl(2\varphi_{LH}\bigr)}},
\label{eq:finalB_LH_1}\\
\mathcal{N}_{n}^{HH} &= \frac{2\sqrt{\lambda_{HH}}}{\sqrt{\left(1-3s_{HH}^2\right)\cos\bigl(2\varphi_{HH}\bigr)-2e^{-2\phi_i^{HH}}+\left(3+2\phi_w^{HH}\right)\left(s_{HH}^2+1\right)-s_{HH}\left(s_{HH}^2-3\right)\sin\bigl(2\varphi_{HH}\bigr)}}.
\label{eq:finalB_HH_1}
\end{align}
We have defined $r=k_{HH}/k_{LH}$, $k_{LH}d_w=\varphi_{LH}$, $rk_{LH}d_w=\varphi_{HH}$, $\lambda_{HH/LH}d_{i,w}=\phi_{i,w}^{HH/LH}$, and $s_{HH/LH}=\lambda_{HH/LH}/k_{HH/LH}$ using the definitions from Appendix~\ref{app:potential_well}. Notice that, to simplify the notation, we have suppressed the explicit vertical quantum-number dependence of $k_{HH/LH}$ and $\lambda_{HH/LH}$ in Eqs.~\eqref{eq:phiz_partial}--\eqref{eq:finalB_HH_1}, although these quantities depend on the corresponding finite-well levels $n$ and $n'$.

\clearpage
\section{Summary of the parameters used in this work}\label{app:parameters}

\begin{table*}[ht]
\centering
\renewcommand{\arraystretch}{1.45}
\begin{tabular}{c|c|p{0.75\textwidth}}
\hline
Quantity & Value & \centering Role \tabularnewline
\hline
$\gamma_1,\gamma_2,\gamma_3$ & $13.38,\,4.24,\,5.69$ & Ge Luttinger--Kohn bulk parameters. \\
$\Delta_{\mathrm{SO}}$ & $0.29~\mathrm{eV}$ & Split-off band separation; justifies restricting the valence-band model to the $J=3/2$ subspace. \\
$\bar{\kappa},q$ & $3.41,\,0.067$ & Isotropic and cubic Zeeman parameters for Ge holes. \\
$a_0$ & $50~\mathrm{nm}$ & Characteristic HH lateral confinement length; the corresponding LH lateral length is $42~\mathrm{nm}$. \\
$a$ & $100~\mathrm{nm}$ & Half interdot separation used in the results; the center-to-center distance is $2a$. \\
$d_w$ & $18~\mathrm{nm}$ & Width of the Ge quantum well in the growth direction. \\
$d_i$ & $54~\mathrm{nm}$ & Extent of the upper barrier region in the growth direction, we chose $d_i=3d_w$. \\
$U_{HH}$ & $100~\mathrm{meV}$ & Effective HH finite-well depth. \\
$U_{LH}$ & $217$ or $232~\mathrm{meV}$ & Effective LH finite-well depth; used to tune the relative HH--LH subband alignment and the low-energy eigenstate character. \\
$\perp$ states & $\mathrm{HH}_{n=2}$, $\mathrm{LH}_{n=1}$ & Opposite-parity HH--LH pair retained in the truncated Hilbert space; the lower HH $n=1$ level is assumed to be filled and inert. \\
$\alpha_R\langle E_z\rangle$ & $12~\mathrm{meV\,nm}$ & Standard Rashba-type SOI prefactor used in the truncated HH--LH envelope-function model. \\
$\omega_c/2\pi$ & $5$--$20~\mathrm{GHz}$ & Cavity-frequency range used when evaluating the transmission response. \\
$g_c/2\pi$ & $50~\mathrm{MHz}$ & Bare charge--photon coupling, taken approximately equal in the HH and LH sectors. \\
$\kappa/2\pi$ & $2~\mathrm{MHz}$ & Cavity linewidth used in the input--output response. \\
$\gamma_{\rm rel}$ & $100~\mathrm{MHz}$ & Environment-induced charge relaxation rate used in the bare decoherence model. \\
$\gamma_\phi^{(0)}$ & $50~\mathrm{MHz}$ & Residual constant pure-dephasing floor. \\
$\gamma_\phi^{(1)}$ & $150~\mathrm{MHz}$ & Maximum detuning-noise contribution to the dephasing, proportional to $\partial(E_m-E_0)/\partial\epsilon$. \\
$\gamma_s$ & $0~\mathrm{MHz}$ & Residual decay rate assigned to predominantly spin-like bare coherences. \\
\hline
\end{tabular}
\caption{Parameters used in the numerical simulations.}
\label{tab:simulation_parameters}
\renewcommand{\arraystretch}{1}
\end{table*} 

\twocolumngrid

\begin{thebibliography}{58}%
\makeatletter
\providecommand \@ifxundefined [1]{%
 \@ifx{#1\undefined}
}%
\providecommand \@ifnum [1]{%
 \ifnum #1\expandafter \@firstoftwo
 \else \expandafter \@secondoftwo
 \fi
}%
\providecommand \@ifx [1]{%
 \ifx #1\expandafter \@firstoftwo
 \else \expandafter \@secondoftwo
 \fi
}%
\providecommand \natexlab [1]{#1}%
\providecommand \enquote  [1]{``#1''}%
\providecommand \bibnamefont  [1]{#1}%
\providecommand \bibfnamefont [1]{#1}%
\providecommand \citenamefont [1]{#1}%
\providecommand \href@noop [0]{\@secondoftwo}%
\providecommand \href [0]{\begingroup \@sanitize@url \@href}%
\providecommand \@href[1]{\@@startlink{#1}\@@href}%
\providecommand \@@href[1]{\endgroup#1\@@endlink}%
\providecommand \@sanitize@url [0]{\catcode `\\12\catcode `\$12\catcode
  `\&12\catcode `\#12\catcode `\^12\catcode `\_12\catcode `\%12\relax}%
\providecommand \@@startlink[1]{}%
\providecommand \@@endlink[0]{}%
\providecommand \url  [0]{\begingroup\@sanitize@url \@url }%
\providecommand \@url [1]{\endgroup\@href {#1}{\urlprefix }}%
\providecommand \urlprefix  [0]{URL }%
\providecommand \Eprint [0]{\href }%
\providecommand \doibase [0]{https://doi.org/}%
\providecommand \selectlanguage [0]{\@gobble}%
\providecommand \bibinfo  [0]{\@secondoftwo}%
\providecommand \bibfield  [0]{\@secondoftwo}%
\providecommand \translation [1]{[#1]}%
\providecommand \BibitemOpen [0]{}%
\providecommand \bibitemStop [0]{}%
\providecommand \bibitemNoStop [0]{.\EOS\space}%
\providecommand \EOS [0]{\spacefactor3000\relax}%
\providecommand \BibitemShut  [1]{\csname bibitem#1\endcsname}%
\let\auto@bib@innerbib\@empty
\bibitem [{\citenamefont {Loss}\ and\ \citenamefont
  {DiVincenzo}(1998)}]{Loss_DiVincenzo_1998}%
  \BibitemOpen
  \bibfield  {author} {\bibinfo {author} {\bibfnamefont {D.}~\bibnamefont
  {Loss}}\ and\ \bibinfo {author} {\bibfnamefont {D.~P.}\ \bibnamefont
  {DiVincenzo}},\ }\bibfield  {title} {\bibinfo {title} {Quantum computation
  with quantum dots},\ }\href {https://doi.org/10.1103/PhysRevA.57.120}
  {\bibfield  {journal} {\bibinfo  {journal} {Phys. Rev. A}\ }\textbf {\bibinfo
  {volume} {57}},\ \bibinfo {pages} {120} (\bibinfo {year} {1998})}\BibitemShut
  {NoStop}%
\bibitem [{\citenamefont {Hanson}\ \emph {et~al.}(2007)\citenamefont {Hanson},
  \citenamefont {Kouwenhoven}, \citenamefont {Petta}, \citenamefont {Tarucha},\
  and\ \citenamefont {Vandersypen}}]{Hanson_RMP_2007}%
  \BibitemOpen
  \bibfield  {author} {\bibinfo {author} {\bibfnamefont {R.}~\bibnamefont
  {Hanson}}, \bibinfo {author} {\bibfnamefont {L.~P.}\ \bibnamefont
  {Kouwenhoven}}, \bibinfo {author} {\bibfnamefont {J.~R.}\ \bibnamefont
  {Petta}}, \bibinfo {author} {\bibfnamefont {S.}~\bibnamefont {Tarucha}},\
  and\ \bibinfo {author} {\bibfnamefont {L.~M.~K.}\ \bibnamefont
  {Vandersypen}},\ }\bibfield  {title} {\bibinfo {title} {Spins in few-electron
  quantum dots},\ }\href {https://doi.org/10.1103/RevModPhys.79.1217}
  {\bibfield  {journal} {\bibinfo  {journal} {Rev. Mod. Phys.}\ }\textbf
  {\bibinfo {volume} {79}},\ \bibinfo {pages} {1217} (\bibinfo {year}
  {2007})}\BibitemShut {NoStop}%
\bibitem [{\citenamefont {Zwanenburg}\ \emph {et~al.}(2013)\citenamefont
  {Zwanenburg}, \citenamefont {Dzurak}, \citenamefont {Morello}, \citenamefont
  {Simmons}, \citenamefont {Hollenberg}, \citenamefont {Klimeck}, \citenamefont
  {Rogge}, \citenamefont {Coppersmith},\ and\ \citenamefont
  {Eriksson}}]{Zwanenburg_RMP_2013}%
  \BibitemOpen
  \bibfield  {author} {\bibinfo {author} {\bibfnamefont {F.~A.}\ \bibnamefont
  {Zwanenburg}}, \bibinfo {author} {\bibfnamefont {A.~S.}\ \bibnamefont
  {Dzurak}}, \bibinfo {author} {\bibfnamefont {A.}~\bibnamefont {Morello}},
  \bibinfo {author} {\bibfnamefont {M.~Y.}\ \bibnamefont {Simmons}}, \bibinfo
  {author} {\bibfnamefont {L.~C.~L.}\ \bibnamefont {Hollenberg}}, \bibinfo
  {author} {\bibfnamefont {G.}~\bibnamefont {Klimeck}}, \bibinfo {author}
  {\bibfnamefont {S.}~\bibnamefont {Rogge}}, \bibinfo {author} {\bibfnamefont
  {S.~N.}\ \bibnamefont {Coppersmith}},\ and\ \bibinfo {author} {\bibfnamefont
  {M.~A.}\ \bibnamefont {Eriksson}},\ }\bibfield  {title} {\bibinfo {title}
  {Silicon quantum electronics},\ }\href
  {https://doi.org/10.1103/RevModPhys.85.961} {\bibfield  {journal} {\bibinfo
  {journal} {Rev. Mod. Phys.}\ }\textbf {\bibinfo {volume} {85}},\ \bibinfo
  {pages} {961} (\bibinfo {year} {2013})}\BibitemShut {NoStop}%
\bibitem [{\citenamefont {Burkard}\ \emph {et~al.}(2023)\citenamefont
  {Burkard}, \citenamefont {Ladd}, \citenamefont {Pan}, \citenamefont
  {Nichol},\ and\ \citenamefont {Petta}}]{Burkard_2023}%
  \BibitemOpen
  \bibfield  {author} {\bibinfo {author} {\bibfnamefont {G.}~\bibnamefont
  {Burkard}}, \bibinfo {author} {\bibfnamefont {T.~D.}\ \bibnamefont {Ladd}},
  \bibinfo {author} {\bibfnamefont {A.}~\bibnamefont {Pan}}, \bibinfo {author}
  {\bibfnamefont {J.~M.}\ \bibnamefont {Nichol}},\ and\ \bibinfo {author}
  {\bibfnamefont {J.~R.}\ \bibnamefont {Petta}},\ }\bibfield  {title} {\bibinfo
  {title} {Semiconductor spin qubits},\ }\href
  {https://doi.org/10.1103/RevModPhys.95.025003} {\bibfield  {journal}
  {\bibinfo  {journal} {Rev. Mod. Phys.}\ }\textbf {\bibinfo {volume} {95}},\
  \bibinfo {pages} {025003} (\bibinfo {year} {2023})}\BibitemShut {NoStop}%
\bibitem [{\citenamefont {Tokura}\ \emph {et~al.}(2006)\citenamefont {Tokura},
  \citenamefont {van~der Wiel}, \citenamefont {Obata},\ and\ \citenamefont
  {Tarucha}}]{Tokura_PRL_2006}%
  \BibitemOpen
  \bibfield  {author} {\bibinfo {author} {\bibfnamefont {Y.}~\bibnamefont
  {Tokura}}, \bibinfo {author} {\bibfnamefont {W.~G.}\ \bibnamefont {van~der
  Wiel}}, \bibinfo {author} {\bibfnamefont {T.}~\bibnamefont {Obata}},\ and\
  \bibinfo {author} {\bibfnamefont {S.}~\bibnamefont {Tarucha}},\ }\bibfield
  {title} {\bibinfo {title} {Coherent single electron spin control in a
  slanting zeeman field},\ }\href
  {https://doi.org/10.1103/PhysRevLett.96.047202} {\bibfield  {journal}
  {\bibinfo  {journal} {Phys. Rev. Lett.}\ }\textbf {\bibinfo {volume} {96}},\
  \bibinfo {pages} {047202} (\bibinfo {year} {2006})}\BibitemShut {NoStop}%
\bibitem [{\citenamefont {Golovach}\ \emph {et~al.}(2006)\citenamefont
  {Golovach}, \citenamefont {Borhani},\ and\ \citenamefont
  {Loss}}]{Golovach_PRB_2006}%
  \BibitemOpen
  \bibfield  {author} {\bibinfo {author} {\bibfnamefont {V.~N.}\ \bibnamefont
  {Golovach}}, \bibinfo {author} {\bibfnamefont {M.}~\bibnamefont {Borhani}},\
  and\ \bibinfo {author} {\bibfnamefont {D.}~\bibnamefont {Loss}},\ }\bibfield
  {title} {\bibinfo {title} {Electric-dipole-induced spin resonance in quantum
  dots},\ }\href {https://doi.org/10.1103/PhysRevB.74.165319} {\bibfield
  {journal} {\bibinfo  {journal} {Phys. Rev. B}\ }\textbf {\bibinfo {volume}
  {74}},\ \bibinfo {pages} {165319} (\bibinfo {year} {2006})}\BibitemShut
  {NoStop}%
\bibitem [{\citenamefont {Pioro-Ladri{\`e}re}\ \emph
  {et~al.}(2008)\citenamefont {Pioro-Ladri{\`e}re}, \citenamefont {Obata},
  \citenamefont {Tokura}, \citenamefont {Shin}, \citenamefont {Kubo},
  \citenamefont {Yoshida}, \citenamefont {Taniyama},\ and\ \citenamefont
  {Tarucha}}]{PioroLadriere_NatPhys_2008}%
  \BibitemOpen
  \bibfield  {author} {\bibinfo {author} {\bibfnamefont {M.}~\bibnamefont
  {Pioro-Ladri{\`e}re}}, \bibinfo {author} {\bibfnamefont {T.}~\bibnamefont
  {Obata}}, \bibinfo {author} {\bibfnamefont {Y.}~\bibnamefont {Tokura}},
  \bibinfo {author} {\bibfnamefont {Y.-S.}\ \bibnamefont {Shin}}, \bibinfo
  {author} {\bibfnamefont {T.}~\bibnamefont {Kubo}}, \bibinfo {author}
  {\bibfnamefont {K.}~\bibnamefont {Yoshida}}, \bibinfo {author} {\bibfnamefont
  {T.}~\bibnamefont {Taniyama}},\ and\ \bibinfo {author} {\bibfnamefont
  {S.}~\bibnamefont {Tarucha}},\ }\bibfield  {title} {\bibinfo {title}
  {Electrically driven single-electron spin resonance in a slanting zeeman
  field},\ }\href {https://doi.org/10.1038/nphys1053} {\bibfield  {journal}
  {\bibinfo  {journal} {Nature Physics}\ }\textbf {\bibinfo {volume} {4}},\
  \bibinfo {pages} {776} (\bibinfo {year} {2008})}\BibitemShut {NoStop}%
\bibitem [{\citenamefont {Medford}\ \emph
  {et~al.}(2013{\natexlab{a}})\citenamefont {Medford}, \citenamefont {Beil},
  \citenamefont {Taylor}, \citenamefont {Bartlett}, \citenamefont {Doherty},
  \citenamefont {Rashba}, \citenamefont {DiVincenzo}, \citenamefont {Lu},
  \citenamefont {Gossard},\ and\ \citenamefont
  {Marcus}}]{Medford_NatNano_2013}%
  \BibitemOpen
  \bibfield  {author} {\bibinfo {author} {\bibfnamefont {J.}~\bibnamefont
  {Medford}}, \bibinfo {author} {\bibfnamefont {J.}~\bibnamefont {Beil}},
  \bibinfo {author} {\bibfnamefont {J.~M.}\ \bibnamefont {Taylor}}, \bibinfo
  {author} {\bibfnamefont {S.~D.}\ \bibnamefont {Bartlett}}, \bibinfo {author}
  {\bibfnamefont {A.~C.}\ \bibnamefont {Doherty}}, \bibinfo {author}
  {\bibfnamefont {E.~I.}\ \bibnamefont {Rashba}}, \bibinfo {author}
  {\bibfnamefont {D.~P.}\ \bibnamefont {DiVincenzo}}, \bibinfo {author}
  {\bibfnamefont {H.}~\bibnamefont {Lu}}, \bibinfo {author} {\bibfnamefont
  {A.~C.}\ \bibnamefont {Gossard}},\ and\ \bibinfo {author} {\bibfnamefont
  {C.~M.}\ \bibnamefont {Marcus}},\ }\bibfield  {title} {\bibinfo {title}
  {Self-consistent measurement and state tomography of an exchange-only spin
  qubit},\ }\href {https://doi.org/10.1038/nnano.2013.168} {\bibfield
  {journal} {\bibinfo  {journal} {Nature Nanotechnology}\ }\textbf {\bibinfo
  {volume} {8}},\ \bibinfo {pages} {654} (\bibinfo {year}
  {2013}{\natexlab{a}})}\BibitemShut {NoStop}%
\bibitem [{\citenamefont {Reed}\ \emph {et~al.}(2016)\citenamefont {Reed},
  \citenamefont {Maune}, \citenamefont {Andrews}, \citenamefont {Borselli},
  \citenamefont {Eng}, \citenamefont {Jura}, \citenamefont {Kiselev},
  \citenamefont {Ladd}, \citenamefont {Merkel}, \citenamefont {Milosavljevic}
  \emph {et~al.}}]{Reed_PRL_2016}%
  \BibitemOpen
  \bibfield  {author} {\bibinfo {author} {\bibfnamefont {M.~D.}\ \bibnamefont
  {Reed}}, \bibinfo {author} {\bibfnamefont {B.~M.}\ \bibnamefont {Maune}},
  \bibinfo {author} {\bibfnamefont {R.~W.}\ \bibnamefont {Andrews}}, \bibinfo
  {author} {\bibfnamefont {M.~G.}\ \bibnamefont {Borselli}}, \bibinfo {author}
  {\bibfnamefont {K.}~\bibnamefont {Eng}}, \bibinfo {author} {\bibfnamefont
  {M.~P.}\ \bibnamefont {Jura}}, \bibinfo {author} {\bibfnamefont {A.~A.}\
  \bibnamefont {Kiselev}}, \bibinfo {author} {\bibfnamefont {T.~D.}\
  \bibnamefont {Ladd}}, \bibinfo {author} {\bibfnamefont {S.~T.}\ \bibnamefont
  {Merkel}}, \bibinfo {author} {\bibfnamefont {I.}~\bibnamefont
  {Milosavljevic}}, \emph {et~al.},\ }\bibfield  {title} {\bibinfo {title}
  {Reduced sensitivity to charge noise in semiconductor spin qubits via
  symmetric operation},\ }\href
  {https://doi.org/10.1103/PhysRevLett.116.110402} {\bibfield  {journal}
  {\bibinfo  {journal} {Phys. Rev. Lett.}\ }\textbf {\bibinfo {volume} {116}},\
  \bibinfo {pages} {110402} (\bibinfo {year} {2016})}\BibitemShut {NoStop}%
\bibitem [{\citenamefont {Medford}\ \emph
  {et~al.}(2013{\natexlab{b}})\citenamefont {Medford}, \citenamefont {Beil},
  \citenamefont {Taylor}, \citenamefont {Rashba}, \citenamefont {Lu},
  \citenamefont {Gossard},\ and\ \citenamefont {Marcus}}]{Medford_prl_2013}%
  \BibitemOpen
  \bibfield  {author} {\bibinfo {author} {\bibfnamefont {J.}~\bibnamefont
  {Medford}}, \bibinfo {author} {\bibfnamefont {J.}~\bibnamefont {Beil}},
  \bibinfo {author} {\bibfnamefont {J.~M.}\ \bibnamefont {Taylor}}, \bibinfo
  {author} {\bibfnamefont {E.~I.}\ \bibnamefont {Rashba}}, \bibinfo {author}
  {\bibfnamefont {H.}~\bibnamefont {Lu}}, \bibinfo {author} {\bibfnamefont
  {A.~C.}\ \bibnamefont {Gossard}},\ and\ \bibinfo {author} {\bibfnamefont
  {C.~M.}\ \bibnamefont {Marcus}},\ }\bibfield  {title} {\bibinfo {title}
  {Quantum-dot-based resonant exchange qubit},\ }\href
  {https://doi.org/10.1103/PhysRevLett.111.050501} {\bibfield  {journal}
  {\bibinfo  {journal} {Phys. Rev. Lett.}\ }\textbf {\bibinfo {volume} {111}},\
  \bibinfo {pages} {050501} (\bibinfo {year} {2013}{\natexlab{b}})}\BibitemShut
  {NoStop}%
\bibitem [{\citenamefont {Croot}\ \emph {et~al.}(2020)\citenamefont {Croot},
  \citenamefont {Mi}, \citenamefont {Putz}, \citenamefont {Benito},
  \citenamefont {Borjans}, \citenamefont {Burkard},\ and\ \citenamefont
  {Petta}}]{Croot_PRR_2020_FloppingEDSR}%
  \BibitemOpen
  \bibfield  {author} {\bibinfo {author} {\bibfnamefont {X.}~\bibnamefont
  {Croot}}, \bibinfo {author} {\bibfnamefont {X.}~\bibnamefont {Mi}}, \bibinfo
  {author} {\bibfnamefont {S.}~\bibnamefont {Putz}}, \bibinfo {author}
  {\bibfnamefont {M.}~\bibnamefont {Benito}}, \bibinfo {author} {\bibfnamefont
  {F.}~\bibnamefont {Borjans}}, \bibinfo {author} {\bibfnamefont
  {G.}~\bibnamefont {Burkard}},\ and\ \bibinfo {author} {\bibfnamefont {J.~R.}\
  \bibnamefont {Petta}},\ }\bibfield  {title} {\bibinfo {title} {Flopping-mode
  electric dipole spin resonance},\ }\href
  {https://doi.org/10.1103/PhysRevResearch.2.012006} {\bibfield  {journal}
  {\bibinfo  {journal} {Phys. Rev. Res.}\ }\textbf {\bibinfo {volume} {2}},\
  \bibinfo {pages} {012006(R)} (\bibinfo {year} {2020})}\BibitemShut {NoStop}%
\bibitem [{\citenamefont {Hajati}\ \emph {et~al.}(2025)\citenamefont {Hajati},
  \citenamefont {Heinz},\ and\ \citenamefont
  {Burkard}}]{Hajati_PRR_2025_Crosstalk}%
  \BibitemOpen
  \bibfield  {author} {\bibinfo {author} {\bibfnamefont {Y.}~\bibnamefont
  {Hajati}}, \bibinfo {author} {\bibfnamefont {I.}~\bibnamefont {Heinz}},\ and\
  \bibinfo {author} {\bibfnamefont {G.}~\bibnamefont {Burkard}},\ }\bibfield
  {title} {\bibinfo {title} {Crosstalk analysis in single hole-spin qubits
  within highly anisotropic $g$-tensors},\ }\href
  {https://doi.org/10.1103/qzn2-l71q} {\bibfield  {journal} {\bibinfo
  {journal} {Phys. Rev. Res.}\ }\textbf {\bibinfo {volume} {7}},\ \bibinfo
  {pages} {023277} (\bibinfo {year} {2025})}\BibitemShut {NoStop}%
\bibitem [{\citenamefont {Benito}\ \emph {et~al.}(2019)\citenamefont {Benito},
  \citenamefont {Croot}, \citenamefont {Adelsberger}, \citenamefont {Putz},
  \citenamefont {Mi}, \citenamefont {Petta},\ and\ \citenamefont
  {Burkard}}]{Benito_PRB_2019_FloppingNoise}%
  \BibitemOpen
  \bibfield  {author} {\bibinfo {author} {\bibfnamefont {M.}~\bibnamefont
  {Benito}}, \bibinfo {author} {\bibfnamefont {X.}~\bibnamefont {Croot}},
  \bibinfo {author} {\bibfnamefont {C.}~\bibnamefont {Adelsberger}}, \bibinfo
  {author} {\bibfnamefont {S.}~\bibnamefont {Putz}}, \bibinfo {author}
  {\bibfnamefont {X.}~\bibnamefont {Mi}}, \bibinfo {author} {\bibfnamefont
  {J.~R.}\ \bibnamefont {Petta}},\ and\ \bibinfo {author} {\bibfnamefont
  {G.}~\bibnamefont {Burkard}},\ }\bibfield  {title} {\bibinfo {title}
  {Electric-field control and noise protection of the flopping-mode spin
  qubit},\ }\href {https://doi.org/10.1103/PhysRevB.100.125430} {\bibfield
  {journal} {\bibinfo  {journal} {Phys. Rev. B}\ }\textbf {\bibinfo {volume}
  {100}},\ \bibinfo {pages} {125430} (\bibinfo {year} {2019})}\BibitemShut
  {NoStop}%
\bibitem [{\citenamefont {Childress}\ \emph {et~al.}(2004)\citenamefont
  {Childress}, \citenamefont {S\o{}rensen},\ and\ \citenamefont
  {Lukin}}]{Childress_pra_2004}%
  \BibitemOpen
  \bibfield  {author} {\bibinfo {author} {\bibfnamefont {L.}~\bibnamefont
  {Childress}}, \bibinfo {author} {\bibfnamefont {A.~S.}\ \bibnamefont
  {S\o{}rensen}},\ and\ \bibinfo {author} {\bibfnamefont {M.~D.}\ \bibnamefont
  {Lukin}},\ }\bibfield  {title} {\bibinfo {title} {Mesoscopic cavity quantum
  electrodynamics with quantum dots},\ }\href
  {https://doi.org/10.1103/PhysRevA.69.042302} {\bibfield  {journal} {\bibinfo
  {journal} {Phys. Rev. A}\ }\textbf {\bibinfo {volume} {69}},\ \bibinfo
  {pages} {042302} (\bibinfo {year} {2004})}\BibitemShut {NoStop}%
\bibitem [{\citenamefont {Wallraff}\ \emph {et~al.}(2004)\citenamefont
  {Wallraff}, \citenamefont {Schuster}, \citenamefont {Blais}, \citenamefont
  {Frunzio}, \citenamefont {Huang}, \citenamefont {Majer}, \citenamefont
  {Kumar}, \citenamefont {Girvin},\ and\ \citenamefont
  {Schoelkopf}}]{Wallraff_Nature_2004}%
  \BibitemOpen
  \bibfield  {author} {\bibinfo {author} {\bibfnamefont {A.}~\bibnamefont
  {Wallraff}}, \bibinfo {author} {\bibfnamefont {D.~I.}\ \bibnamefont
  {Schuster}}, \bibinfo {author} {\bibfnamefont {A.}~\bibnamefont {Blais}},
  \bibinfo {author} {\bibfnamefont {L.}~\bibnamefont {Frunzio}}, \bibinfo
  {author} {\bibfnamefont {R.-S.}\ \bibnamefont {Huang}}, \bibinfo {author}
  {\bibfnamefont {J.}~\bibnamefont {Majer}}, \bibinfo {author} {\bibfnamefont
  {S.}~\bibnamefont {Kumar}}, \bibinfo {author} {\bibfnamefont {S.~M.}\
  \bibnamefont {Girvin}},\ and\ \bibinfo {author} {\bibfnamefont {R.~J.}\
  \bibnamefont {Schoelkopf}},\ }\bibfield  {title} {\bibinfo {title} {Strong
  coupling of a single photon to a superconducting qubit using circuit quantum
  electrodynamics},\ }\href {https://doi.org/10.1038/nature02851} {\bibfield
  {journal} {\bibinfo  {journal} {Nature}\ }\textbf {\bibinfo {volume} {431}},\
  \bibinfo {pages} {162} (\bibinfo {year} {2004})}\BibitemShut {NoStop}%
\bibitem [{\citenamefont {Blais}\ \emph {et~al.}(2021)\citenamefont {Blais},
  \citenamefont {Grimsmo}, \citenamefont {Girvin},\ and\ \citenamefont
  {Wallraff}}]{Blais_RMP_2021}%
  \BibitemOpen
  \bibfield  {author} {\bibinfo {author} {\bibfnamefont {A.}~\bibnamefont
  {Blais}}, \bibinfo {author} {\bibfnamefont {A.~L.}\ \bibnamefont {Grimsmo}},
  \bibinfo {author} {\bibfnamefont {S.~M.}\ \bibnamefont {Girvin}},\ and\
  \bibinfo {author} {\bibfnamefont {A.}~\bibnamefont {Wallraff}},\ }\bibfield
  {title} {\bibinfo {title} {Circuit quantum electrodynamics},\ }\href
  {https://doi.org/10.1103/RevModPhys.93.025005} {\bibfield  {journal}
  {\bibinfo  {journal} {Rev. Mod. Phys.}\ }\textbf {\bibinfo {volume} {93}},\
  \bibinfo {pages} {025005} (\bibinfo {year} {2021})}\BibitemShut {NoStop}%
\bibitem [{\citenamefont {Frey}\ \emph {et~al.}(2012)\citenamefont {Frey},
  \citenamefont {Leek}, \citenamefont {Beck}, \citenamefont {Blais},
  \citenamefont {Ihn}, \citenamefont {Ensslin},\ and\ \citenamefont
  {Wallraff}}]{Frey_PRL_2012}%
  \BibitemOpen
  \bibfield  {author} {\bibinfo {author} {\bibfnamefont {T.}~\bibnamefont
  {Frey}}, \bibinfo {author} {\bibfnamefont {P.~J.}\ \bibnamefont {Leek}},
  \bibinfo {author} {\bibfnamefont {M.}~\bibnamefont {Beck}}, \bibinfo {author}
  {\bibfnamefont {A.}~\bibnamefont {Blais}}, \bibinfo {author} {\bibfnamefont
  {T.}~\bibnamefont {Ihn}}, \bibinfo {author} {\bibfnamefont {K.}~\bibnamefont
  {Ensslin}},\ and\ \bibinfo {author} {\bibfnamefont {A.}~\bibnamefont
  {Wallraff}},\ }\bibfield  {title} {\bibinfo {title} {Dipole coupling of a
  double quantum dot to a microwave resonator},\ }\href
  {https://doi.org/10.1103/PhysRevLett.108.046807} {\bibfield  {journal}
  {\bibinfo  {journal} {Phys. Rev. Lett.}\ }\textbf {\bibinfo {volume} {108}},\
  \bibinfo {pages} {046807} (\bibinfo {year} {2012})}\BibitemShut {NoStop}%
\bibitem [{\citenamefont {Petersson}\ \emph {et~al.}(2012)\citenamefont
  {Petersson}, \citenamefont {McFaul}, \citenamefont {Schroer}, \citenamefont
  {Jung}, \citenamefont {Taylor}, \citenamefont {Houck},\ and\ \citenamefont
  {Petta}}]{Petersson_Nature_2012}%
  \BibitemOpen
  \bibfield  {author} {\bibinfo {author} {\bibfnamefont {K.~D.}\ \bibnamefont
  {Petersson}}, \bibinfo {author} {\bibfnamefont {L.~W.}\ \bibnamefont
  {McFaul}}, \bibinfo {author} {\bibfnamefont {M.~D.}\ \bibnamefont {Schroer}},
  \bibinfo {author} {\bibfnamefont {M.}~\bibnamefont {Jung}}, \bibinfo {author}
  {\bibfnamefont {J.~M.}\ \bibnamefont {Taylor}}, \bibinfo {author}
  {\bibfnamefont {A.~A.}\ \bibnamefont {Houck}},\ and\ \bibinfo {author}
  {\bibfnamefont {J.~R.}\ \bibnamefont {Petta}},\ }\bibfield  {title} {\bibinfo
  {title} {Circuit quantum electrodynamics with a spin qubit},\ }\href
  {https://doi.org/10.1038/nature11559} {\bibfield  {journal} {\bibinfo
  {journal} {Nature}\ }\textbf {\bibinfo {volume} {490}},\ \bibinfo {pages}
  {380} (\bibinfo {year} {2012})}\BibitemShut {NoStop}%
\bibitem [{\citenamefont {Viennot}\ \emph {et~al.}(2015)\citenamefont
  {Viennot}, \citenamefont {Dartiailh}, \citenamefont {Cottet},\ and\
  \citenamefont {Kontos}}]{Viennot_Science_2015}%
  \BibitemOpen
  \bibfield  {author} {\bibinfo {author} {\bibfnamefont {J.~J.}\ \bibnamefont
  {Viennot}}, \bibinfo {author} {\bibfnamefont {M.~C.}\ \bibnamefont
  {Dartiailh}}, \bibinfo {author} {\bibfnamefont {A.}~\bibnamefont {Cottet}},\
  and\ \bibinfo {author} {\bibfnamefont {T.}~\bibnamefont {Kontos}},\
  }\bibfield  {title} {\bibinfo {title} {Coherent coupling of a single spin to
  microwave cavity photons},\ }\href {https://doi.org/10.1126/science.aaa3786}
  {\bibfield  {journal} {\bibinfo  {journal} {Science}\ }\textbf {\bibinfo
  {volume} {349}},\ \bibinfo {pages} {408} (\bibinfo {year}
  {2015})}\BibitemShut {NoStop}%
\bibitem [{\citenamefont {Mi}\ \emph {et~al.}(2017)\citenamefont {Mi},
  \citenamefont {Cady}, \citenamefont {Zajac}, \citenamefont {Deelman},\ and\
  \citenamefont {Petta}}]{Mi_Science_2017}%
  \BibitemOpen
  \bibfield  {author} {\bibinfo {author} {\bibfnamefont {X.}~\bibnamefont
  {Mi}}, \bibinfo {author} {\bibfnamefont {J.~V.}\ \bibnamefont {Cady}},
  \bibinfo {author} {\bibfnamefont {D.~M.}\ \bibnamefont {Zajac}}, \bibinfo
  {author} {\bibfnamefont {P.~W.}\ \bibnamefont {Deelman}},\ and\ \bibinfo
  {author} {\bibfnamefont {J.~R.}\ \bibnamefont {Petta}},\ }\bibfield  {title}
  {\bibinfo {title} {Strong coupling of a single electron in silicon to a
  microwave photon},\ }\href {https://doi.org/10.1126/science.aal2469}
  {\bibfield  {journal} {\bibinfo  {journal} {Science}\ }\textbf {\bibinfo
  {volume} {355}},\ \bibinfo {pages} {156} (\bibinfo {year}
  {2017})}\BibitemShut {NoStop}%
\bibitem [{\citenamefont {Stockklauser}\ \emph {et~al.}(2017)\citenamefont
  {Stockklauser}, \citenamefont {Maisi}, \citenamefont {Basset}, \citenamefont
  {Cujia}, \citenamefont {Reichl}, \citenamefont {Wegscheider},\ and\
  \citenamefont {Wallraff}}]{Stockklauser_PRX_2017}%
  \BibitemOpen
  \bibfield  {author} {\bibinfo {author} {\bibfnamefont {A.}~\bibnamefont
  {Stockklauser}}, \bibinfo {author} {\bibfnamefont {V.~F.}\ \bibnamefont
  {Maisi}}, \bibinfo {author} {\bibfnamefont {J.}~\bibnamefont {Basset}},
  \bibinfo {author} {\bibfnamefont {K.~S.}\ \bibnamefont {Cujia}}, \bibinfo
  {author} {\bibfnamefont {C.}~\bibnamefont {Reichl}}, \bibinfo {author}
  {\bibfnamefont {W.}~\bibnamefont {Wegscheider}},\ and\ \bibinfo {author}
  {\bibfnamefont {A.}~\bibnamefont {Wallraff}},\ }\bibfield  {title} {\bibinfo
  {title} {Strong coupling cavity qed with gate-defined double quantum dots
  enabled by a high impedance resonator},\ }\href
  {https://doi.org/10.1103/PhysRevX.7.011030} {\bibfield  {journal} {\bibinfo
  {journal} {Phys. Rev. X}\ }\textbf {\bibinfo {volume} {7}},\ \bibinfo {pages}
  {011030} (\bibinfo {year} {2017})}\BibitemShut {NoStop}%
\bibitem [{\citenamefont {Benito}\ \emph {et~al.}(2017)\citenamefont {Benito},
  \citenamefont {Mi}, \citenamefont {Taylor}, \citenamefont {Petta},\ and\
  \citenamefont {Burkard}}]{input_output_benito_prb}%
  \BibitemOpen
  \bibfield  {author} {\bibinfo {author} {\bibfnamefont {M.}~\bibnamefont
  {Benito}}, \bibinfo {author} {\bibfnamefont {X.}~\bibnamefont {Mi}}, \bibinfo
  {author} {\bibfnamefont {J.~M.}\ \bibnamefont {Taylor}}, \bibinfo {author}
  {\bibfnamefont {J.~R.}\ \bibnamefont {Petta}},\ and\ \bibinfo {author}
  {\bibfnamefont {G.}~\bibnamefont {Burkard}},\ }\bibfield  {title} {\bibinfo
  {title} {Input-output theory for spin-photon coupling in si double quantum
  dots},\ }\href {https://doi.org/10.1103/PhysRevB.96.235434} {\bibfield
  {journal} {\bibinfo  {journal} {Phys. Rev. B}\ }\textbf {\bibinfo {volume}
  {96}},\ \bibinfo {pages} {235434} (\bibinfo {year} {2017})}\BibitemShut
  {NoStop}%
\bibitem [{\citenamefont {Samkharadze}\ \emph {et~al.}(2018)\citenamefont
  {Samkharadze}, \citenamefont {Zheng}, \citenamefont {Kalhor}, \citenamefont
  {Brousse}, \citenamefont {Sammak}, \citenamefont {Mendes}, \citenamefont
  {Blais},\ and\ \citenamefont {Vandersypen}}]{Samkharadze_Science_2018}%
  \BibitemOpen
  \bibfield  {author} {\bibinfo {author} {\bibfnamefont {N.}~\bibnamefont
  {Samkharadze}}, \bibinfo {author} {\bibfnamefont {G.}~\bibnamefont {Zheng}},
  \bibinfo {author} {\bibfnamefont {N.}~\bibnamefont {Kalhor}}, \bibinfo
  {author} {\bibfnamefont {D.}~\bibnamefont {Brousse}}, \bibinfo {author}
  {\bibfnamefont {A.}~\bibnamefont {Sammak}}, \bibinfo {author} {\bibfnamefont
  {U.~C.}\ \bibnamefont {Mendes}}, \bibinfo {author} {\bibfnamefont
  {A.}~\bibnamefont {Blais}},\ and\ \bibinfo {author} {\bibfnamefont
  {L.~M.~K.}\ \bibnamefont {Vandersypen}},\ }\bibfield  {title} {\bibinfo
  {title} {Strong spin-photon coupling in silicon},\ }\href
  {https://doi.org/10.1126/science.aar4054} {\bibfield  {journal} {\bibinfo
  {journal} {Science}\ }\textbf {\bibinfo {volume} {359}},\ \bibinfo {pages}
  {1123} (\bibinfo {year} {2018})}\BibitemShut {NoStop}%
\bibitem [{\citenamefont {Mi}\ \emph {et~al.}(2018)\citenamefont {Mi},
  \citenamefont {Benito}, \citenamefont {Putz}, \citenamefont {Zajac},
  \citenamefont {Taylor}, \citenamefont {Burkard},\ and\ \citenamefont
  {Petta}}]{Mi_Nature_2018_SpinPhoton}%
  \BibitemOpen
  \bibfield  {author} {\bibinfo {author} {\bibfnamefont {X.}~\bibnamefont
  {Mi}}, \bibinfo {author} {\bibfnamefont {M.}~\bibnamefont {Benito}}, \bibinfo
  {author} {\bibfnamefont {S.}~\bibnamefont {Putz}}, \bibinfo {author}
  {\bibfnamefont {D.~M.}\ \bibnamefont {Zajac}}, \bibinfo {author}
  {\bibfnamefont {J.~M.}\ \bibnamefont {Taylor}}, \bibinfo {author}
  {\bibfnamefont {G.}~\bibnamefont {Burkard}},\ and\ \bibinfo {author}
  {\bibfnamefont {J.~R.}\ \bibnamefont {Petta}},\ }\bibfield  {title} {\bibinfo
  {title} {A coherent spin-photon interface in silicon},\ }\href
  {https://doi.org/10.1038/nature25769} {\bibfield  {journal} {\bibinfo
  {journal} {Nature}\ }\textbf {\bibinfo {volume} {555}},\ \bibinfo {pages}
  {599} (\bibinfo {year} {2018})}\BibitemShut {NoStop}%
\bibitem [{\citenamefont {Landig}\ \emph {et~al.}(2018)\citenamefont {Landig},
  \citenamefont {Koski}, \citenamefont {Scarlino}, \citenamefont {Mendes},
  \citenamefont {Blais}, \citenamefont {Reichl}, \citenamefont {Wegscheider},
  \citenamefont {Wallraff}, \citenamefont {Ensslin},\ and\ \citenamefont
  {Ihn}}]{Landig_Nature_2018}%
  \BibitemOpen
  \bibfield  {author} {\bibinfo {author} {\bibfnamefont {A.~J.}\ \bibnamefont
  {Landig}}, \bibinfo {author} {\bibfnamefont {J.~V.}\ \bibnamefont {Koski}},
  \bibinfo {author} {\bibfnamefont {P.}~\bibnamefont {Scarlino}}, \bibinfo
  {author} {\bibfnamefont {U.~C.}\ \bibnamefont {Mendes}}, \bibinfo {author}
  {\bibfnamefont {A.}~\bibnamefont {Blais}}, \bibinfo {author} {\bibfnamefont
  {C.}~\bibnamefont {Reichl}}, \bibinfo {author} {\bibfnamefont
  {W.}~\bibnamefont {Wegscheider}}, \bibinfo {author} {\bibfnamefont
  {A.}~\bibnamefont {Wallraff}}, \bibinfo {author} {\bibfnamefont
  {K.}~\bibnamefont {Ensslin}},\ and\ \bibinfo {author} {\bibfnamefont
  {T.}~\bibnamefont {Ihn}},\ }\bibfield  {title} {\bibinfo {title} {Coherent
  spin--photon coupling using a resonant exchange qubit},\ }\href
  {https://doi.org/10.1038/s41586-018-0365-y} {\bibfield  {journal} {\bibinfo
  {journal} {Nature}\ }\textbf {\bibinfo {volume} {560}},\ \bibinfo {pages}
  {179} (\bibinfo {year} {2018})}\BibitemShut {NoStop}%
\bibitem [{\citenamefont {Borjans}\ \emph {et~al.}(2020)\citenamefont
  {Borjans}, \citenamefont {Croot}, \citenamefont {Mi}, \citenamefont
  {Gullans},\ and\ \citenamefont {Petta}}]{Borjans_Nature_2020}%
  \BibitemOpen
  \bibfield  {author} {\bibinfo {author} {\bibfnamefont {F.}~\bibnamefont
  {Borjans}}, \bibinfo {author} {\bibfnamefont {X.~G.}\ \bibnamefont {Croot}},
  \bibinfo {author} {\bibfnamefont {X.}~\bibnamefont {Mi}}, \bibinfo {author}
  {\bibfnamefont {M.~J.}\ \bibnamefont {Gullans}},\ and\ \bibinfo {author}
  {\bibfnamefont {J.~R.}\ \bibnamefont {Petta}},\ }\bibfield  {title} {\bibinfo
  {title} {Resonant microwave-mediated interactions between distant electron
  spins},\ }\href {https://doi.org/10.1038/s41586-019-1867-y} {\bibfield
  {journal} {\bibinfo  {journal} {Nature}\ }\textbf {\bibinfo {volume} {577}},\
  \bibinfo {pages} {195} (\bibinfo {year} {2020})}\BibitemShut {NoStop}%
\bibitem [{\citenamefont {Harvey-Collard}\ \emph {et~al.}(2022)\citenamefont
  {Harvey-Collard}, \citenamefont {D'Anjou}, \citenamefont {Rudolph},
  \citenamefont {Jacobson}, \citenamefont {Dominguez}, \citenamefont {Mounce},
  \citenamefont {Mu{\"n}ks}, \citenamefont {Coop}, \citenamefont {Ward},
  \citenamefont {Friesen}, \citenamefont {Eriksson},\ and\ \citenamefont
  {Petta}}]{HarveyCollard_PRX_2022}%
  \BibitemOpen
  \bibfield  {author} {\bibinfo {author} {\bibfnamefont {P.}~\bibnamefont
  {Harvey-Collard}}, \bibinfo {author} {\bibfnamefont {B.}~\bibnamefont
  {D'Anjou}}, \bibinfo {author} {\bibfnamefont {M.}~\bibnamefont {Rudolph}},
  \bibinfo {author} {\bibfnamefont {N.~T.}\ \bibnamefont {Jacobson}}, \bibinfo
  {author} {\bibfnamefont {J.}~\bibnamefont {Dominguez}}, \bibinfo {author}
  {\bibfnamefont {A.~M.}\ \bibnamefont {Mounce}}, \bibinfo {author}
  {\bibfnamefont {M.}~\bibnamefont {Mu{\"n}ks}}, \bibinfo {author}
  {\bibfnamefont {S.}~\bibnamefont {Coop}}, \bibinfo {author} {\bibfnamefont
  {D.~R.}\ \bibnamefont {Ward}}, \bibinfo {author} {\bibfnamefont
  {M.}~\bibnamefont {Friesen}}, \bibinfo {author} {\bibfnamefont {M.~A.}\
  \bibnamefont {Eriksson}},\ and\ \bibinfo {author} {\bibfnamefont {J.~R.}\
  \bibnamefont {Petta}},\ }\bibfield  {title} {\bibinfo {title} {Coherent
  spin-spin coupling mediated by virtual microwave photons},\ }\href
  {https://doi.org/10.1103/PhysRevX.12.021026} {\bibfield  {journal} {\bibinfo
  {journal} {Phys. Rev. X}\ }\textbf {\bibinfo {volume} {12}},\ \bibinfo
  {pages} {021026} (\bibinfo {year} {2022})}\BibitemShut {NoStop}%
\bibitem [{\citenamefont {D'Anjou}\ and\ \citenamefont
  {Burkard}(2019)}]{DAnjou_Burkard_PRB_2019_OptimalReadout}%
  \BibitemOpen
  \bibfield  {author} {\bibinfo {author} {\bibfnamefont {B.}~\bibnamefont
  {D'Anjou}}\ and\ \bibinfo {author} {\bibfnamefont {G.}~\bibnamefont
  {Burkard}},\ }\bibfield  {title} {\bibinfo {title} {Optimal dispersive
  readout of a spin qubit with a microwave resonator},\ }\href
  {https://doi.org/10.1103/PhysRevB.100.245427} {\bibfield  {journal} {\bibinfo
   {journal} {Phys. Rev. B}\ }\textbf {\bibinfo {volume} {100}},\ \bibinfo
  {pages} {245427} (\bibinfo {year} {2019})}\BibitemShut {NoStop}%
\bibitem [{\citenamefont {Kloeffel}\ and\ \citenamefont
  {Loss}(2013)}]{Kloeffel_ARCMP_2013}%
  \BibitemOpen
  \bibfield  {author} {\bibinfo {author} {\bibfnamefont {C.}~\bibnamefont
  {Kloeffel}}\ and\ \bibinfo {author} {\bibfnamefont {D.}~\bibnamefont
  {Loss}},\ }\bibfield  {title} {\bibinfo {title} {Prospects for spin-based
  quantum computing in quantum dots},\ }\href
  {https://doi.org/10.1146/annurev-conmatphys-030212-184248} {\bibfield
  {journal} {\bibinfo  {journal} {Annual Review of Condensed Matter Physics}\
  }\textbf {\bibinfo {volume} {4}},\ \bibinfo {pages} {51} (\bibinfo {year}
  {2013})}\BibitemShut {NoStop}%
\bibitem [{\citenamefont {Scappucci}\ \emph {et~al.}(2021)\citenamefont
  {Scappucci}, \citenamefont {Kloeffel}, \citenamefont {Zwanenburg},
  \citenamefont {Loss}, \citenamefont {Myronov}, \citenamefont {Zhang},
  \citenamefont {De~Franceschi}, \citenamefont {Katsaros},\ and\ \citenamefont
  {Veldhorst}}]{Scappucci_NatRevMater_2021}%
  \BibitemOpen
  \bibfield  {author} {\bibinfo {author} {\bibfnamefont {G.}~\bibnamefont
  {Scappucci}}, \bibinfo {author} {\bibfnamefont {C.}~\bibnamefont {Kloeffel}},
  \bibinfo {author} {\bibfnamefont {F.~A.}\ \bibnamefont {Zwanenburg}},
  \bibinfo {author} {\bibfnamefont {D.}~\bibnamefont {Loss}}, \bibinfo {author}
  {\bibfnamefont {M.}~\bibnamefont {Myronov}}, \bibinfo {author} {\bibfnamefont
  {J.-J.}\ \bibnamefont {Zhang}}, \bibinfo {author} {\bibfnamefont
  {S.}~\bibnamefont {De~Franceschi}}, \bibinfo {author} {\bibfnamefont
  {G.}~\bibnamefont {Katsaros}},\ and\ \bibinfo {author} {\bibfnamefont
  {M.}~\bibnamefont {Veldhorst}},\ }\bibfield  {title} {\bibinfo {title} {The
  germanium quantum information route},\ }\href
  {https://doi.org/10.1038/s41578-020-00262-z} {\bibfield  {journal} {\bibinfo
  {journal} {Nature Reviews Materials}\ }\textbf {\bibinfo {volume} {6}},\
  \bibinfo {pages} {926} (\bibinfo {year} {2021})}\BibitemShut {NoStop}%
\bibitem [{\citenamefont {Sammak}\ \emph {et~al.}(2019)\citenamefont {Sammak},
  \citenamefont {Sabbagh}, \citenamefont {Hendrickx}, \citenamefont {Lodari},
  \citenamefont {Paquelet~Wuetz}, \citenamefont {Tosato}, \citenamefont {Yeoh},
  \citenamefont {Bollani}, \citenamefont {Virgilio}, \citenamefont {Schubert},
  \citenamefont {Zaumseil}, \citenamefont {Capellini}, \citenamefont
  {Veldhorst},\ and\ \citenamefont {Scappucci}}]{Sammak_AdvFunctMater_2019}%
  \BibitemOpen
  \bibfield  {author} {\bibinfo {author} {\bibfnamefont {A.}~\bibnamefont
  {Sammak}}, \bibinfo {author} {\bibfnamefont {D.}~\bibnamefont {Sabbagh}},
  \bibinfo {author} {\bibfnamefont {N.~W.}\ \bibnamefont {Hendrickx}}, \bibinfo
  {author} {\bibfnamefont {M.}~\bibnamefont {Lodari}}, \bibinfo {author}
  {\bibfnamefont {B.}~\bibnamefont {Paquelet~Wuetz}}, \bibinfo {author}
  {\bibfnamefont {A.}~\bibnamefont {Tosato}}, \bibinfo {author} {\bibfnamefont
  {L.}~\bibnamefont {Yeoh}}, \bibinfo {author} {\bibfnamefont {M.}~\bibnamefont
  {Bollani}}, \bibinfo {author} {\bibfnamefont {M.}~\bibnamefont {Virgilio}},
  \bibinfo {author} {\bibfnamefont {M.~A.}\ \bibnamefont {Schubert}}, \bibinfo
  {author} {\bibfnamefont {P.}~\bibnamefont {Zaumseil}}, \bibinfo {author}
  {\bibfnamefont {G.}~\bibnamefont {Capellini}}, \bibinfo {author}
  {\bibfnamefont {M.}~\bibnamefont {Veldhorst}},\ and\ \bibinfo {author}
  {\bibfnamefont {G.}~\bibnamefont {Scappucci}},\ }\bibfield  {title} {\bibinfo
  {title} {Shallow and undoped germanium quantum wells: A playground for spin
  and hybrid quantum technology},\ }\href
  {https://doi.org/https://doi.org/10.1002/adfm.201807613} {\bibfield
  {journal} {\bibinfo  {journal} {Advanced Functional Materials}\ }\textbf
  {\bibinfo {volume} {29}},\ \bibinfo {pages} {1807613} (\bibinfo {year}
  {2019})}\BibitemShut {NoStop}%
\bibitem [{\citenamefont {Lodari}\ \emph {et~al.}(2019)\citenamefont {Lodari},
  \citenamefont {Tosato}, \citenamefont {Sabbagh}, \citenamefont {Schubert},
  \citenamefont {Capellini}, \citenamefont {Sammak}, \citenamefont
  {Veldhorst},\ and\ \citenamefont {Scappucci}}]{Lodari_PRB_2019}%
  \BibitemOpen
  \bibfield  {author} {\bibinfo {author} {\bibfnamefont {M.}~\bibnamefont
  {Lodari}}, \bibinfo {author} {\bibfnamefont {A.}~\bibnamefont {Tosato}},
  \bibinfo {author} {\bibfnamefont {D.}~\bibnamefont {Sabbagh}}, \bibinfo
  {author} {\bibfnamefont {M.~A.}\ \bibnamefont {Schubert}}, \bibinfo {author}
  {\bibfnamefont {G.}~\bibnamefont {Capellini}}, \bibinfo {author}
  {\bibfnamefont {A.}~\bibnamefont {Sammak}}, \bibinfo {author} {\bibfnamefont
  {M.}~\bibnamefont {Veldhorst}},\ and\ \bibinfo {author} {\bibfnamefont
  {G.}~\bibnamefont {Scappucci}},\ }\bibfield  {title} {\bibinfo {title} {Light
  effective hole mass in undoped ge/sige quantum wells},\ }\href
  {https://doi.org/10.1103/PhysRevB.100.041304} {\bibfield  {journal} {\bibinfo
   {journal} {Phys. Rev. B}\ }\textbf {\bibinfo {volume} {100}},\ \bibinfo
  {pages} {041304} (\bibinfo {year} {2019})}\BibitemShut {NoStop}%
\bibitem [{\citenamefont {Hendrickx}\ \emph {et~al.}(2020)\citenamefont
  {Hendrickx}, \citenamefont {Franke}, \citenamefont {Sammak}, \citenamefont
  {Scappucci},\ and\ \citenamefont {Veldhorst}}]{Hendrickx_Nature_2020}%
  \BibitemOpen
  \bibfield  {author} {\bibinfo {author} {\bibfnamefont {N.~W.}\ \bibnamefont
  {Hendrickx}}, \bibinfo {author} {\bibfnamefont {D.~P.}\ \bibnamefont
  {Franke}}, \bibinfo {author} {\bibfnamefont {A.}~\bibnamefont {Sammak}},
  \bibinfo {author} {\bibfnamefont {G.}~\bibnamefont {Scappucci}},\ and\
  \bibinfo {author} {\bibfnamefont {M.}~\bibnamefont {Veldhorst}},\ }\bibfield
  {title} {\bibinfo {title} {Fast two-qubit logic with holes in germanium},\
  }\href {https://doi.org/10.1038/s41586-019-1919-3} {\bibfield  {journal}
  {\bibinfo  {journal} {Nature}\ }\textbf {\bibinfo {volume} {577}},\ \bibinfo
  {pages} {487} (\bibinfo {year} {2020})}\BibitemShut {NoStop}%
\bibitem [{\citenamefont {Hendrickx}\ \emph {et~al.}(2021)\citenamefont
  {Hendrickx}, \citenamefont {Lawrie}, \citenamefont {Russ}, \citenamefont {van
  Riggelen}, \citenamefont {de~Snoo}, \citenamefont {Schouten}, \citenamefont
  {Sammak}, \citenamefont {Scappucci},\ and\ \citenamefont
  {Veldhorst}}]{Hendrickx_Nature_2021}%
  \BibitemOpen
  \bibfield  {author} {\bibinfo {author} {\bibfnamefont {N.~W.}\ \bibnamefont
  {Hendrickx}}, \bibinfo {author} {\bibfnamefont {W.~I.~L.}\ \bibnamefont
  {Lawrie}}, \bibinfo {author} {\bibfnamefont {M.}~\bibnamefont {Russ}},
  \bibinfo {author} {\bibfnamefont {F.}~\bibnamefont {van Riggelen}}, \bibinfo
  {author} {\bibfnamefont {S.~L.}\ \bibnamefont {de~Snoo}}, \bibinfo {author}
  {\bibfnamefont {R.~N.}\ \bibnamefont {Schouten}}, \bibinfo {author}
  {\bibfnamefont {A.}~\bibnamefont {Sammak}}, \bibinfo {author} {\bibfnamefont
  {G.}~\bibnamefont {Scappucci}},\ and\ \bibinfo {author} {\bibfnamefont
  {M.}~\bibnamefont {Veldhorst}},\ }\bibfield  {title} {\bibinfo {title} {A
  four-qubit germanium quantum processor},\ }\href
  {https://doi.org/10.1038/s41586-021-03332-6} {\bibfield  {journal} {\bibinfo
  {journal} {Nature}\ }\textbf {\bibinfo {volume} {591}},\ \bibinfo {pages}
  {580} (\bibinfo {year} {2021})}\BibitemShut {NoStop}%
\bibitem [{\citenamefont {Wang}\ \emph {et~al.}(2024)\citenamefont {Wang},
  \citenamefont {Ercan}, \citenamefont {Gyure}, \citenamefont {Scappucci},
  \citenamefont {Veldhorst},\ and\ \citenamefont {Rimbach-Russ}}]{Wang2024}%
  \BibitemOpen
  \bibfield  {author} {\bibinfo {author} {\bibfnamefont {C.-A.}\ \bibnamefont
  {Wang}}, \bibinfo {author} {\bibfnamefont {H.~E.}\ \bibnamefont {Ercan}},
  \bibinfo {author} {\bibfnamefont {M.~F.}\ \bibnamefont {Gyure}}, \bibinfo
  {author} {\bibfnamefont {G.}~\bibnamefont {Scappucci}}, \bibinfo {author}
  {\bibfnamefont {M.}~\bibnamefont {Veldhorst}},\ and\ \bibinfo {author}
  {\bibfnamefont {M.}~\bibnamefont {Rimbach-Russ}},\ }\bibfield  {title}
  {\bibinfo {title} {Modeling of planar germanium hole qubits in electric and
  magnetic fields},\ }\href {https://doi.org/10.1038/s41534-024-00897-8}
  {\bibfield  {journal} {\bibinfo  {journal} {npj Quantum Information}\
  }\textbf {\bibinfo {volume} {10}},\ \bibinfo {pages} {102} (\bibinfo {year}
  {2024})}\BibitemShut {NoStop}%
\bibitem [{\citenamefont {Mutter}\ and\ \citenamefont
  {Burkard}(2021)}]{Mutter_PRR_2021_Natural_HHs}%
  \BibitemOpen
  \bibfield  {author} {\bibinfo {author} {\bibfnamefont {P.~M.}\ \bibnamefont
  {Mutter}}\ and\ \bibinfo {author} {\bibfnamefont {G.}~\bibnamefont
  {Burkard}},\ }\bibfield  {title} {\bibinfo {title} {Natural heavy-hole
  flopping mode qubit in germanium},\ }\href
  {https://doi.org/10.1103/PhysRevResearch.3.013194} {\bibfield  {journal}
  {\bibinfo  {journal} {Phys. Rev. Res.}\ }\textbf {\bibinfo {volume} {3}},\
  \bibinfo {pages} {013194} (\bibinfo {year} {2021})}\BibitemShut {NoStop}%
\bibitem [{\citenamefont {Hajati}\ and\ \citenamefont
  {Burkard}(2024)}]{Hajati_Burkard_PRB_2024_DynamicSweetSpot}%
  \BibitemOpen
  \bibfield  {author} {\bibinfo {author} {\bibfnamefont {Y.}~\bibnamefont
  {Hajati}}\ and\ \bibinfo {author} {\bibfnamefont {G.}~\bibnamefont
  {Burkard}},\ }\bibfield  {title} {\bibinfo {title} {Dynamic sweet spot of
  driven flopping-mode spin qubits in planar quantum dots},\ }\href
  {https://doi.org/10.1103/PhysRevB.110.245301} {\bibfield  {journal} {\bibinfo
   {journal} {Phys. Rev. B}\ }\textbf {\bibinfo {volume} {110}},\ \bibinfo
  {pages} {245301} (\bibinfo {year} {2024})}\BibitemShut {NoStop}%
\bibitem [{\citenamefont {Winkler}(2003)}]{Winkler_2003_SpinOrbit}%
  \BibitemOpen
  \bibfield  {author} {\bibinfo {author} {\bibfnamefont {R.}~\bibnamefont
  {Winkler}},\ }\href@noop {} {\emph {\bibinfo {title} {Spin-Orbit Coupling
  Effects in Two-Dimensional Electron and Hole Systems}}},\ \bibinfo {series}
  {Springer Tracts in Modern Physics}, Vol.\ \bibinfo {volume} {191}\ (\bibinfo
   {publisher} {Springer},\ \bibinfo {address} {Berlin},\ \bibinfo {year}
  {2003})\BibitemShut {NoStop}%
\bibitem [{\citenamefont {Bulaev}\ and\ \citenamefont
  {Loss}(2007)}]{Bulaev_Loss_PRB_2007}%
  \BibitemOpen
  \bibfield  {author} {\bibinfo {author} {\bibfnamefont {D.~V.}\ \bibnamefont
  {Bulaev}}\ and\ \bibinfo {author} {\bibfnamefont {D.}~\bibnamefont {Loss}},\
  }\bibfield  {title} {\bibinfo {title} {Electric dipole spin resonance for
  heavy holes in quantum dots},\ }\href
  {https://doi.org/10.1103/PhysRevLett.98.097202} {\bibfield  {journal}
  {\bibinfo  {journal} {Phys. Rev. Lett.}\ }\textbf {\bibinfo {volume} {98}},\
  \bibinfo {pages} {097202} (\bibinfo {year} {2007})}\BibitemShut {NoStop}%
\bibitem [{\citenamefont {Kloeffel}\ \emph {et~al.}(2013)\citenamefont
  {Kloeffel}, \citenamefont {Trif}, \citenamefont {Stano},\ and\ \citenamefont
  {Loss}}]{Kloeffel_PRB_2013_CircuitQED}%
  \BibitemOpen
  \bibfield  {author} {\bibinfo {author} {\bibfnamefont {C.}~\bibnamefont
  {Kloeffel}}, \bibinfo {author} {\bibfnamefont {M.}~\bibnamefont {Trif}},
  \bibinfo {author} {\bibfnamefont {P.}~\bibnamefont {Stano}},\ and\ \bibinfo
  {author} {\bibfnamefont {D.}~\bibnamefont {Loss}},\ }\bibfield  {title}
  {\bibinfo {title} {Circuit qed with hole-spin qubits in ge/si nanowire
  quantum dots},\ }\href {https://doi.org/10.1103/PhysRevB.88.241405}
  {\bibfield  {journal} {\bibinfo  {journal} {Phys. Rev. B}\ }\textbf {\bibinfo
  {volume} {88}},\ \bibinfo {pages} {241405} (\bibinfo {year}
  {2013})}\BibitemShut {NoStop}%
\bibitem [{\citenamefont {Kloeffel}\ \emph {et~al.}(2018)\citenamefont
  {Kloeffel}, \citenamefont {Ran{\v c}i{\'c}},\ and\ \citenamefont
  {Loss}}]{Loss_direct_rashba_2018}%
  \BibitemOpen
  \bibfield  {author} {\bibinfo {author} {\bibfnamefont {C.}~\bibnamefont
  {Kloeffel}}, \bibinfo {author} {\bibfnamefont {M.~J.}\ \bibnamefont {Ran{\v
  c}i{\'c}}},\ and\ \bibinfo {author} {\bibfnamefont {D.}~\bibnamefont
  {Loss}},\ }\bibfield  {title} {\bibinfo {title} {Direct rashba spin-orbit
  interaction in si and ge nanowires with different growth directions},\ }\href
  {https://doi.org/10.1103/PhysRevB.97.235422} {\bibfield  {journal} {\bibinfo
  {journal} {Phys. Rev. B}\ }\textbf {\bibinfo {volume} {97}},\ \bibinfo
  {pages} {235422} (\bibinfo {year} {2018})}\BibitemShut {NoStop}%
\bibitem [{\citenamefont {Adelsberger}\ \emph {et~al.}(2022)\citenamefont
  {Adelsberger}, \citenamefont {Benito}, \citenamefont {Bosco}, \citenamefont
  {Klinovaja},\ and\ \citenamefont {Loss}}]{Adelsberger_Benito_PRB_2022}%
  \BibitemOpen
  \bibfield  {author} {\bibinfo {author} {\bibfnamefont {C.}~\bibnamefont
  {Adelsberger}}, \bibinfo {author} {\bibfnamefont {M.}~\bibnamefont {Benito}},
  \bibinfo {author} {\bibfnamefont {S.}~\bibnamefont {Bosco}}, \bibinfo
  {author} {\bibfnamefont {J.}~\bibnamefont {Klinovaja}},\ and\ \bibinfo
  {author} {\bibfnamefont {D.}~\bibnamefont {Loss}},\ }\bibfield  {title}
  {\bibinfo {title} {Hole-spin qubits in ge nanowire quantum dots: Interplay of
  orbital magnetic field, strain, and growth direction},\ }\href
  {https://doi.org/10.1103/PhysRevB.105.075308} {\bibfield  {journal} {\bibinfo
   {journal} {Phys. Rev. B}\ }\textbf {\bibinfo {volume} {105}},\ \bibinfo
  {pages} {075308} (\bibinfo {year} {2022})}\BibitemShut {NoStop}%
\bibitem [{\citenamefont {Luttinger}\ and\ \citenamefont
  {Kohn}(1955)}]{Luttinger_Kohn_1955}%
  \BibitemOpen
  \bibfield  {author} {\bibinfo {author} {\bibfnamefont {J.~M.}\ \bibnamefont
  {Luttinger}}\ and\ \bibinfo {author} {\bibfnamefont {W.}~\bibnamefont
  {Kohn}},\ }\bibfield  {title} {\bibinfo {title} {Motion of electrons and
  holes in perturbed periodic fields},\ }\href
  {https://doi.org/10.1103/PhysRev.97.869} {\bibfield  {journal} {\bibinfo
  {journal} {Physical Review}\ }\textbf {\bibinfo {volume} {97}},\ \bibinfo
  {pages} {869} (\bibinfo {year} {1955})}\BibitemShut {NoStop}%
\bibitem [{\citenamefont {Vurgaftman}\ \emph {et~al.}(2001)\citenamefont
  {Vurgaftman}, \citenamefont {Meyer},\ and\ \citenamefont
  {Ram-Mohan}}]{Vurgaftman_JAP_2001}%
  \BibitemOpen
  \bibfield  {author} {\bibinfo {author} {\bibfnamefont {I.}~\bibnamefont
  {Vurgaftman}}, \bibinfo {author} {\bibfnamefont {J.~R.}\ \bibnamefont
  {Meyer}},\ and\ \bibinfo {author} {\bibfnamefont {L.~R.}\ \bibnamefont
  {Ram-Mohan}},\ }\bibfield  {title} {\bibinfo {title} {Band parameters for
  {III--V} compound semiconductors and their alloys},\ }\href
  {https://doi.org/10.1063/1.1368156} {\bibfield  {journal} {\bibinfo
  {journal} {Journal of Applied Physics}\ }\textbf {\bibinfo {volume} {89}},\
  \bibinfo {pages} {5815} (\bibinfo {year} {2001})}\BibitemShut {NoStop}%
\bibitem [{\citenamefont {Nichele}\ \emph {et~al.}(2014)\citenamefont
  {Nichele}, \citenamefont {Chesi}, \citenamefont {Hennel}, \citenamefont
  {Wittmann}, \citenamefont {Gerl}, \citenamefont {Wegscheider}, \citenamefont
  {Loss}, \citenamefont {Ihn},\ and\ \citenamefont
  {Ensslin}}]{Fabrizio_SOI_GaAs_HH_prl_2014}%
  \BibitemOpen
  \bibfield  {author} {\bibinfo {author} {\bibfnamefont {F.}~\bibnamefont
  {Nichele}}, \bibinfo {author} {\bibfnamefont {S.}~\bibnamefont {Chesi}},
  \bibinfo {author} {\bibfnamefont {S.}~\bibnamefont {Hennel}}, \bibinfo
  {author} {\bibfnamefont {A.}~\bibnamefont {Wittmann}}, \bibinfo {author}
  {\bibfnamefont {C.}~\bibnamefont {Gerl}}, \bibinfo {author} {\bibfnamefont
  {W.}~\bibnamefont {Wegscheider}}, \bibinfo {author} {\bibfnamefont
  {D.}~\bibnamefont {Loss}}, \bibinfo {author} {\bibfnamefont {T.}~\bibnamefont
  {Ihn}},\ and\ \bibinfo {author} {\bibfnamefont {K.}~\bibnamefont {Ensslin}},\
  }\bibfield  {title} {\bibinfo {title} {Characterization of spin-orbit
  interactions of gaas heavy holes using a quantum point contact},\ }\href
  {https://doi.org/10.1103/PhysRevLett.113.046801} {\bibfield  {journal}
  {\bibinfo  {journal} {Phys. Rev. Lett.}\ }\textbf {\bibinfo {volume} {113}},\
  \bibinfo {pages} {046801} (\bibinfo {year} {2014})}\BibitemShut {NoStop}%
\bibitem [{\citenamefont {Philippopoulos}\ \emph {et~al.}(2020)\citenamefont
  {Philippopoulos}, \citenamefont {Chesi}, \citenamefont {Culcer},\ and\
  \citenamefont {Coish}}]{Philippopoulos_PRB_2020}%
  \BibitemOpen
  \bibfield  {author} {\bibinfo {author} {\bibfnamefont {P.}~\bibnamefont
  {Philippopoulos}}, \bibinfo {author} {\bibfnamefont {S.}~\bibnamefont
  {Chesi}}, \bibinfo {author} {\bibfnamefont {D.}~\bibnamefont {Culcer}},\ and\
  \bibinfo {author} {\bibfnamefont {W.~A.}\ \bibnamefont {Coish}},\ }\bibfield
  {title} {\bibinfo {title} {Pseudospin-electric coupling for holes beyond the
  envelope-function approximation},\ }\href
  {https://doi.org/10.1103/PhysRevB.102.075310} {\bibfield  {journal} {\bibinfo
   {journal} {Physical Review B}\ }\textbf {\bibinfo {volume} {102}},\ \bibinfo
  {pages} {075310} (\bibinfo {year} {2020})}\BibitemShut {NoStop}%
\bibitem [{\citenamefont {Luo}\ \emph {et~al.}(2010)\citenamefont {Luo},
  \citenamefont {Chantis}, \citenamefont {van Schilfgaarde}, \citenamefont
  {Bester},\ and\ \citenamefont {Zunger}}]{Luo_PRL_2010}%
  \BibitemOpen
  \bibfield  {author} {\bibinfo {author} {\bibfnamefont {J.-W.}\ \bibnamefont
  {Luo}}, \bibinfo {author} {\bibfnamefont {A.~N.}\ \bibnamefont {Chantis}},
  \bibinfo {author} {\bibfnamefont {M.}~\bibnamefont {van Schilfgaarde}},
  \bibinfo {author} {\bibfnamefont {G.}~\bibnamefont {Bester}},\ and\ \bibinfo
  {author} {\bibfnamefont {A.}~\bibnamefont {Zunger}},\ }\bibfield  {title}
  {\bibinfo {title} {Discovery of a novel linear-in-\(k\) spin splitting for
  holes in the 2d gaas/alas system},\ }\href
  {https://doi.org/10.1103/PhysRevLett.104.066405} {\bibfield  {journal}
  {\bibinfo  {journal} {Physical Review Letters}\ }\textbf {\bibinfo {volume}
  {104}},\ \bibinfo {pages} {066405} (\bibinfo {year} {2010})}\BibitemShut
  {NoStop}%
\bibitem [{\citenamefont {Rodr{\'i}guez-Mena}\ \emph
  {et~al.}(2023)\citenamefont {Rodr{\'i}guez-Mena}, \citenamefont
  {Abadillo-Uriel}, \citenamefont {Veste}, \citenamefont {Martinez},
  \citenamefont {Li}, \citenamefont {Skl{\'e}nard},\ and\ \citenamefont
  {Niquet}}]{RodriguezMena_PRB_2023}%
  \BibitemOpen
  \bibfield  {author} {\bibinfo {author} {\bibfnamefont {E.~A.}\ \bibnamefont
  {Rodr{\'i}guez-Mena}}, \bibinfo {author} {\bibfnamefont {J.~C.}\ \bibnamefont
  {Abadillo-Uriel}}, \bibinfo {author} {\bibfnamefont {G.}~\bibnamefont
  {Veste}}, \bibinfo {author} {\bibfnamefont {B.}~\bibnamefont {Martinez}},
  \bibinfo {author} {\bibfnamefont {J.}~\bibnamefont {Li}}, \bibinfo {author}
  {\bibfnamefont {B.}~\bibnamefont {Skl{\'e}nard}},\ and\ \bibinfo {author}
  {\bibfnamefont {Y.-M.}\ \bibnamefont {Niquet}},\ }\bibfield  {title}
  {\bibinfo {title} {Linear-in-momentum spin-orbit interactions in planar
  ge/gesi heterostructures and spin qubits},\ }\href
  {https://doi.org/10.1103/PhysRevB.108.205416} {\bibfield  {journal} {\bibinfo
   {journal} {Physical Review B}\ }\textbf {\bibinfo {volume} {108}},\ \bibinfo
  {pages} {205416} (\bibinfo {year} {2023})}\BibitemShut {NoStop}%
\bibitem [{\citenamefont {Lawrie}\ \emph {et~al.}(2020)\citenamefont {Lawrie},
  \citenamefont {Hendrickx}, \citenamefont {van Riggelen}, \citenamefont
  {Russ}, \citenamefont {Petit}, \citenamefont {Sammak}, \citenamefont
  {Scappucci},\ and\ \citenamefont {Veldhorst}}]{Lawrie_NanoLett_2020}%
  \BibitemOpen
  \bibfield  {author} {\bibinfo {author} {\bibfnamefont {W.~I.~L.}\
  \bibnamefont {Lawrie}}, \bibinfo {author} {\bibfnamefont {N.~W.}\
  \bibnamefont {Hendrickx}}, \bibinfo {author} {\bibfnamefont {F.}~\bibnamefont
  {van Riggelen}}, \bibinfo {author} {\bibfnamefont {M.}~\bibnamefont {Russ}},
  \bibinfo {author} {\bibfnamefont {L.}~\bibnamefont {Petit}}, \bibinfo
  {author} {\bibfnamefont {A.}~\bibnamefont {Sammak}}, \bibinfo {author}
  {\bibfnamefont {G.}~\bibnamefont {Scappucci}},\ and\ \bibinfo {author}
  {\bibfnamefont {M.}~\bibnamefont {Veldhorst}},\ }\bibfield  {title} {\bibinfo
  {title} {Spin relaxation benchmarks and individual qubit addressability for
  holes in quantum dots},\ }\href
  {https://doi.org/10.1021/acs.nanolett.0c02589} {\bibfield  {journal}
  {\bibinfo  {journal} {Nano Letters}\ }\textbf {\bibinfo {volume} {20}},\
  \bibinfo {pages} {7237} (\bibinfo {year} {2020})}\BibitemShut {NoStop}%
\bibitem [{\citenamefont {Liles}\ \emph {et~al.}(2018)\citenamefont {Liles},
  \citenamefont {Li}, \citenamefont {Yang}, \citenamefont {Hudson},
  \citenamefont {Veldhorst}, \citenamefont {Dzurak},\ and\ \citenamefont
  {Hamilton}}]{Liles_NatCommun_2018}%
  \BibitemOpen
  \bibfield  {author} {\bibinfo {author} {\bibfnamefont {S.~D.}\ \bibnamefont
  {Liles}}, \bibinfo {author} {\bibfnamefont {R.}~\bibnamefont {Li}}, \bibinfo
  {author} {\bibfnamefont {C.~H.}\ \bibnamefont {Yang}}, \bibinfo {author}
  {\bibfnamefont {F.~E.}\ \bibnamefont {Hudson}}, \bibinfo {author}
  {\bibfnamefont {M.}~\bibnamefont {Veldhorst}}, \bibinfo {author}
  {\bibfnamefont {A.~S.}\ \bibnamefont {Dzurak}},\ and\ \bibinfo {author}
  {\bibfnamefont {A.~R.}\ \bibnamefont {Hamilton}},\ }\bibfield  {title}
  {\bibinfo {title} {Spin and orbital structure of the first six holes in a
  silicon metal-oxide-semiconductor quantum dot},\ }\href
  {https://doi.org/10.1038/s41467-018-05700-9} {\bibfield  {journal} {\bibinfo
  {journal} {Nature Communications}\ }\textbf {\bibinfo {volume} {9}},\
  \bibinfo {pages} {3255} (\bibinfo {year} {2018})}\BibitemShut {NoStop}%
\bibitem [{\citenamefont {John}\ \emph {et~al.}(2025)\citenamefont {John},
  \citenamefont {Yu}, \citenamefont {van Straaten}, \citenamefont
  {Rodr{\'i}guez-Mena}, \citenamefont {Rodr{\'i}guez}, \citenamefont
  {Oosterhout}, \citenamefont {Stehouwer}, \citenamefont {Scappucci},
  \citenamefont {Rimbach-Russ}, \citenamefont {Bosco}, \citenamefont {Borsoi},
  \citenamefont {Niquet},\ and\ \citenamefont
  {Veldhorst}}]{John_NatCommun_2025_10SpinGe}%
  \BibitemOpen
  \bibfield  {author} {\bibinfo {author} {\bibfnamefont {V.}~\bibnamefont
  {John}}, \bibinfo {author} {\bibfnamefont {C.~X.}\ \bibnamefont {Yu}},
  \bibinfo {author} {\bibfnamefont {B.}~\bibnamefont {van Straaten}}, \bibinfo
  {author} {\bibfnamefont {E.~A.}\ \bibnamefont {Rodr{\'i}guez-Mena}}, \bibinfo
  {author} {\bibfnamefont {M.}~\bibnamefont {Rodr{\'i}guez}}, \bibinfo {author}
  {\bibfnamefont {S.~D.}\ \bibnamefont {Oosterhout}}, \bibinfo {author}
  {\bibfnamefont {L.~E.~A.}\ \bibnamefont {Stehouwer}}, \bibinfo {author}
  {\bibfnamefont {G.}~\bibnamefont {Scappucci}}, \bibinfo {author}
  {\bibfnamefont {M.}~\bibnamefont {Rimbach-Russ}}, \bibinfo {author}
  {\bibfnamefont {S.}~\bibnamefont {Bosco}}, \bibinfo {author} {\bibfnamefont
  {F.}~\bibnamefont {Borsoi}}, \bibinfo {author} {\bibfnamefont {Y.-M.}\
  \bibnamefont {Niquet}},\ and\ \bibinfo {author} {\bibfnamefont
  {M.}~\bibnamefont {Veldhorst}},\ }\bibfield  {title} {\bibinfo {title}
  {Robust and localised control of a 10-spin qubit array in germanium},\ }\href
  {https://doi.org/10.1038/s41467-025-65577-3} {\bibfield  {journal} {\bibinfo
  {journal} {Nature Communications}\ }\textbf {\bibinfo {volume} {16}},\
  \bibinfo {pages} {10560} (\bibinfo {year} {2025})}\BibitemShut {NoStop}%
\bibitem [{\citenamefont {Burkard}\ and\ \citenamefont
  {Petta}(2016)}]{Guido_prb_2016_valley}%
  \BibitemOpen
  \bibfield  {author} {\bibinfo {author} {\bibfnamefont {G.}~\bibnamefont
  {Burkard}}\ and\ \bibinfo {author} {\bibfnamefont {J.~R.}\ \bibnamefont
  {Petta}},\ }\bibfield  {title} {\bibinfo {title} {Dispersive readout of
  valley splittings in cavity-coupled silicon quantum dots},\ }\href
  {https://doi.org/10.1103/PhysRevB.94.195305} {\bibfield  {journal} {\bibinfo
  {journal} {Phys. Rev. B}\ }\textbf {\bibinfo {volume} {94}},\ \bibinfo
  {pages} {195305} (\bibinfo {year} {2016})}\BibitemShut {NoStop}%
\bibitem [{\citenamefont {Gardiner}\ and\ \citenamefont
  {Collett}(1985)}]{Gardiner_Collett_1985}%
  \BibitemOpen
  \bibfield  {author} {\bibinfo {author} {\bibfnamefont {C.~W.}\ \bibnamefont
  {Gardiner}}\ and\ \bibinfo {author} {\bibfnamefont {M.~J.}\ \bibnamefont
  {Collett}},\ }\bibfield  {title} {\bibinfo {title} {Input and output in
  damped quantum systems: Quantum stochastic differential equations and the
  master equation},\ }\href {https://doi.org/10.1103/PhysRevA.31.3761}
  {\bibfield  {journal} {\bibinfo  {journal} {Physical Review A}\ }\textbf
  {\bibinfo {volume} {31}},\ \bibinfo {pages} {3761} (\bibinfo {year}
  {1985})}\BibitemShut {NoStop}%
\bibitem [{\citenamefont {Gardiner}\ and\ \citenamefont
  {Zoller}(2004)}]{Gardiner_Zoller_2004}%
  \BibitemOpen
  \bibfield  {author} {\bibinfo {author} {\bibfnamefont {C.~W.}\ \bibnamefont
  {Gardiner}}\ and\ \bibinfo {author} {\bibfnamefont {P.}~\bibnamefont
  {Zoller}},\ }\href@noop {} {\emph {\bibinfo {title} {Quantum Noise: A
  Handbook of Markovian and Non-Markovian Quantum Stochastic Methods with
  Applications to Quantum Optics}}},\ \bibinfo {edition} {3rd}\ ed.\ (\bibinfo
  {publisher} {Springer},\ \bibinfo {address} {Berlin},\ \bibinfo {year}
  {2004})\BibitemShut {NoStop}%
\bibitem [{\citenamefont {Feshbach}(1958)}]{Feshbach_1958}%
  \BibitemOpen
  \bibfield  {author} {\bibinfo {author} {\bibfnamefont {H.}~\bibnamefont
  {Feshbach}},\ }\bibfield  {title} {\bibinfo {title} {Unified theory of
  nuclear reactions},\ }\href {https://doi.org/10.1016/0003-4916(58)90007-1}
  {\bibfield  {journal} {\bibinfo  {journal} {Annals of Physics}\ }\textbf
  {\bibinfo {volume} {5}},\ \bibinfo {pages} {357} (\bibinfo {year}
  {1958})}\BibitemShut {NoStop}%
\bibitem [{\citenamefont {Bastard}(1988)}]{bastard1988wave}%
  \BibitemOpen
  \bibfield  {author} {\bibinfo {author} {\bibfnamefont {G.}~\bibnamefont
  {Bastard}},\ }\href@noop {} {\emph {\bibinfo {title} {Wave Mechanics Applied
  to Semiconductor Heterostructures}}}\ (\bibinfo  {publisher} {Les
  {\'E}ditions de Physique},\ \bibinfo {address} {Les Ulis},\ \bibinfo {year}
  {1988})\BibitemShut {NoStop}%
\bibitem [{\citenamefont {Davies}(1998)}]{davies1998physics}%
  \BibitemOpen
  \bibfield  {author} {\bibinfo {author} {\bibfnamefont {J.~H.}\ \bibnamefont
  {Davies}},\ }\href@noop {} {\emph {\bibinfo {title} {The Physics of
  Low-Dimensional Semiconductors: An Introduction}}}\ (\bibinfo  {publisher}
  {Cambridge University Press},\ \bibinfo {address} {Cambridge},\ \bibinfo
  {year} {1998})\BibitemShut {NoStop}%
\bibitem [{\citenamefont {BenDaniel}\ and\ \citenamefont
  {Duke}(1966)}]{bendandaniel1966space}%
  \BibitemOpen
  \bibfield  {author} {\bibinfo {author} {\bibfnamefont {D.~J.}\ \bibnamefont
  {BenDaniel}}\ and\ \bibinfo {author} {\bibfnamefont {C.~B.}\ \bibnamefont
  {Duke}},\ }\bibfield  {title} {\bibinfo {title} {Space-charge effects on
  electron tunneling},\ }\href {https://doi.org/10.1103/PhysRev.152.683}
  {\bibfield  {journal} {\bibinfo  {journal} {Physical Review}\ }\textbf
  {\bibinfo {volume} {152}},\ \bibinfo {pages} {683} (\bibinfo {year}
  {1966})}\BibitemShut {NoStop}%
\end{thebibliography}
%

\end{document}